\pdfoutput=1
\documentclass[11pt]{article}
\usepackage{caption}
\usepackage{subcaption}
\usepackage{graphicx}
\newcommand{\FMDG}[1]{\includegraphics{#1}}

\usepackage{amsfonts}
\usepackage{amsmath,amssymb,amsthm}
\usepackage{mathrsfs}
\usepackage{float}
\usepackage{booktabs}
\usepackage{mathtools}

\usepackage[utf8]{inputenc}
\usepackage{caption,xspace}

\numberwithin{equation}{section}

\usepackage[sort&compress,numbers,colon]{natbib}
\usepackage{multirow}
\usepackage{color}
\usepackage[colorlinks,citecolor=blue]{hyperref}

\usepackage{verbatim}
\usepackage{float}
\usepackage{tikz}
\usetikzlibrary{shapes,arrows}
\newcommand{\be}{\begin{equation}}
\newcommand{\ee}{\end{equation}}
\newcommand{\bea}{\begin{eqnarray}}
\newcommand{\eea}{\end{eqnarray}}

\def\4vol{{\int d^4x \sqrt{-g}}}

\def\beq{\begin{equation}}
\def\eeq{\end{equation}}
\def\bea{\begin{eqnarray}}
\def\eea{\end{eqnarray}}
\def\bitem{\begin{itemize}}
\def\eitem{\end{itemize}}

\newcommand{\eV}{\ensuremath{\,\mathrm{eV}}}

\newcommand{\MeV}{\ensuremath{\,\mathrm{MeV}}}
\newcommand{\GeV}{\ensuremath{\,\mathrm{GeV}}}
\newcommand{\TeV}{\ensuremath{\,\mathrm{TeV}}}

\newcommand{\nc}{\newcommand}

\definecolor{MS}{rgb}{1,0,0}

\nc{\nt}{\tilde{N}}
\nc{\ra}{\rightarrow}
\nc{\lsim}{\begin{array}{c}\,\sim\vspace{-21pt}\\< \end{array}}
\nc{\gsim}{\begin{array}{c}\sim\vspace{-21pt}\\> \end{array}}
\nc{\tnt}{\tilde{N}}
\nc{\tst}{\tilde{t}}

\nc{\LL}{L}
\nc{\vv}{\tilde{v}}
\newcommand{\footlabel}[2]{%
    \addtocounter{footnote}{1}%
    \footnotetext[\thefootnote]{%
        \addtocounter{footnote}{-1}%
        \refstepcounter{footnote}\label{#1}%
        #2%
    }%
    $^{\ref{#1}}$%
}

\newcommand{\footref}[1]{%
    $^{\ref{#1}}$%
}

\title{}

\begin{document}
\allowdisplaybreaks[1]

\begin{titlepage}
\rightline{\sc\small UMN-TH-3419/15}
\vspace{1cm}

\begin{center}
{\LARGE\textbf{SUSY Implications from WIMP Annihilation \\into Scalars at the Galactic Centre}}
\\[10mm]
\vspace{.2cm}
{\normalsize\sc
Tony Gherghetta$^{a,}$\footnote{\texttt{tgher@umn.edu}},
Benedict von Harling$^{b,}$\footnote{\texttt{bharling@sissa.it}},
Anibal D.~Medina$^{c,}$\footnote{\texttt{anibal.medina@unimelb.edu.au}},\\
Michael A.~Schmidt$^{d,}$\footnote{\texttt{m.schmidt@physics.usyd.edu.au}}
 and Timothy Trott$^{c,}$\footnote{\texttt{t.trott@student.unimelb.edu.au}}}
\\[5mm]
{\small\textit{
$^{a}$School of Physics $\&$ Astronomy, University of Minnesota, Minneapolis, MN 55455, USA\\
$^{b}$SISSA and INFN, Via Bonomea 265, 34136 Trieste, Italy\\
$^{c}$ARC Centre of Excellence for Particle Physics at the Terascale,\\
School of Physics, The University of Melbourne, Victoria 3010, Australia\\
$^{d}$ARC Centre of Excellence for Particle Physics at the Terascale,\\
School of Physics, The University of Sydney, NSW 2006, Australia }
}
\end{center}

\vspace*{0.2cm}
\date{\today}

\begin{abstract}
\noindent An excess in $\gamma$-rays emanating from the galactic centre has recently been observed in the Fermi-LAT data. 
This signal can be interpreted as resulting from WIMP annihilation, with the spectrum well-fit by dark matter annihilating dominantly into either bottom-quark or Higgs pairs. 
Supersymmetric models provide a well-motivated framework to study the implications of this signal in these channels. With a neutralino dark matter candidate, the $\gamma$-ray excess cannot be easily accommodated in the minimal supersymmetric model, which in any case requires tuning below the percent level to explain the observed Higgs mass. Instead we are naturally led to consider the next-to-minimal model with a singlet superfield. 
This not only allows for the annihilation channel into bottom-quark pairs to be implemented, but also provides new possibilities for annihilation into Higgs-pseudoscalar pairs.
We show that the fit to the $\gamma$-ray excess for the Higgs-pseudoscalar channel can be just as good as for annihilation into bottom-quark pairs.
Moreover, in the parameter range of interest, the next-to-minimal supersymmetric model solves the $\mu$-problem and can explain the 125 GeV Higgs mass with improved naturalness. We also consider an extension by adding a right-handed neutrino superfield with the right-handed sneutrino acting as a dark matter candidate. Interestingly, this allows for the annihilation into pseudoscalar pairs which also provide a good fit to the $\gamma$-ray excess. Furthermore, in the case of a neutralino LSP, the late decay of a sneutrino NLSP can non-thermally produce the observed relic abundance. Finally, the WIMP annihilation into scalar pairs allows for the possibility of detecting the Higgs or pseudoscalar decay into two photons, providing a smoking-gun signal of the model.

\end{abstract}

\end{titlepage}

\setcounter{footnote}{0}

\section{Introduction}

Cosmic-ray experiments are a promising way to search for dark matter (DM). In particular, the satellite-based experiment Fermi-LAT is able to measure the $\gamma$-ray sky with unprecedented precision. If DM annihilates into photons with energies from $20 \MeV$ to $300 \GeV$, an imprint can be left on these measurements. However, the astrophysical $\gamma$-ray background has to be carefully modelled in order to disentangle a putative DM signal. 
To be consistent with DM, any excess over this background needs to have a morphology which is  consistent with the range of DM density profiles indicated by observations and simulations of structure formation.

Intriguingly, an excess from the galactic centre with such a morphology has been identified in the Fermi-LAT data~\cite{Goodenough:2009gk,Hooper:2010mq,Hooper:2011ti,Abazajian:2012pn,Hooper:2013rwa,Huang:2013pda,Gordon:2013vta,Abazajian:2014fta,Daylan:2014rsa,Calore:2014xka}. 
The spectrum of this Galactic Center Excess (GCE) peaks at energies of about $3 \GeV$. It can be well fit by DM with mass in the range $30 - 70 \GeV$ which annihilates dominantly into $b \bar b$-pairs (or lighter quarks).
Interestingly, the annihilation cross section that is then required to explain the GCE is of the right size to account for the DM density from thermal freeze-out.
The systematic uncertainties of the spectrum from modelling the astrophysical $\gamma$-ray background have recently been studied in Refs.~\cite{Calore:2014xka, Calore:2014nla}.  It was found that, taking the estimated uncertainty in the high-energy tail of the spectrum into account, DM annihilating to Higgs pairs close to threshold also gives a good fit~\cite{Calore:2014nla,Agrawal:2014oha}.  Moreover, the Fermi-LAT collaboration has recently confirmed the GCE and presented preliminary spectra (although without information on the errors)~\cite{MurgiaTalk:2014FermiSymposium}. It was subsequently pointed out~\cite{Agrawal:2014oha} that this may also allow for annihilation into top-, $W$- and $Z$-pairs to reproduce the spectrum well.
However, care must be taken in interpreting these results as they are still preliminary and little detail of the modelling and uncertainties have been publicly released. We will therefore focus on the best-fit channels, to $b \bar b$- and Higgs-pairs (and possibly other scalars), for the spectrum from Ref.~\cite{Calore:2014nla}. It is then interesting to study well-motivated DM models in which these channels arise naturally (see e.g.~Refs.~\cite{Huang:2013apa,Modak:2013jya,Alves:2014yha,Cheung:2014lqa,Huang:2014cla,Cahill-Rowley:2014ora,Ghorbani:2014qpa,Modak:2015uda} for earlier studies of DM models to explain the GCE).

A weakly interacting massive particle (WIMP) has the right properties to account for the observed DM from thermal freeze-out and is naturally obtained in supersymmetric models with $R$-parity. 
An excellent DM candidate is the lightest neutralino. The annihilation cross section of neutralinos into Higgs-pairs is $p$-wave suppressed~\cite{Griest:1989zh}. Due to the small velocities in our galaxy, this is therefore not large enough to generate the GCE. We will instead consider the related annihilation channel of a Higgs and a Higgs-sector pseudoscalar, which proceeds at the $s$-wave level (for two pseudoscalars in the final state, the cross section is again $p$-wave suppressed). Since this channel has not been considered before (with the exception of Ref.~\cite{Berlin:2014pya} which, however, focused on a singlet-like scalar instead of the Higgs in the final state), we have performed a fit using the spectra and error estimates from Ref.~\cite{Calore:2014xka}. For comparison, we have also performed fits for the annihilation into $b \bar b$- and Higgs-pairs. We find that an acceptable fit is obtained for pseudoscalar masses up to around 150 $\GeV$. In particular, in the region where the pseudoscalar is lighter than the Higgs, the fit  
is improved compared to the annihilation channel into Higgs-pairs. For a sufficiently light pseudoscalar, the fit becomes even better than that for the $b \bar b$-channel.

Similarly a light pseudoscalar is required to mediate the $b \bar b$-channel. Even though the annihilation of neutralinos into quarks can be mediated by $t/u$-channel exchange of squarks, the cross section for this process is typically not large enough to generate the GCE due to collider constraints on the squark masses~\cite{Han:2014nba}. In the $s$-channel, possible mediators are the $Z$-boson and $CP$-even and $CP$-odd scalars from the Higgs sector. The $Z$-boson and $CP$-even scalars lead to annihilation cross sections that are respectively helicity and $p$-wave suppressed and can thus not significantly contribute to the GCE either. Higgs-sector pseudoscalars will therefore be the dominant mediators in the $s$-channel. As was previously observed~\cite{Cheung:2014lqa,Berlin:2015wwa}, the corresponding annihilation cross section needs to be resonantly enhanced in order to generate a $\gamma$-ray flux consistent with the GCE. Given that the DM mass that best fits the GCE for the $b \bar b$-channel is in the range $30 - 70 \GeV$~\cite{Goodenough:2009gk,Hooper:2010mq,Hooper:2011ti,Abazajian:2012pn,Hooper:2013rwa,Huang:2013pda,Gordon:2013vta,Abazajian:2014fta,Daylan:2014rsa,Calore:2014xka}, this means that the pseudoscalar mediator needs to be lighter than about $140 \GeV$.

The light pseudoscalar required for the Higgs-pseudoscalar and $b \bar b$-channels is, however, difficult to obtain in the minimal supersymmetric standard model (MSSM).\footnote{The $WW$-channel was recently studied in the context of the MSSM in Ref.~\cite{Caron:2015wda}.
} 
Indeed, CMS excludes an MSSM Higgs sector with a pseudoscalar mass below $ 180 \GeV$ for values of $\tan \beta$ required to obtain the observed Higgs mass for stop masses around $ 1 \TeV$~\cite{Khachatryan:2014wca}. Even though this bound may be relaxed for heavier stops (see, however, Ref.~\cite{Djouadi:2015jea}), this leads to excessive fine-tuning (which is already high in the MSSM). In addition, a light pseudoscalar implies that also the second $CP$-even Higgs and the charged Higgs are light. This results in large deviations in the Higgs couplings from standard model (SM) values (see e.g.~Refs.~\cite{Gherghetta:2014xea,Carena:2014nza}) and large contributions to flavour-changing processes (see e.g.~Ref.~\cite{Eriksson:2008cx}), respectively, both in conflict with experiment.
There may be ways out to avoid these constraints but they would require accidental cancellations and thus be likely contrived.

Given these problems in the MSSM, we will extend the Higgs sector by a singlet superfield and consider the next-to-minimal supersymmetric standard model (NMSSM). The lightest pseudoscalar is then generically an admixture of the MSSM pseudoscalar and the $CP$-odd component of the singlet. The mass of the lightest pseudoscalar can then be decoupled from the masses of the (doublet-like) $CP$-even and charged Higgs states, allowing the collider constraints in the MSSM to be sufficiently 
relaxed. In addition, the coupling of the Higgs to two pseudoscalars has new contributions in the NMSSM, making it easier to obtain a sufficiently large annihilation cross section for the Higgs-pseudoscalar channel. 
The NMSSM has two further advantages. Firstly, we will consider a superpotential with only dimensionless couplings. The $\mu$-term is then dynamically generated when the singlet scalar obtains a vacuum expectation value (VEV), solving the $\mu$-problem. Secondly, the superpotential coupling between the Higgs and the singlet gives an additional contribution to the Higgs quartic coupling that raises the Higgs mass at tree level. This alleviates the need for large stop-sector soft masses to increase the Higgs mass, improving the naturalness of the model. In order to obtain the observed Higgs mass, the Higgs-singlet coupling needs to be quite large, $\lambda \gtrsim 0.7$. We will focus on this regime, although smaller values of $\lambda$ can also provide different accounts of the GCE (at the expense of naturalness).

Generically, the lightest neutralino in the NMSSM is an admixture of electroweak gauginos, Higgsinos and the singlino. Higgsino- and wino-dominated lightest supersymmetric particles (LSPs) annihilate too efficiently to be produced as thermal relics via freeze-out in the mass range of interest for the GCE, leaving only bino- and singlino-dominated compositions as viable. The Higgsino fractions of these species are strongly constrained 
by direct-detection experiments.
For light enough neutralinos, these are similarly constrained by the invisible decay width of the Higgs. We will derive approximate analytical expressions for the resulting bounds on the Higgsino fractions. 
In order to explore the region of parameter space where the NMSSM can explain the GCE, we have performed numerical scans using the programs \texttt{NMSSMTools\;4.4.0}~\cite{Ellwanger:2004xm,Ellwanger:2005dv,Belanger:2005kh} and \texttt{micrOMEGAs\;3.0}~\cite{Belanger:2005kh}. For the $b \bar b$-channel, we focus on bino-dominated neutralinos. Anticipating that an enhancement of the cross section is required to generate a sufficiently large $\gamma$-ray flux, the search is optimised for pseudoscalar mediators which are close to resonance in the $s$-channel. For the Higgs-pseudoscalar channel, we consider both bino- and singlino-dominated neutralinos. Our scans find regions of parameter space where the GCE can be generated via these channels. 
We study how these regions will be tested by upcoming direct-detection experiments and improvements in the measurement of the invisible Higgs decay width.

We also consider a model that gives rise to DM annihilation into pairs of $CP$-even or $CP$-odd scalars. 
A simple extension of the NMSSM is the introduction of right-handed neutrino superfields and a scale-invariant term in the superpotential that couples them to the singlet~\cite{Cerdeno:2009dv}. We first consider the lightest right-handed sneutrino as the DM candidate (the corresponding cross sections are no longer $p$-wave suppressed because the DM is now a scalar particle). 
The superpotential term gives rise to a four-point scalar coupling that allows for the direct annihilation of the sneutrino into two Higgs or Higgs-sector pseudoscalars, in addition to $s$- and $t/u$-channel mediated processes. We find that the annihilation into two Higgs is typically subdominant compared to two pseudoscalars. We have performed a fit to the GCE for the latter final state, and find that the fit is again improved for pseudoscalar masses below the Higgs mass compared to the annihilation channel to Higgs-pairs.
The sneutrino also offers the possibility of annihilation into light quarks as has been discussed in Ref.~\cite{Cerdeno:2014cda} but we will not consider this case further. Our work complements the recent analysis in Ref.~\cite{Cerdeno:2015ega}, 
which performed a scan in a similar right-handed neutrino model, but for smaller values of $\lambda$.

Furthermore, we consider the possibility that the neutralino is the LSP and the sneutrino is the next-to-LSP (NLSP). As mentioned previously, the neutralino annihilation needs to occur on resonance for the $b \bar b$-channel in order to obtain a sufficiently large annihilation cross section. Due to thermal broadening of the resonance at high temperatures, the cross sections at freeze-out and in our galaxy are often vastly different. For many points in parameter space, the abundance of the neutralino is then too small to account for the DM. Due to its larger mass and interactions, the sneutrino freezes out before the neutralino. If the sneutrino is metastable and decays long after the neutralino freezes out, then the sneutrino decays can replenish the neutralino abundance. This decouples the generation of the observed DM abundance from the explanation of the GCE. Points in parameter space with an otherwise underabundant LSP may therefore become viable once these NLSP decays are included in the model.

The WIMP annihilation in our non-minimal supersymmetric model includes the new Higgs-pseudoscalar and pseudoscalar-pseudoscalar channels that can provide just as good a fit to the GCE as the Higgs- and $b \bar b$-channels. Both the Higgs and the pseudoscalar can decay into two photons. Since they need to be produced close to threshold in order to give a good fit, this may lead to detectable $\gamma$-ray lines in the spectrum (as previously described in Refs.~\cite{Calore:2014nla,Agrawal:2014oha,Cerdeno:2015ega}). The branching fraction of the pseudoscalar into two photons is suppressed compared to that for the Higgs. Although we find that the $\gamma$-ray line from pseudoscalar decays is distinguishable from the continuum only in an optimal-case scenario, more sensitive $\gamma$-ray experiments in the future may be able to detect it. Searching for these lines in the $\gamma$-ray spectrum could provide a smoking-gun signal of the new channels.

The paper is organised as follows. In Sec.~\ref{GCE-section} we summarise the different analyses of the GCE and perform our own fit for the Higgs-pseudoscalar and pseudoscalar-pseudoscalar channels. After a discussion of the LSP composition and suitable mediators for the DM annihilation in Sec.~\ref{CGENMSSM}, we present the results of our parameter scans in Sec.~\ref{ParameterScans}. In Sec.~\ref{NLSPdecays} we introduce an extension of the NMSSM with right-handed neutrinos. We first discuss how sneutrino LSPs can explain the GCE and then consider an interesting non-thermal production mechanism of neutralino LSPs from late decays of sneutrino NLSPs.
Finally, we conclude in Sec.~\ref{Conclusions} and list the expressions of a few relevant couplings in the Appendix.

\section{The $\gamma$-ray excess from the galactic centre}\label{CGESection}
\label{GCE-section}

Several analyses~\cite{Goodenough:2009gk,Hooper:2010mq,Hooper:2011ti,Abazajian:2012pn,Hooper:2013rwa,Huang:2013pda,Gordon:2013vta,Abazajian:2014fta,Daylan:2014rsa,Calore:2014xka} of the Fermi-LAT data have found an excess in $\gamma$-rays over known astrophysical backgrounds which originates from the galactic centre and peaks at energies 1-3 GeV. This excess has recently also been confirmed by the Fermi-LAT collaboration~\cite{MurgiaTalk:2014FermiSymposium}. 

There have been several hypothesised explanations for this excess which have a purely astrophysical origin. 
One possible source is a previously unidentified population of millisecond pulsars in the galactic centre. 
The resulting spectrum is described by a power-law with an exponential cutoff and was found to fit the GCE well in Refs.~\cite{Gordon:2013vta,Abazajian:2014fta,Yuan:2014rca}. However, this explanation was disfavoured in the more recent analysis of Ref.~\cite{Calore:2014xka} compared to DM annihilation (see below) or a source with a broken power-law spectrum. Furthermore, Ref.~\cite{Hooper:2013nhl} 
found that it is difficult to explain the spatial extent of the GCE with a population of millisecond pulsars while Ref.~\cite{Cholis:2014lta} argued that they can only account for 5\% of the GCE. Another proposed explanation is that the signal originates from a transient cosmic-ray injection~\cite{Carlson:2014cwa,Petrovic:2014uda}. Even though astrophysics may ultimately be responsible for the excess, we will not consider these possibilities further.

We will instead consider the exciting possibility that the GCE is 
produced by the annihilation of DM. As we have mentioned in the introduction, we will focus on the annihilation channels of $b \bar b$-pairs,  Higgs-pseudoscalar pairs and pseudoscalar pairs, where the pseudoscalars subsequently decay into $b \bar b$-pairs. The latter two channels arise quite naturally in the NMSSM and the NMSSM with right-handed neutrinos, respectively. In Table~\ref{tab:data}, we summarise the best-fit values for the DM mass and annihilation cross section for the $b \bar b$-channel from different analyses\footlabel{RoddTalk}{The inner-galaxy analysis in the original version of Ref.~\cite{Daylan:2014rsa} was subject to an error. In the updated analysis, the best-fit mass and cross-section for the $b \bar b$-channel shift by $\sim 10 \%$ to $40 \GeV$ and $1.8 \times 10^{-26} \text{cm}^3/\text{s}$, respectively, with $\gamma=1.28$ \cite{RoddTalk:2014TeVPA}. Since the shift is small and the updated best-fit regions are still preliminary, we will list the regions from the original version of \cite{Daylan:2014rsa} in Table~\ref{tab:data}.} of the Fermi-LAT data (see \cite{Goodenough:2009gk,Hooper:2010mq,Hooper:2011ti,Abazajian:2012pn} for earlier studies). Here and below, we denote the annihilation cross section to particles $x$ and $y$ in our galaxy (i.e.~at small velocities) by $\langle \sigma_{xy} v \rangle_0$.
Note that assumptions about the parameters of the DM halo profile differ for the different analyses and thus the masses and cross sections cannot be directly compared. These values are nevertheless indicative of the target masses and cross sections that DM models must satisfy in order to explain the GCE.

\begin{table}[btp]\centering
\begin{tabular}{lccccc}
\toprule
& $m_{_{\rm DM}}$ [GeV] & $\langle \sigma_{b \bar b} v \rangle_0$ [$10^{-26} \frac{\mathrm{cm}^3}{\mathrm{s}}$] & $\gamma$ & $\rho_\odot$ [$\frac{\mathrm{GeV}}{\mathrm{cm}^3}$] & $R_s$ [kpc]\\
\midrule
Hooper+\cite{Hooper:2013rwa} & $\sim 40-50$ & $\sim 0.8$ & $1.2$ & $0.4$ & $20$\\
Huang+\cite{Huang:2013pda} & $61.8^{+6.9}_{-4.9}$ & $3.30^{+0.69}_{-0.49}$& $1.2$ & $0.4$ & $20$\\
Gordon+\cite{Gordon:2013vta} & $34.1^{+4.0}_{-3.5}$ & $2.47^{+0.28}_{-0.25}$ & $1.2$ & $0.36$ & $23.1$\\
Abazajian+\cite{Abazajian:2014fta} & $39.4^{+3.7}_{-2.9}\pm7.9$ (sys) & $5.1\pm2.4$ & $1.1$ & $0.3$ & $23.1$\\
Daylan+\cite{Daylan:2014rsa}\footref{RoddTalk} & $31-40$ & $0.7-3.9$ & $1.26$ &$0.3$& $20$\\
Calore+\cite{Calore:2014xka} & $49^{+6.4}_{-5.4}$ & $1.76^{+0.28}_{-0.27}$ & $1.2$  & $0.4$ & $20$\\
\bottomrule
\end{tabular}
\caption{Results from different fits to the GCE assuming a 100\% branching fraction 
to $b \bar b$-pairs and the generalised NFW profile~\cite{Navarro:1995iw,Klypin:2001xu} (see~Eq.~\eqref{genNFW}) with the listed parameters. We only give results for the ``inner galaxy'' analysis of \cite{Daylan:2014rsa}. Caution should be taken in comparing results as different methods have been employed in each study for background modelling, data selection and the combining of uncertainties. Note that $\gamma$ is fit to the spectrum in \cite{Abazajian:2014fta}, \cite{Daylan:2014rsa} and \cite{Calore:2014xka}, while it has been simply fixed in the earlier studies. For \cite{Calore:2014xka}, however, $\gamma$ listed here is the fixed value for which the fit to the mass and cross-section has been performed, while the authors find that $\gamma=1.28$ provides a better fit for fixed $m_{_{\rm DM}}=49\,\GeV$. Uncertainties listed here for \cite{Huang:2013pda}, \cite{Gordon:2013vta} and \cite{Calore:2014xka} do not include those arising from the DM halo distribution.}
\label{tab:data}
\end{table}

We next discuss the annihilation channels to a Higgs-pseudoscalar pair and two pseudoscalars in the context of 
a two-Higgs-doublet model of type II with an arbitrary number of additional singlets. This includes the NMSSM which we focus on later. Annihilations into scalar-pseudoscalar pairs from the Higgs sector of the generalised NMSSM were previously considered in  Ref.~\cite{Berlin:2014pya}, where the scalar was singlet-dominated. We will instead be interested in the case where the scalar is the Higgs which we will denote as $h$. We fix its mass to $125 \GeV$ and its couplings to SM values, as implemented in \texttt{PYTHIA 8.201}~\cite{Sjostrand:2014zea}. We denote the pseudoscalar as $a$ and set $\tan\beta=3$ to fix its couplings (which is within the region of small $\tan \beta$ that we focus on later). Furthermore, we assume that neither the $a$ nor the $h$ can decay into other scalars. We have then performed our own fits to the reduced spectra of \cite{Calore:2014xka}. For completeness, we have also performed fits for the $b \bar b$- and $hh$-channels. We assume a generalised NFW profile~\cite{Navarro:1995iw,Klypin:2001xu}
\begin{equation}
\rho(r)=\rho_\odot \left(\frac{r}{r_\odot}\right)^{-\gamma}\left(\frac{1+r_\odot/R_s}{1+r/R_s}\right)^{3-\gamma}
\label{genNFW}
\end{equation}
with slope parameter $\gamma=1.26$, scale radius $R_s=20$ kpc and the DM density $\rho_\odot=0.4\, \mathrm{GeV}/\mathrm{cm}^3$ at the radial distance of the sun from the galactic centre $r_\odot$.
We use the prompt photon spectrum, $dN_\gamma/dE$, for annihilations into $b\bar b$ from \texttt{PPPC4MID}~\cite{Cirelli:2010xx} including electroweak corrections~\cite{Ciafaloni:2010ti} and find good agreement with our own simulation using \texttt{PYTHIA 8.201}~\cite{Sjostrand:2014zea}.  
We only consider the dominant decay channels of the pseudoscalar, to $b\bar b$, $\tau^+\tau^-$, $c\bar c$, photons and gluons, 
and simulate the resulting prompt photon spectra for $hh$, $ha$, and $aa$ final states using \texttt{PYTHIA 8.201}~\cite{Sjostrand:2014zea}. 
Note that these spectra are unaffected by a possible singlet admixture of the Higgs or pseudoscalar. Indeed, such an admixture reduces their total decay widths but the branching fractions remain unaffected (to leading order).
The differential flux measured by Fermi-LAT is given by
\begin{equation}\label{eq:dNdENFW}
\frac{dN}{dE}=\frac{\langle\sigma v\rangle_0}{8\pi \, m_{_{\rm DM}}^2} \frac{dN_\gamma^f}{dE} \int_{\rm l.o.s.} \hspace{-.2cm} ds\, \rho^2(r(s,\psi))
\end{equation}
with the line-of-sight integral $\int_{\rm l.o.s.}\hspace{-.1cm} ds$ over the squared DM density. The coordinate $r$ is centred on the galactic centre and can be expressed as $r^2(s,\psi)=r_\odot^2+s^2-2 r_\odot s \cos\psi$, where $s$ is the line-of-sight distance and $\psi$ is the aperture angle between the axis connecting the earth with the galactic centre and the line-of-sight. If DM annihilates into multiple final states, the different fluxes are summed over.
 
\begin{figure}[tbp]\centering
\begin{subfigure}{0.49\linewidth}
\includegraphics[width=\linewidth]{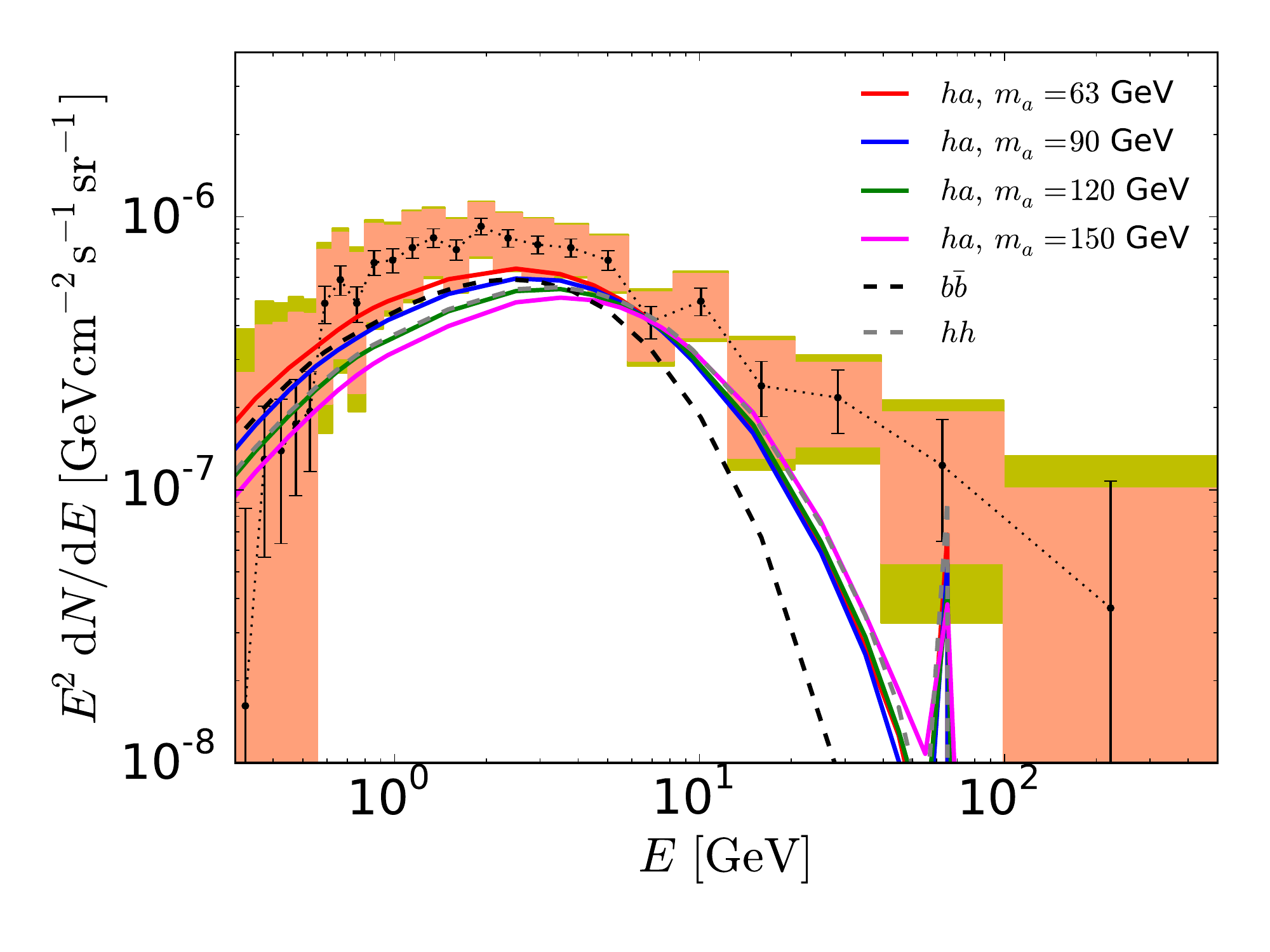}
\caption{\label{fig:fit-spectrum}} 
\end{subfigure}
\hfill
\begin{subfigure}{0.49\linewidth}
\includegraphics[width=\linewidth]{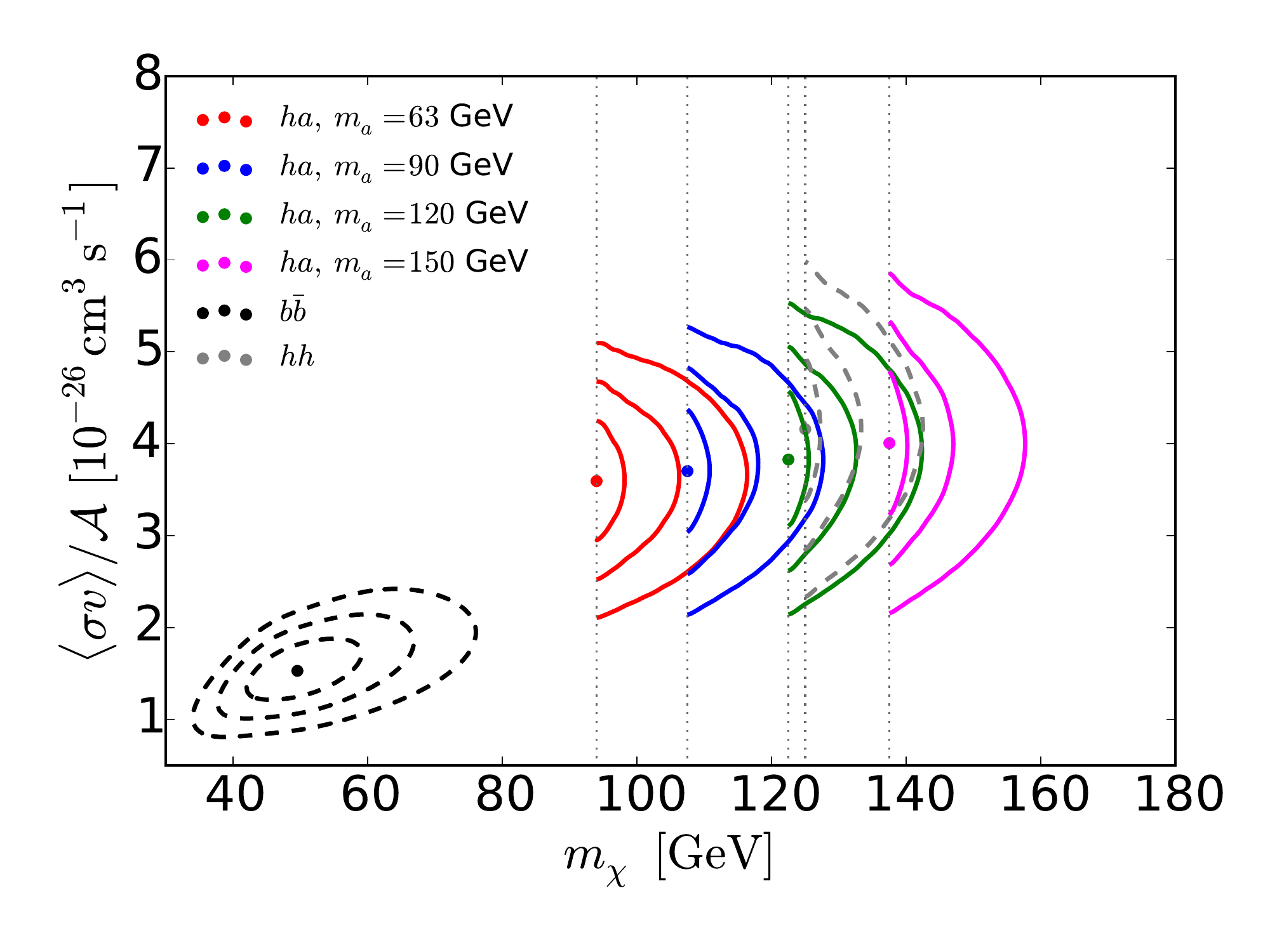}
\caption{\label{fig:fit-regions}} 
\end{subfigure}
\begin{subfigure}{0.49\linewidth}
\includegraphics[width=\linewidth]{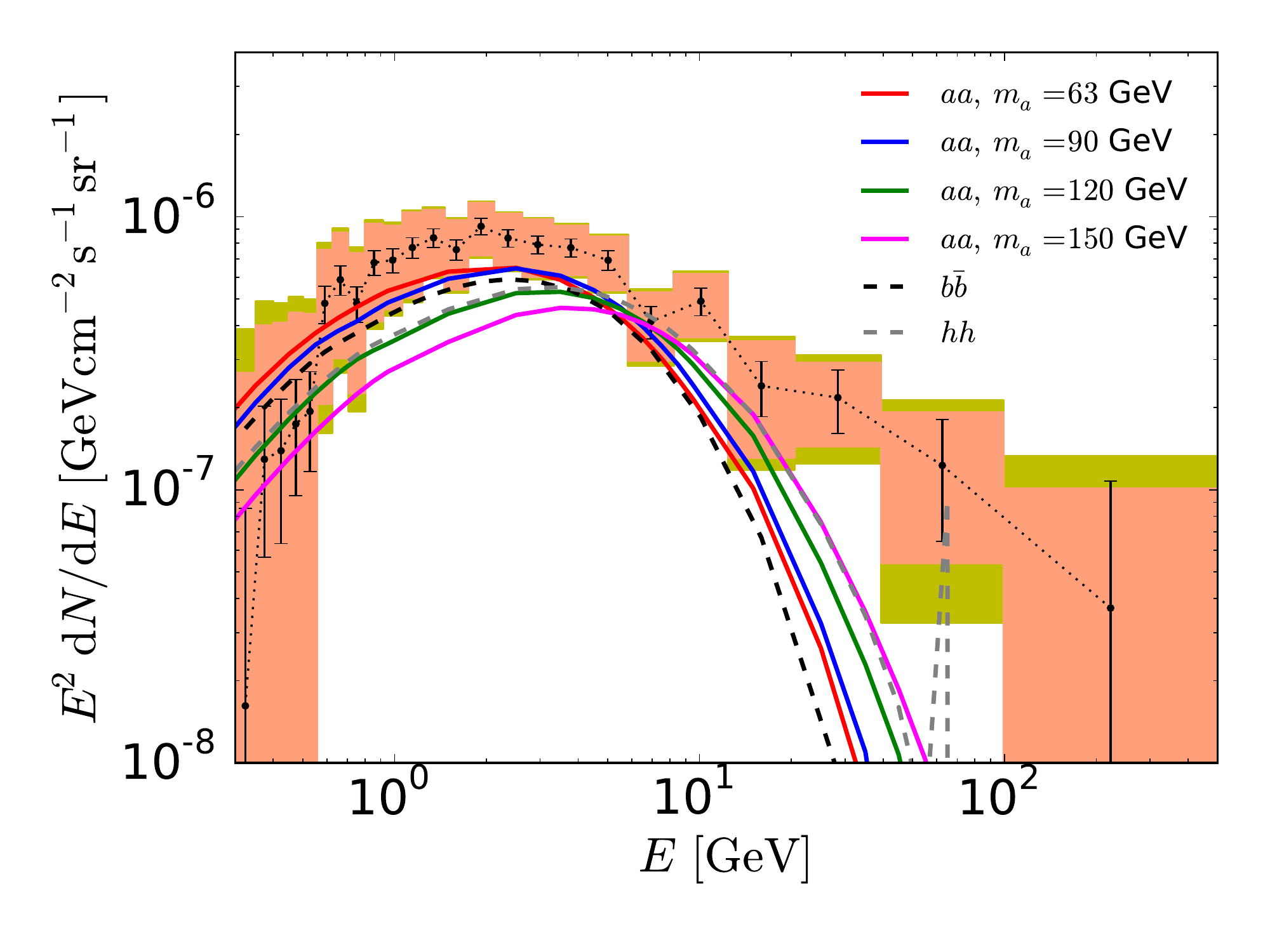}
\caption{\label{fig:fit-spectrum-aa}} 
\end{subfigure}
\hfill
\begin{subfigure}{0.49\linewidth}
\includegraphics[width=\linewidth]{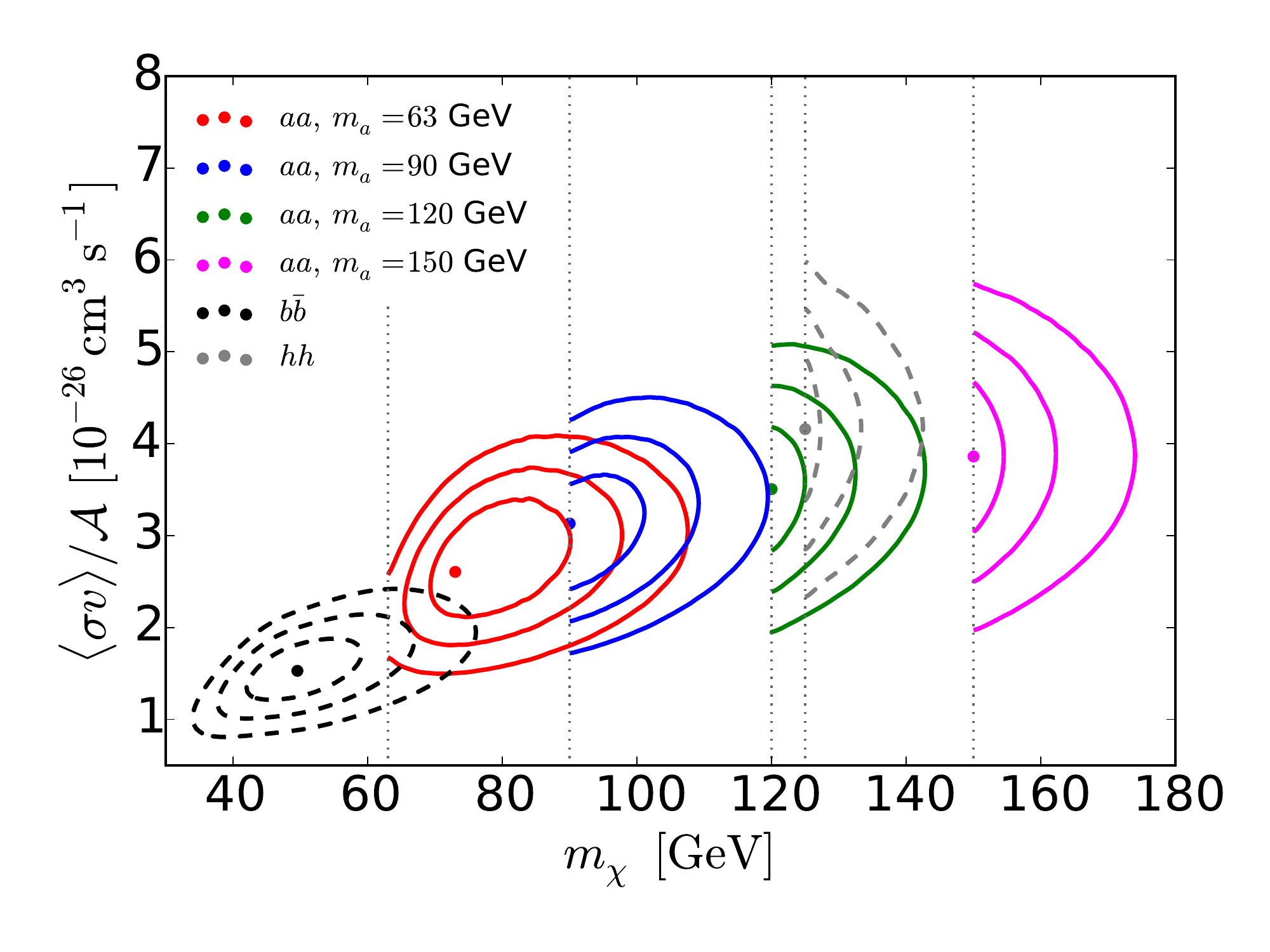}
\caption{\label{fig:fit-regions-aa}} 
\end{subfigure}
\caption{(a) Spectrum of the GCE from Ref.~\cite{Calore:2014xka} and best-fit spectra for DM annihilation to a pseudoscalar and a Higgs for different values of the pseudoscalar mass. For comparison, we also show the spectra for the $b \bar b$- and $h h$-channels. The salmon-colored boxes depict the empirical model systematics as described in Sec.~4.2 of Ref.~\cite{Calore:2014xka}, the error bars correspond to the statistical errors, and the yellow boxes are the combination of the statistical errors, empirical model systematics and other systematics modelled as $dN/dE_{\rm res} = 6 \times 10^{-8}\, \mathrm{GeV}^{-1} \mathrm{cm}^{-2} \mathrm{s}^{-1} \mathrm{sr}^{-1} (E/1 \mathrm{GeV})^{-3}$. See Sec.~4 in Ref.~\cite{Calore:2014xka} for a detailed discussion. (b) Best-fit regions in the DM mass and cross section for $m_a=63, 90, 120$  and $150 \GeV$ (from left to right). For comparison, we also show the best-fit regions for the $b \bar b$- and $hh$-channel with $m_h=125$ GeV. In all regions, the inner, middle and outer contours delimit the $1\sigma$-, $2\sigma$- and $3\sigma$-regions, respectively. Figures~(c) and (d) are the same as (a) and (b) but for DM annihilation to two pseudoscalars.}
\end{figure}

We use the reduced spectrum of the GCE from Ref.~\cite{Calore:2014xka} and the corresponding covariance matrix of the flux uncertainties including statistical and systematic errors which is publicly available. For simplicity we keep the pseudoscalar mass fixed and perform a two-parameter fit in the DM mass and annihilation cross section. In Figs.~\ref{fig:fit-spectrum} and \ref{fig:fit-spectrum-aa}, we show the resulting best-fit spectra from DM annihilation together with the spectrum of the GCE for different values of the pseudoscalar mass and the different annihilation channels. The smallest pseudoscalar mass in these figures is chosen such that Higgs decays to pseudoscalars are kinematically forbidden, i.e. $m_a > m_h/2$. The salmon-colored boxes depict the empirical model systematics as described in Sec.~4.2 of Ref.~\cite{Calore:2014xka}, the error bars correspond to the statistical errors, and the yellow boxes are the combination of the statistical errors, empirical model systematics and other systematics modelled as $dN/dE_{\rm res} = 6 \times 10^{-8}\, \mathrm{GeV}^{-1} \mathrm{cm}^{-2} \mathrm{s}^{-1} \mathrm{sr}^{-1} (E/1 \mathrm{GeV})^{-3}$. See Sec.~4 in Ref.~\cite{Calore:2014xka} for more details. 

Notice that, similar to the $hh$-channel, the spectra for the $ha$-channel in Fig.~\ref{fig:fit-spectrum} have a peak at energies $m_h/2$ which is produced from on-shell decays of the Higgs to two photons. The peak is less pronounced than for the $hh$-channel because there is only one Higgs in the final state. There is only mild line broadening for the best-fit masses, since the $hh$- or $ha$-pair is produced close to threshold (see Fig.~\ref{fig:fit-regions} discussed below). If the DM mass is somewhat above threshold, however, the line broadening efficiently smears out the peak. 
Notice that the spectra have no visible peak at energies $m_a/2$ from pseudoscalar decays to two photons. This can be understood as follows: In a two-Higgs-doublet model, the coupling of top quarks to the pseudoscalar compared to a Higgs with SM couplings is suppressed for $\tan \beta > 1$ as $g_{at\bar{t}}/g_{h_{\rm SM} t\bar{t}}= 1/\tan\beta$, whereas for bottom quarks it is enhanced as $g_{ab\bar{b}}/g_{h_{\rm SM} b\bar{b}}= \tan\beta$. For intermediate values of $\tan\beta$, $\tan\beta \lesssim 5$, the top-quark loop dominates the pseudoscalar decay to two photons. For the Higgs, neglecting interference effects for simplicity, the relative strength of the top-quark loop compared to the dominant $W$-boson loop is $\sim 1/10$. Given that decays into bottom-quarks dominate the pseudoscalar decay width, we can then estimate the ratio of branching fractions for the decays into two photons of the pseudoscalar and the Higgs as
${\rm Br} (a\to\gamma\gamma)/{\rm Br} (h_{_{\rm SM}}\to\gamma\gamma)\sim 1/(10 \times \tan^4\beta)$.  Even modest values of $\tan\beta$ (we consider $\tan \beta=3$) result in the peak being undetectable. Indeed, the peak in Fig.~\ref{fig:fit-spectrum} for the $ha$-channel is purely due to Higgs decays. 
Using the Fermi-LAT limits on $\gamma$-ray lines~\cite{Ackermann:2013uma}, Ref.~\cite{Calore:2014xka} estimated that the line strength from the decay $hh\to 4\gamma$ is just below current limits and may be detected (or excluded) in the near future. In contrast, the intensity of the line produced by pseudoscalar decays is too weak to be in tension with line searches. We find that only in the best-case scenario of $m_a\sim 150$ GeV and $\tan\beta\sim 1$ is the peak barely distinguishable from the continuum. More sensitive $\gamma$-ray experiments may be able to detect the photon peak for lower $m_a$ and larger $\tan\beta$ in the future.
The line at $m_h/2$ from Higgs decays may allow different channels to be distinguished because it has a different height relative to the continuum for the annihilation channels to Higgs and Higgs-pseudoscalar pairs  and vanishes altogether for the pseudoscalar-pseudoscalar and $b \bar b$-channels.

In Figs.~\ref{fig:fit-regions} and \ref{fig:fit-regions-aa}, we show the best-fit regions in the DM mass and cross section for the $ha$- and $aa$-channel, respectively. 
For comparison, we also show the best-fit regions that we find for the $b \bar b$- and $hh$-channels. These regions agree well with those in Ref.~\cite{Calore:2014xka}. 
Similar to Refs.~\cite{Agrawal:2014oha,Calore:2014nla}, we will assume a multiplicative astrophysical-uncertainty factor $\mathcal{A}$ for our best-fit cross-sections.
This factor takes into account the uncertainties in the local DM density $\rho_\odot$, the scale radius $R_s$ and the slope parameter $\gamma$. 
We have used the same reference values for $\rho_\odot$, $R_s$ and $\gamma$ for our fits as Ref.~\cite{Calore:2014nla} and will therefore use their estimate for the range of the astrophysical-uncertainty factor, $\mathcal{A} \in [0.17,5.3]$.
Notice from Figs.~\ref{fig:fit-regions} and \ref{fig:fit-regions-aa} that the best-fit regions for the $ha$- and $aa$-channel 
lie very close to threshold (except for the case $m_a=63 \GeV$ for the $aa$-channel). The best-fit cross sections are fairly independent of the pseudoscalar and DM mass for the $ha$-channel, while the cross sections tend to increase with the pseudoscalar mass for the $aa$-channel.  
In Table~\ref{tab:bestfit}, we show the values and 1$\sigma$-regions for the best-fit DM mass and cross section and the associated $\chi^2$ and $p$-values for the different annihilation channels. Notice that smaller pseudoscalar masses lead to a better fit, i.e.~a smaller $\chi^2$ and thus a larger $p$-value. In particular, for $m_a\lesssim 120 \GeV$, the fit for the $ha$- and $aa$-channels is better than for $hh$ final states and for $m_a=63 \GeV$ (slightly above $m_h/2$), the fit becomes better than for $b\bar{b}$ final states. 
Furthermore, notice that the best-fit cross sections for the scalar channels are larger than for the $b \bar b$-channel. 
This is because the heavier DM requires larger cross sections to obtain the same photon flux due to the smaller number density (cf.~Eq.~\eqref{eq:dNdENFW}; this is partly counteracted by the larger number of photons per decay since the scalar channels dominantly produce four $b$-quarks).

\begin{table}[btp]\centering
\begin{tabular}{lccccc}
\toprule
channel& $m_a$ [GeV] & $m_{_{\rm DM}}$ [GeV] & $\langle\sigma v\rangle_0 \, [10^{-26}\mathrm{cm^3}/\mathrm{s}]$ & $\chi^2_{\mathrm{min}}$ & $p$-value\\
\midrule 
$b\bar b$ & & $49.6_{-6.3}^{+8.1}$ & $1.5_{-0.2}^{+0.3}$ &  $24.5$ & $0.32$ \\
$hh$ & & $125.0_{-0.0}^{+2.3}$ & $4.2_{-0.8}^{+0.8}$ &  $30.0$ & $0.12$ \\
\midrule
\multirow{4}{*}{$ha$}
 &$ 63$ & $94.0_{-0.0}^{+4.2}$ & $3.6_{-0.6}^{+0.7}$ &  $22.4$ & $0.43$ \\
 &$ 90$ & $107.5_{-0.0}^{+3.4}$ & $3.7_{-0.7}^{+0.7}$ &  $25.3$ & $0.28$ \\
 &$120$ & $122.5_{-0.0}^{+3.0}$ & $3.8_{-0.7}^{+0.8}$ &  $30.3$ & $0.11$ \\
 &$150$ & $137.5_{-0.0}^{+2.7}$ & $4.0_{-0.8}^{+0.8}$ &  $36.0$ & $0.03$ \\
\midrule
\multirow{4}{*}{$aa$}
 &$ 63$ & $73.0_{-10.0}^{+15.4}$ & $2.6_{-0.5}^{+0.5}$ &  $24.3$ & $0.33$ \\
 &$ 90$ & $90.0_{-0.0}^{+10.9}$ & $3.1_{-0.7}^{+0.4}$ &  $24.4$ & $0.33$ \\
 &$120$ & $120.0_{-0.0}^{+4.9}$ & $3.5_{-0.7}^{+0.7}$ &  $31.0$ & $0.10$ \\
 &$150$ & $150.0_{-0.0}^{+4.4}$ & $3.9_{-0.8}^{+0.8}$ &  $41.4$ & $0.01$ \\
\bottomrule
\end{tabular}
\caption{
Best-fit values and $1\sigma$-regions from our fits for the $b \bar b$, $hh$, $ha$ and $aa$ annihilation channels, assuming a SM-like Higgs and $\tan\beta=3$ for pseudoscalars in the final state.}
\label{tab:bestfit}
\end{table}

Constraints on the DM annihilation cross section can be derived from $\gamma$-ray observations of dwarf spheroidal galaxies, which are relatively clean of astrophysical backgrounds compared to the Milky Way. For DM which annihilates with a 100\% branching fraction into $b \bar b$-pairs, the Fermi-LAT collaboration finds the limits $\langle \sigma_{b \bar b} v \rangle_0 < 5.00\,(7.91)\times 10^{-26}\,\mathrm{cm^3}/\mathrm{s}$ for $m_{_{\rm DM}}=25\, (50)$ GeV \cite{Ackermann:2013yva} (with the exclusion limit growing with increasing DM masses; see their Table~VI). The cross sections in Table~\ref{tab:data} are consistent with these upper bounds. Recently, the Fermi-LAT collaboration has presented updated limits from dwarf spheroidals that would rule out cross sections $\langle \sigma_{b \bar b} v \rangle_0 \gtrsim 1.5\times 10^{-26}\,\mathrm{cm^3}/\mathrm{s}$ 
for $m_{_{\rm DM}} \sim 30 - 50 \GeV$~\cite{AndersonTalk:2014FermiSymposium}. However, given that these limits are still preliminary, we will impose the limits from~\cite{Ackermann:2013yva} but will also illustrate how our findings are affected if the updated limits are imposed as well. 

There is no dedicated analysis for annihilation into $hh$, $ha$ and $aa$ for either the published nor the latest, preliminary dwarf-spheroidal searches \cite{Ackermann:2013yva,AndersonTalk:2014FermiSymposium}. However, $b$-quarks are still the dominant decay product for these channels. 
We can therefore use the limits for the $b \bar b$-channel by observing that DM of mass $m_{_{\rm DM}}$ annihilating into four $ b$-quarks behaves kinematically similar to DM of mass $m_{_{\rm DM}}/2$ annihilating into two $ b$-quarks. We assume a 100\% branching fraction for the decay $h,a \rightarrow b \bar b$. Using the preliminary limit for the $b \bar b$-channel for a $50 \GeV$ DM mass from \cite{AndersonTalk:2014FermiSymposium} and accounting for a factor of 2 due to the smaller DM density and larger number of $b \bar b$-pairs in our case, we obtain the rough estimate $\langle \sigma  v \rangle_0  \lesssim 3\times 10^{-26}\text{cm}^3/\text{s}$ for DM with $m_{_{\rm DM}} \sim 100 \GeV$ annihilating into $hh$ or $aa$ close to threshold. Using a more sophisticated argument, we have checked that this also applies for the $ha$ channel and depends relatively weakly on the pseudoscalar mass in the range of interest here. For heavier DM masses the bound on $\langle \sigma  v \rangle_0$ is weaker.

The DM interpretation of the GCE is also constrained by other astrophysical observations, in particular by measurements of the anti-proton flux and radio signals from the galactic centre. Reference~\cite{Bringmann:2014lpa} found that the corresponding limits are in strong tension with the required annihilation cross section for the $c \bar c$- and $b \bar b$-channel.
However, it was subsequently argued that systematic uncertainties in the modelling of both the expected anti-proton flux \cite{Cirelli:2014lwa} and the radio signals \cite{Cholis:2014fja} were underestimated in Ref.~\cite{Bringmann:2014lpa}. Taking these uncertainties into account, the required cross sections may still be consistent.

For the $b \bar b$-channel, we will study DM particles with masses in the range $m_{_{\rm DM}} \in [30,50]$ GeV and annihilation cross sections to $b \bar b$-pairs in our galaxy in the range $\langle \sigma_{b \bar b} v \rangle_0 \in [0.5,5] \times 10^{-26}\,\mathrm{cm}^3/\mathrm{s}$. These ranges are motivated by the best-fit regions in Tables \ref{tab:data} and \ref{tab:bestfit} after taking the astrophysical-uncertainty factor $\mathcal{A}$ into account. The upper limit on the cross section is determined by the dwarf spheroidal limit~\cite{Ackermann:2013yva}. For the $ha$-channel, on the other hand, we will consider masses  
$m_{_{\rm DM}} \in [\frac34 m_h,150\, \GeV]$ and 
require that $m_a+m_h\leq2\,m_{_{\rm DM}}\leq m_a+m_h+40$ GeV, as well as cross sections $\langle \sigma_{ah} v \rangle_0 \in [0.4,30] \times 10^{-26}\,\mathrm{cm}^3/\mathrm{s}$. 
Finally, for the $aa$-channel, we allow for masses $m_{_{\rm DM}} \in [\frac12 m_h,150 \GeV]$ with $m_a\leq m_{_{\rm DM}}\leq m_a+30\GeV$ and cross sections $\langle\sigma_{aa} v \rangle_0 \in [0.3,30]\times 10^{-26}\,\mathrm{cm}^3/\mathrm{s}$. These regions are obtained from Table \ref{tab:bestfit}, again taking the possible range of $\mathcal{A}$ into account. We will comment below
on how the upper limits of the cross sections for the $ha$- and $aa$-channels are affected by our estimate of the dwarf spheroidal limit (the vast majority of points from our scan will be found to lie below this).

\section{
Implications of the $\gamma$-ray excess for SUSY
}\label{CGENMSSM}

\subsection{The natural NMSSM}

The absence of new particles beyond the SM in collider searches is undermining the naturalness of the MSSM. 
The recent discovery~\cite{ATLAS:2012gk,CMS:2012gu} of the Higgs with mass $m_h \approx 125$ GeV and SM-like couplings poses additional challenges for naturalness.\footnote{See~\cite{Arvanitaki:2011ck,Craig:2013xia,Farina:2013ssa,Farina:2013fsa,Gherghetta:2014xea,Fan:2014txa} for the connection between the Higgs couplings and naturalness.} Indeed, in order to reproduce the measured value for the Higgs mass in the MSSM, large soft-breaking masses in the stop sector, $|m_{Q_3}|, |m_{u_3}|, |A_t| \gg v$, are necessary. These raise the quartic Higgs coupling in the effective Higgs potential and thus increase the Higgs mass. However, the same soft-breaking masses provide large corrections to the Higgs-sector mass terms via renormalization group running.
The soft parameters at the messenger scale then have to be fine-tuned in order to stabilize the Higgs VEV at the electroweak scale.
This tuning is reaching a level of 1 in 1000 or worse in the MSSM (see e.g.~\cite{CahillRowley:2012rv}). Much of the viable parameter space of the MSSM thus faces the little hierarchy problem.

Naturalness is improved in the NMSSM \cite{Fayet:1974pd,Fayet:1977yc,Fayet:1976cr} which adds a singlet superfield $S$ to the particle content of the MSSM. We will assume the superpotential
\be
W \, \supset \, \lambda S \,  H_u H_d \, + \, \frac{\kappa}{3} S^3 
\label{superpotential}
\ee 
for the Higgs sector, where $H_u$ and $H_d$ are the two Higgs doublets of the MSSM. The coupling $\lambda$ can be chosen real and positive by appropriate field redefinitions, whereas $\kappa$ is real but can have both signs. 
Indeed, the superpotential term $\lambda S H_u H_d$ gives an additional contribution of order $\lambda^2$ to the Higgs quartic coupling. This raises the Higgs mass at tree-level, circumventing the necessity of heavy stop-sector soft masses and thus relieving the tuning. In addition, the $\lambda^2$-contribution to the quartic coupling makes the Higgs VEV less sensitive to destabilising loop corrections, as follows from the minimisation conditions for the Higgs potential~\cite{Hall:2011aa}. Due to this effect, the required amount of tuning for the Higgs VEV decreases with increasing $\lambda$. It was pointed out in \cite{Gherghetta:2012gb}, however, that the Higgs mass in turn grows and eventually overshoots the observed value for increasing $\lambda$. Loop corrections and mixing effects are then required to bring the Higgs mass back down. This is an additional type of tuning that should be taken into account when assessing the naturalness of the model 
and a combined tuning measure was defined in \cite{Gherghetta:2012gb} as the product of the tunings in the Higgs VEV and the Higgs mass. This combined tuning was found to be minimised for intermediate values $\lambda \sim 1$.

The superpotential \eqref{superpotential} does not have any dimensionful parameters. The $\mu$-term is instead dynamically generated from the term $\lambda S H_u H_d$ once the scalar component of $S$ obtains a VEV. This provides a solution to the $\mu$-problem of the MSSM. A parameter scan of this model was performed in Ref.~\cite{Gherghetta:2012gb}, with a focus on regions with $\lambda \sim 1$ in order to minimise the fine-tuning. Two model-building assumptions were made to improve naturalness further. The first of these is a split spectrum, where the first-and-second generation squarks and all sleptons are much heavier than the remaining superpartners, which allows collider constraints to be satisfied while keeping the superpartners most relevant for naturalness relatively light. Secondly, a small messenger scale of $\mathcal{O}(20 \TeV)$ reduces the size of loop corrections during the renormalization group evolution and therefore the required amount of fine-tuning.
Such a low scale is further motivated by the fact that, for $\lambda \gtrsim 0.7$, the coupling runs fast enough to reach a Landau pole below the GUT scale. Some UV completion for the Higgs sector is then required to emerge below this energy scale. 
In order to ensure that the low-energy description of the theory is valid up to the messenger scale, one has to demand that no Landau pole appears at lower energies. Since the energy scale at which the Landau pole arises decreases with increasing $\lambda$, a low messenger scale allows for a larger range of $\lambda$. 
Imposing the relevant limits on superpartner masses, electroweak precision tests and flavour violation, Ref.~\cite{Gherghetta:2012gb} found large regions of parameter space for which the combined tuning in the Higgs VEV and Higgs mass is still better than 5\%. These regions had parameters in the ranges 
$0.8 \lesssim \lambda \lesssim 1.3$, $2.5\lesssim \tan\beta \lesssim 4.2$, 
and $100 \GeV \lesssim |\mu_{\rm eff}|\lesssim 600 \GeV$, where $\mu_{\rm eff} \equiv \lambda v_s$ is the effective $\mu$-term after the scalar component of $S$ obtains a VEV $v_s$.

Motivated by naturalness, we focus on the region $\lambda \sim 1$ for the NMSSM. In order to avoid the $\mu$-problem, we consider the scale-invariant superpotential \eqref{superpotential}. For definiteness, we furthermore assume a split sparticle spectrum. We discuss this spectrum in more detail in  Sec.~\ref{ParameterScans}. We emphasise, however, that our results generically hold for non-split spectra. The messenger scale, on the other hand, does not play a role in our analysis because we specify all parameters near the electroweak scale.

\subsection{Composition of the LSP}
\label{CGENMSSM:Composition}

We will focus on neutralinos as the LSP. The symmetric neutralino mass matrix in the basis $(\tilde{B},\tilde{W}^3,\tilde{h}^0_d,\tilde{h}^0_u, \tilde{s})$ is given by
\begin{eqnarray}
M_{\tilde{\chi}}&=&\left(
\begin{array}{cccccc}
M_{1} & 0  & -\frac{g_1}{\sqrt{2}} v \cos\beta & \frac{g_1}{\sqrt{2}} v \sin\beta & 0\\
.  & M_2   &  \frac{g_2}{\sqrt{2}} v \cos\beta  &   -\frac{g_2}{\sqrt{2}} v \sin\beta & 0\\
. & . & 0 & -\mu_{\rm eff} & -\lambda v  \sin \beta \\
. & . & . & 0 & -\lambda v  \cos \beta \\
. & . & . & . & 2 \kappa v_s
\end{array}
\right)\, , 
\label{chi0massmatrix}
\end{eqnarray}
where $M_1, M_2$ are the gaugino masses and $g_1,g_2$ are the gauge couplings corresponding to U(1)$_Y$ and SU(2)$_L$, respectively. 
We denote the neutral scalar components of the superfields $H_u$, $H_d$ and $S$ by $h_u^0$, $h_d^0$ and $s$, respectively. The electroweak scale is then given by $v \equiv \sqrt{\langle h_u^0 \rangle^2 +  \langle h_d^0 \rangle^2}=174 \GeV$, while $\tan \beta \equiv \langle h_u^0 \rangle / \langle h_d^0 \rangle$ and $v_s \equiv \langle s \rangle$.

A Higgsino- or wino-dominated neutralino with the required properties for the DM and to account for the GCE is problematic. The LEP bounds on charginos require that $M_2, \mu_{\rm eff}\gtrsim 100\,\text{GeV}$~\cite{PDB}, which restricts the possibility of a Higgsino- or wino-dominated LSP in the mass range relevant for explaining the GCE via the $b\overline{b}$-channel. Furthermore, neutralinos with a significant admixture of Higgsino states are severely constrained by direct detection~\cite{Perelstein:2012qg,Cheung:2012qy}. Although very pure Higgsino-like neutralinos avoid these limits, in our mass range of interest they are in tension with indirect-detection bounds~\cite{Cheung:2012qy} and have annihilation cross-sections that are too large for thermal production (the latter being generically true for Higgsinos of mass $\lesssim 1 \TeV$). This problem also arises for winos with masses below $2.7\, \TeV$, which are also in strong tension with indirect-detection bounds~\cite{Cohen:2013ama,Fan:2013faa,Cheung:2012qy}. This leaves bino- or singlino-dominated LSPs. For simplicity, we will decouple the wino, as it need not be lighter than several TeV in a natural superpartner spectrum.

Both a bino- and a singlino-dominated LSP will generically have a non-negligible Higgsino admixture, while the admixture between the bino and the singlino is naturally suppressed, as follows from the mass matrix \eqref{chi0massmatrix}. Both direct-detection experiments and constraints on the invisible decay width of the Higgs restrict the size of the Higgsino admixture. In the scale-invariant NMSSM, these bounds are readily satisfied if $\mu_{\rm eff} = \lambda v_s$ is sufficiently large to suppress the Higgsino mixing. For the bino-dominated case, the mixing matrix elements with Higgsinos in \eqref{chi0massmatrix} are relatively suppressed by the gauge coupling $g_1$, so smaller values of $\mu_{\rm eff}$ are more tolerable. For the singlino-dominated case, however, the Higgsino mixing is greater because $\lambda \sim 1$. The latter case therefore prefers a larger $\mu_{\rm eff}$, which in turn increases the fine-tuning if too large.
Furthermore, simultaneously keeping the singlino mass (the $(5,5)$-element in \eqref{chi0massmatrix}) small, and hence the singlino-dominated LSP light, requires a relatively small singlet self-coupling $\kappa \sim 0.1$. These features make the parameter space with a viable singlino-like LSP more restrictive. For the bino-dominated LSP, the direct-detection bounds (in conjunction with the other experimental constraints on the model) are most easily accommodated in the region of $\mu_{\rm eff}<0$, as observed previously in~\cite{Cheung:2012qy}.

To obtain the constraints on the Higgsino fraction from direct detection and the invisible decay width of the Higgs, we first derive the resulting bounds on the coupling $c_{h \tilde{\chi} \tilde{\chi}}$ of the Higgs to LSPs.
The coupling is given by \cite{Ellwanger:2009dp}
\begin{multline}
c_{h \tilde{\chi} \tilde{\chi}}  \, =  \,  g_1 \, \mathcal{N}_{11} (\mathcal{S}_{11} \, \mathcal{N}_{13}  -  \mathcal{S}_{12} \, \mathcal{N}_{14}) \, - \,g_2 \, \mathcal{N}_{12} (\mathcal{S}_{11} \, \mathcal{N}_{13}  -  \mathcal{S}_{12} \, \mathcal{N}_{14})  \\
    - \, \sqrt{2} \kappa \, \mathcal{S}_{13} \, \mathcal{N}_{15}^2 \, + \, \sqrt{2}\lambda \, (\mathcal{S}_{11} \, \mathcal{N}_{14} \, \mathcal{N}_{15}  +  \mathcal{S}_{12} \, \mathcal{N}_{13} \, \mathcal{N}_{15}  +  \mathcal{S}_{13} \, \mathcal{N}_{13}\, \mathcal{N}_{14}) \, ,
\label{hXX}
\end{multline}
where $\mathcal{S}$ and $\mathcal{N}$ are diagonalisation matrices that relate the mass eigenstates $h_i$ and $\tilde{\chi}_i$ of the $CP$-even Higgses and the neutralinos, respectively, to the interaction eigenstates according to 
\begin{equation}
\left( \begin{matrix}
  h_1 \\  h_2 \\  h_3
 \end{matrix}
\right) \, = \, \mathcal{S} \, 
\left( \begin{matrix}
  h_d^0 \\  h_u^0 \\  s  
 \end{matrix}
\right)  ,
\qquad \qquad 
\left( \begin{matrix}
  \tilde{\chi}_1 \\  \tilde{\chi}_2 \\  \tilde{\chi}_3 \\ \tilde{\chi}_4 \\ \tilde{\chi}_5
 \end{matrix}
\right) \, = \, \mathcal{N} \, 
\left( \begin{matrix}
  \tilde{B} \\ \tilde{W}^3 \\ \tilde{h}^0_d \\ \tilde{h}^0_u \\ \tilde{s} 
 \end{matrix}
\right)\, .
\end{equation}
We identify the lightest $CP$-even state $h_1$ with the Higgs observed at the LHC. 
For simplicity, this state will be abbreviated as $h$ and the LSP $\tilde{\chi}_1$ as $\tilde{\chi}$.

The Higgs contributes to spin-independent DM-nucleon scattering. Electroweak gauge bosons and pseudoscalars, on the other hand, contribute only to spin-dependent scattering for which the limits are much less stringent. The direct-detection experiment LUX provides the strongest constraints on spin-independent DM-nucleon scattering to date (which are also strongest in the mass range relevant for the GCE) \cite{Akerib:2013tjd}. For DM masses $\sim 40 \, (200)$ GeV, the bound is $\sigma_{p}^{\rm SI}\lesssim 7 \, (25)\times 10^{-46}$ $\text{cm}^2$. In the absence of light scalars that could interfere with LSP-nucleon scattering (coloured sparticles or the other $CP$-even states from the Higgs sector), this directly provides a bound on the coupling $c_{h \tilde{\chi} \tilde{\chi}}$. 
The spin-independent cross section for LSP-nucleon scattering is given by (assuming isospin symmetry for simplicity) \cite{Jungman:1995df}
\be\label{eq:DD}
\sigma_{p}^{\rm SI} \, \simeq \, \frac{4}{\pi} \, \mu_N^2 f_p^2 \, ,
\ee
where $\mu_N\approx m_p$ is the LSP-nucleon reduced mass and 
\bea\label{eq:fp}
f_{p}=m_p \, \Big( \hspace{-.2cm} \sum_{q=u,d,s}f^{p}_{T_q}\frac{a_q}{m_q}+\frac{2}{27} f^{p}_{T_G}\sum_{q=c,b,t}\frac{a_q}{m_q} \, \Big)
\eea
is the proton form-factor with $f^{p}_{T_q}=\{0.0153,0.0191,0.0447\}$
and $f^{p}_{T_G}=1-f^{p}_{T_u}-f^{p}_{T_d}-f^{p}_{T_s}$, while $m_p$ is the proton mass and $m_q$ denotes the quark masses. For tree-level parton scattering mediated only by the Higgs $h$, 
\begin{equation}
a_{u} \, = \, -\frac{m_u}{2\sqrt{2}v} \frac{\mathcal{S}_{12}}{ \sin \beta}\frac{c_{h \tilde{\chi} \tilde{\chi} }}{m_h^2} \qquad \text{and} \qquad a_{d} \, = \, -\frac{m_d}{2\sqrt{2} v} \frac{\mathcal{S}_{11}}{\cos \beta}\frac{c_{h \tilde{\chi} \tilde{\chi}}}{m_h^2} \, ,
\end{equation}
and analogously for the other up- and down-type quarks. 
For a Higgs with SM-like couplings (as observed at the LHC), we have $\mathcal{S}_{11} \approx \cos\beta$ and ${\mathcal{S}_{12}\approx \sin\beta}$. The bound on the cross-section from LUX then gives the approximate bound $|c_{h \tilde{\chi} \tilde{\chi}}|\lesssim 0.04 \, (0.07)$ for $m_{\tilde{\chi}}=40 \, (200) \GeV$. In this derivation, we are crucially assuming that the other $CP$-even Higgses are sufficiently heavy to neglect their contribution to the scattering cross-section. This is justified for most of the points in our analysis and was found to hold over much of the space uncovered by Ref.~\cite{Gherghetta:2012gb}.

If Higgs decays to LSPs are kinematically allowed, the coupling $c_{h \tilde{\chi} \tilde{\chi}}$ is also constrained by limits on the invisible branching fraction of the Higgs provided by fits of the Higgs couplings to LHC data.
These require that ${\rm Br}(h\rightarrow {\rm inv}) \lesssim 0.24$~\cite{Belanger:2013xza,Bechtle:2014ewa}, taking the fit to the Higgs couplings with assumptions of SM-couplings to gluons and photons and couplings to weak vector bosons of at most SM-strength (the latter assumption being always satisfied by a general two-Higgs-doublet model with additional singlets as in the case of the NMSSM). The decay width of the Higgs to LSPs is given by
\bea
\Gamma_{h\rightarrow \tilde{\chi} \tilde{\chi}} \, = \, |c_{h\tilde{\chi}\tilde{\chi}}|^2 \, \frac{m_h}{16\pi} \, \left(1-\frac{4 \, m_{\tilde{\chi}}^2}{m_h^2}\right)^\frac{3}{2}
\eea
at tree-level. For a Higgs with SM-like couplings, the total decay width to SM particles is $\Gamma_{\rm SM} \approx 4.07 \MeV$ \cite{Barger:2012hv}. For $m_{\tilde{\chi}} = 40\,\text{GeV}$, the bound on the invisible branching fraction then gives $|c_{h\tilde{\chi}\tilde{\chi}}|\lesssim 0.03$ which is slightly more stringent than the bound from direct detection, although this weakens for masses closer to threshold (for $m_{\tilde{\chi}} = 50\,\text{GeV}$, $|c_{h\tilde{\chi}\tilde{\chi}}|\lesssim 0.04$). As DM-nucleon scattering mediated by the Higgs may be subject to interference from amplitudes mediated by other light scalars, the bound from the invisible branching fraction is more robust provided that the DM is light enough.

These limits on the coupling $c_{h\tilde{\chi}\tilde{\chi}}$ translate into limits on the Higgsino fraction of the LSP under simplifying assumptions. Let us define $c_{\rm max} \equiv 0.03\; (0.07)$  as the maximum allowed Higgs-LSP coupling for $m_{\chi}=40\; (200) \GeV$. 
We first consider a bino-dominated LSP.  From the neutralino mass matrix \eqref{chi0massmatrix}, we see that the mixing matrix element between the bino and singlino vanishes and thus ${\mathcal{N}_{15} \ll 1}$, while we assume that the wino is decoupled so that $\mathcal{N}_{12} \ll 1$. The Higgs-LSP coupling is then approximately given by
\be
c_{h \tilde{\chi} \tilde{\chi}}  \, \approx \,   g_1 \, \mathcal{N}_{11} (\mathcal{S}_{11} \, \mathcal{N}_{13}-\mathcal{S}_{12} \, \mathcal{N}_{14}) \, + \, \sqrt{2}\lambda \, \mathcal{S}_{13} \, \mathcal{N}_{13} \, \mathcal{N}_{14} \, .
\label{hXXapprox1}
\ee
Anticipating a small Higgsino fraction, we set $\mathcal{N}_{11}\approx 1$. Furthermore, we use $\mathcal{S}_{11}\approx \cos\beta$ and $\mathcal{S}_{12}\approx \sin\beta$ for a SM-like Higgs.
Then ignoring the possibility of blind spots and applying the limits on $|c_{h \tilde{\chi} \tilde{\chi}}|$ to the first and second term in (\ref{hXXapprox1}) separately gives the approximate bounds $\smash{|\mathcal{N}_{13}|\lesssim c_{\rm max} /(g_1 \cos\beta)}$  and $\smash{|\mathcal{N}_{14}|\lesssim c_{\rm max}/(g_1 \sin\beta)}$.\footnote{Using these bounds, the third term in \eqref{hXXapprox1} satisfies e.g.~$\sqrt{2} \lambda |\, \mathcal{S}_{13} \, \mathcal{N}_{13} \, \mathcal{N}_{14}| \lesssim  0.7 \,c_{\rm max} \, \lambda / \sin 2 \beta$ for $m_{\tilde{\chi}}=40 \GeV$. Since $\lambda \sim 1$ and $\tan \beta \lesssim 4$, this term at most marginally violates the limit on $c_{h \tilde{\chi} \tilde{\chi}}$ and can thus give only a slightly more stringent limit on the product $\mathcal{N}_{13} \, \mathcal{N}_{14}$ than the first two terms.
}

Let us now consider a singlino-dominated LSP. 
Since the mass matrix elements mixing the gauginos and the singlino vanish, we have $\mathcal{N}_{11},\mathcal{N}_{12} \ll 1$. The Higgs-LSP coupling then simplifies to
\be
c_{h \tilde{\chi} \tilde{\chi}}  \, \approx  \, \sqrt{2} \lambda \, (\mathcal{S}_{11} \, \mathcal{N}_{14} \, \mathcal{N}_{15}  +  \mathcal{S}_{12} \, \mathcal{N}_{13} \, \mathcal{N}_{15}  +  \mathcal{S}_{13} \,\mathcal{N}_{13} \, \mathcal{N}_{14}) \, - \, \sqrt{2} \kappa \, \mathcal{S}_{13} \, \mathcal{N}_{15}^2 \, .
\label{hXXapprox2}
\ee
Anticipating a small Higgsino fraction, we set $\mathcal{N}_{15}\approx 1$. We again ignore possible blind spots and apply the limits on $c_{h \tilde{\chi} \tilde{\chi}}$ to the first two terms in \eqref{hXXapprox2} separately.
This yields the approximate bounds $\smash{|\mathcal{N}_{13}|\lesssim  c_{\rm max} /(\sqrt{2} \lambda \sin \beta)}$  and $\smash{|\mathcal{N}_{14}|\lesssim  c_{\rm max} /(\sqrt{2} \lambda \cos\beta)} $.\footnote{Ignoring possible cancellations among the terms in \eqref{hXXapprox2}, the last term translates to a bound on the singlet fraction of the Higgs, $|\mathcal{S}_{13}|\lesssim  c_{\rm max}/ (\sqrt{2} |\kappa|)$. 
This is, however, not constraining in the limit of small $\kappa$ (as is required for a singlino-dominated LSP). 
But even for $\mathcal{S}_{13}\sim 1$ (which would typically be in conflict with the measured Higgs couplings), the third term in \eqref{hXXapprox2} satisfies e.g.~$\sqrt{2} \lambda |\mathcal{S}_{13} \, \mathcal{N}_{13} \, \mathcal{N}_{14}| \lesssim 0.04 \, c_{\rm max}/ (\lambda \sin 2 \beta$) for $m_{\tilde{\chi}} = 40 \GeV$.  Since $\lambda \sim 1$ and $\tan \beta \lesssim 4$, this term is always negligible when deriving bounds on the Higgsino fractions.
}

We caution the reader that the above analysis is quite simplified and that the bounds derived are only 
approximately valid.
In particular, blind spots, as defined in~\cite{Cheung:2012qy}, can potentially be important 
and open regions of the natural parameter space that would otherwise be ruled-out.
These involve the partial cancellation of the terms in the couplings and allow for larger Higgsino components of the neutralino than naively derived above. 
Furthermore, a singlino-dominated LSP requires small $\kappa$ in the scale-invariant NMSSM and in this region a $CP$-even singlet-dominated state from the Higgs sector is typically light and can contribute to LSP-nucleon scattering, thereby complicating the analysis. However, we still find that our estimates above are approximately consistent with the results from our scans below. For the singlino-dominated LSP, 
we find that typically $|\mathcal{N}_{14}|\sim 0.3-0.4$ and $|\mathcal{N}_{13}| \sim 0.05$ for $m_{\tilde{\chi}}\sim 100 \GeV$, while for the bino-dominated LSP, $|\mathcal{N}_{14}|\lesssim 0.01$ and $|\mathcal{N}_{13}|\sim 0.1 \, (0.2)$ for $m_{\tilde{\chi}}\sim 50\, (100) \GeV$. Given the range of $\lambda$ and $\tan \beta$ that we scan over, these values are approximately consistent with our bounds (although $\mathcal{N}_{13}$ can be up to a factor of $\sim 2$ larger for the singlinos).

\subsection{Mediators for the annihilation of the LSP}
\label{CGENMSSM:Mediators}

We will next consider possible mediators for the annihilation of DM. Due to the hierarchy in the Yukawa couplings, annihilations via the Higgs naturally give a large branching fraction to $b \bar b$-pairs for an LSP within the best-fit mass range for the $b \bar b$-channel. However, annihilations of Majorana fermions via scalars are $p$-wave suppressed and the annihilation cross section via the Higgs (and the other two $CP$-even Higgses) in our galaxy is consequently too small to account for the GCE.  
Annihilations into fermions mediated by the $Z$-boson are helicity-suppressed in the $s$-wave and are also too small to significantly contribute to the GCE.
In the early universe, however, the velocity suppression of the $p$-wave component mediated by the $Z$-boson is lifted and can enhance the total annihilation cross section during freeze-out.

For bino-dominated LSPs, $t$-channel exchange of sbottoms is another possible annihilation process to $b \bar b$-pairs. However, the cross section is typically too small because of the small hypercharge coupling and collider limits on the sbottom mass. It was found in Ref.~\cite{Han:2014nba} that a highly degenerate sbottom-neutralino pair with mass splitting $2 \GeV <m_{\tilde{b}_1}-m_{\tilde{\chi}}< m_b$ and large mixing in the sbottom sector can have a sufficiently large annihilation cross section and may avoid direct-detection constraints.

As we will now discuss, $CP$-odd scalars from the Higgs sector are suitable mediators. The annihilation of Majorana fermions via these states at small velocities is dominantly $s$-wave and thus not suppressed. In addition, $CP$-odd scalars contribute only to spin-dependent DM-nucleon scattering for which the bounds are much weaker. 
One linear combination of the $CP$-odd components of the Higgs doublets $h^0_u$ and $h^0_d$ gets eaten by the $Z$-boson. We denote the orthogonal linear combination of the doublets by $H_I$ and the $CP$-odd component of the singlet scalar by $s_I$. In the basis $(H_I,s_I)$, the mass matrix then reads 
\begin{eqnarray}
M^2_{\text{$CP$-odd}} \, = \, \left(
\begin{array}{ccc}
m_A^2  & \lambda v(A_{\lambda}-2\kappa v_s)  \\
.  & \lambda \, v^2 \sin 2\beta \left(\frac{A_{\lambda}}{2 v_s}+2\kappa\right)-3\kappa A_{\kappa} v_s
\end{array}
\right) \, ,
\label{CPoddmatrix}
\end{eqnarray}
where $m_A^2 \equiv 2\lambda v_s(A_{\lambda} +\kappa v_s)/\sin2\beta$. We write the lightest $CP$-odd state as $a =\cos\theta_A  H_I+ \sin\theta_A  s_I$. 
The coupling of $a$ to an LSP pair then reads
\begin{multline}
c_{a \tilde{\chi} \tilde{\chi}}\, =\, i \, \cos\theta_A \left[  g_2 \, \mathcal{N}_{12}  (\mathcal{N}_{13} \, \sin\beta \, - \, \mathcal{N}_{14} \, \cos\beta ) \,  - \, g_1 \, \mathcal{N}_{11}  (\mathcal{N}_{13} \, \sin\beta \, - \, \mathcal{N}_{14} \, \cos\beta )\right. \\ \left. \, + \,\sqrt{2} \lambda \, \mathcal{N}_{15} (\mathcal{N}_{13} \, \cos\beta+ \mathcal{N}_{14} \, \sin\beta)\right] 
+ \, i \,\sqrt{2}  \sin\theta_A (\lambda \, \mathcal{N}_{13} \,\mathcal{N}_{14} \, -\, \kappa\, \mathcal{N}_{15}^2) \, .
\label{ga1neut}
\end{multline}

We can use the upper bounds on the Higgsino admixture of the LSP, $\mathcal{N}_{13}$ and $\mathcal{N}_{14}$, derived in the last section to bound the size of the coupling $c_{a \tilde{\chi} \tilde{\chi}}$. For the bino-dominated case, using $\mathcal{N}_{11} \approx 1$ and $\mathcal{N}_{12},\mathcal{N}_{15} \ll 1$ as before and the triangle inequality, we find that
\be
|c_{a \tilde{\chi} \tilde{\chi}}| \, \lesssim\,2 \, c_{\rm max} \, |\cos \theta_A|/ \sin 2 \beta  \, + \, 2\,\sqrt{2} \, \lambda \, c_{\rm max}^2 \,  |\sin \theta_A| / (g_1^2 \sin 2 \beta) \, .
\label{CaChiChiLimitBino}
\ee 
For the singlino-dominated case, on the other hand, again using $\mathcal{N}_{15}\approx 1$ and $\mathcal{N}_{11},\mathcal{N}_{12} \ll 1$,
\be
|c_{a \tilde{\chi} \tilde{\chi}}| \, \lesssim \,2 \, c_{\rm max} \, |\cos \theta_A|/  \sin 2 \beta  \,  + \, \sqrt{2} \,  |\sin \theta_A| \left(|\kappa| +c_{\rm max}^2/ (\lambda \sin 2 \beta) \right) \, .
\label{CaChiChiLimitSinglino}
\ee
From these bounds, we may estimate the available parameter space that can explain the GCE for the annihilation channels to $b \bar b$ and to Higgs-pseudoscalar. To this end, we will discuss these two channels separately.

\subsubsection{Annihilation into $b \bar b$} 
\label{AnalyticalArgumentBB}

\begin{figure}[bt]\centering
\begin{subfigure}{0.3\linewidth}\centering
\FMDG{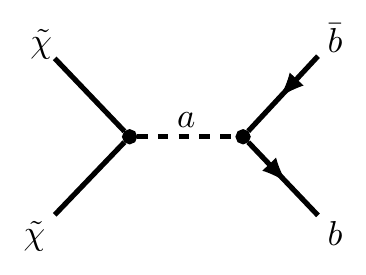}
\end{subfigure}
\hspace{1ex}
\begin{subfigure}{0.3\linewidth}\centering
\FMDG{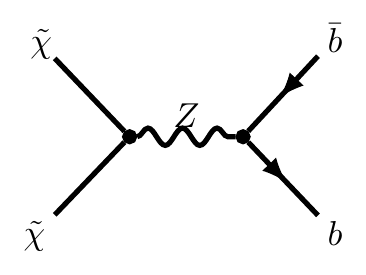}
\end{subfigure}
\caption{Feynman diagrams contributing to neutralino annihilation to $b \bar b$-pairs.}
\label{fig:XXbbdiag}
\end{figure}

The most important processes for annihilation into $b \bar b$-pairs are shown in Fig.~\ref{fig:XXbbdiag}.
The $CP$-odd scalars from the Higgs sector couple to the SM via the usual Yukawa couplings and can thus have large branching fractions to $b \bar b$-pairs. 
The coupling of the lightest $CP$-odd Higgs to a pair of bottom quarks is given by
\begin{equation} 
c_{a b\bar{b}} \, = \, i\,\frac{y_b}{\sqrt{2}}\cos\theta_A\tan\beta \, ,
\label{ga1bb}
\end{equation}
where $y_b \equiv m_b/v$ is the SM Yukawa coupling to bottom quarks.

The annihilation cross section  
for the process $\tilde{\chi} \tilde{\chi} \rightarrow a^* \rightarrow b \bar b$ at small velocities is given by \cite{Cheung:2014lqa}
\be
 \langle \sigma v \rangle_0 \, \approx \, 2 \times 10^{-26} \, \frac{\text{cm}^3}{\text{s}} \left|\frac{c_{a b\bar{b}}}{y_b}\right|^2 \left|\frac{c_{a \tilde{\chi} \tilde{\chi}}}{0.5}\right|^2 \left(\frac{m_{\tilde{\chi}}}{40 \GeV}\right)^2  \left(\frac{(120 \GeV)^2-4 (40 \GeV)^2}{m_{a}^2-4 m_{\tilde{\chi}}^2+ m_a^2 \Gamma_a^2 }\right)^2 \, ,
\label{crosssectionNR}
\ee 
where $\Gamma_a$ is the total decay width of the pseudoscalar. 
Let us for the moment assume that the annihilation is not resonantly enhanced, i.e.~that $m_{a}$ is not very close to $2 m_{\tilde{\chi}}$. The $\Gamma_a$-dependent term in the denominator is then negligible and the last two factors in the formula are of order one in the mass range of interest for the GCE.
In order to explain the GCE and avoid constraints from additional annihilation channels, the process $\tilde{\chi} \tilde{\chi} \rightarrow a^* \rightarrow b \bar b$ should dominate the total annihilation cross section at the present time. 
In the absence of additional important annihilation channels during freeze-out (e.g.~mediated by the $Z$-boson) and coannihilation, the cross section at freeze-out is of the same size as \eqref{crosssectionNR}. 
Both the required amount of DM and the GCE can then be accommodated if the remaining factors in (\ref{crosssectionNR}) are close to one.

From our upper estimates of the coupling $c_{a \tilde{\chi} \tilde{\chi}}$, we can obtain an upper bound on these factors and the cross section. Note that these estimates and the coupling $c_{a b\bar{b}}$ grow with $\tan \beta$ (cf.~Eqs.~\eqref{CaChiChiLimitBino}-\eqref{ga1bb}) and that $c_{a b\bar{b}}$ vanishes if $a$ consists purely of $s_I$. 
In order to obtain a conservative bound, we therefore set $\tan \beta = 3$, corresponding to the largest value for which we find points in our scan with a significant $H_I$-admixture to $a$. 
We assume $m_{\tilde{\chi}} = 40 \GeV$ (which is within the mass range of interest for the $b \bar b$-channel) and distinguish between the two cases of $a$ being dominantly $H_I$ and an equal admixture of $H_I$ and $s_I$ (the bounds become stronger if $a$ has an even larger $s_I$-admixture).
Let us first consider the bino-dominated case. The second and third factor in \eqref{crosssectionNR} can be estimated as
\be
\left|\frac{c_{a b\bar{b}}}{y_b}\right|^2 \left|\frac{c_{a \tilde{\chi} \tilde{\chi}}}{0.5}\right|^2 \, \lesssim \, 
\begin{cases}
 0.2 \, \,  & \text{for } \cos \theta_A \approx 1 \\
(0.2 + 0.07 \, \lambda)^2  \, \, & \text{for } \cos \theta_A \approx 1/\sqrt{2} \, .
\end{cases}
\label{factors}
\ee
Similarly for the singlino-dominated case, setting again $\tan \beta = 3$, we can estimate the second and third factor in \eqref{crosssectionNR} as
\be \label{FactorSinglino}
\left|\frac{c_{a b\bar{b}}}{y_b}\right|^2 \left|\frac{c_{a \tilde{\chi} \tilde{\chi}}}{0.5}\right|^2 \, \lesssim \, 
\begin{cases}
 0.2 \, \,  & \text{for } \cos \theta_A \approx 1 \\
 (0.2 +3 \, |\kappa|  + 5 \times 10^{-3} / \lambda )^2 \, \, & \text{for } \cos \theta_A \approx 1/\sqrt{2} \, .
\end{cases}
\ee
This factor for either the bino- or singlino-dominated case needs to be not much smaller than one for a sufficiently large annihilation cross section \eqref{crosssectionNR}. 
As we have discussed in the last section, $\kappa \ll 1$ is required in the scale-invariant NMSSM for a light singlino-dominated LSP with a small Higgsino fraction. This means that the $\kappa$-dependent term in \eqref{FactorSinglino} does not help. 
Even though the case $\cos \theta_A \approx 1$ seems marginally consistent with the required cross section for the GCE when taking the astrophysical uncertainty factor $\mathcal{A}$ into account, the freeze-out cross section is then typically too small to generate the required amount of DM. This would require additional contributions to raise the freeze-out cross section, e.g.~mediated by the $Z$-boson.
In addition, we find that $\tan \beta < 3$ for $\cos \theta_A \approx 1$. This is because $\cos \theta_A$ is correlated with $\tan\beta$. Indeed, the mass term of the singlet $s_I$, the $(2,2)$-element of the $CP$-odd mass matrix \eqref{CPoddmatrix}, is suppressed while the mass term of the doublet $H_I$, the $(1,1)$-element, is enhanced by $\tan\beta$.\footnote{An exception occurs if the first term in the $(2,2)$-element is negative and cancels against the second term in which case the $(2,2)$-element is enhanced by $\tan \beta$ too.
} This means that the admixture of $H_I$ to $a$ and thus $\cos \theta_A$ decreases with increasing $\tan \beta$. This dependence is indeed what we observe in our scan. 
For $\tan \beta \approx 3$, we find points only up to $|\cos \theta_A| \approx 0.75$ (for a bino-like LSP; $\cos \theta_A$ is even smaller for larger $\tan \beta$). This means that the bound for the case $\cos \theta_A \approx 1$ in \eqref{factors} and \eqref{FactorSinglino} is actually more stringent. We will therefore need an enhancement of the cross section. 
To this end, we will rely on the regime $m_{a} \simeq 2 m_{\tilde{\chi}}$ in which case the cross section is resonantly enhanced and the last factor in \eqref{crosssectionNR} is much larger than unity.
This is in contrast to Ref.~\cite{Cheung:2014lqa}, which considers smaller $\lambda$ and can thus tolerate large $\tan \beta$, allowing for larger couplings $c_{a b\bar{b}}$ and $c_{a \tilde{\chi} \tilde{\chi}}$ and a sufficiently large cross section without resonant enhancement for the bino-dominated case (though the smaller $\lambda$ is at the expense of increased fine-tuning due to the Higgs mass).

\subsubsection{Annihilation into $ha$}

\begin{figure}[bt]\centering
\begin{subfigure}{0.3\linewidth}
\FMDG{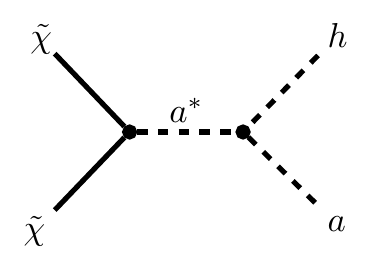}
\end{subfigure}
\hspace{1ex}
\begin{subfigure}{0.3\linewidth}
\FMDG{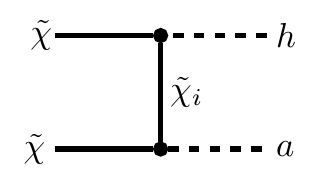}
\end{subfigure}
\hspace{1ex}
\caption{Feynman diagrams contributing to neutralino annihilation to $ha$-pairs. Note that there is a $u$-channel diagram in addition to the $t$-channel one.}
\label{fig:XXhadiag}
\end{figure}

The annihilation of neutralinos is $p$-wave suppressed if the final state is even under $CP$, such as for Higgs or pseudoscalar pairs~\cite{Griest:1989zh}. Therefore it is only through the Higgs-pseudoscalar channel that neutralino annihilation into scalars can account for the GCE. The graphs that contribute to the corresponding cross section are the $s$-channel exchange of the two pseudoscalars and the $t/u$-channel exchange of the neutralinos. These are shown in Fig.~\ref{fig:XXhadiag}. As follows from Fig.~\ref{fig:fit-regions}, a good fit to the GCE requires that the annihilation into $ha$ happens close to threshold. Let us define the dimensionless ratio $\delta \equiv (2 m_{\tilde{\chi}} - (m_a + m_h))/2 m_{\tilde{\chi}}$ so that $\delta =0$ corresponds to the annihilation being precisely at threshold. Expanding for $\delta\ll 1$, the cross section for the process $\tilde{\chi} \tilde{\chi}  \rightarrow h \, a$ is given by~\cite{Griest:1989zh}
\begin{multline}
 \langle \sigma_{ha} v \rangle_0  \, \approx \,   \frac{1}{8\pi}\left(\frac{m_h}{m_{\tilde{\chi}}}\right)^{1/2}\left(1-\frac{m_h}{2m_{\tilde{\chi}}}\right)^{1/2} \sqrt{\delta} \\ 
\times \, \left[\frac{c_{aah} c_{a\tilde{\chi} \tilde{\chi}}}{m_h(4 m_{\tilde{\chi}}-m_h)}+\frac{c_{a_2 ah} c_{a_2 \tilde{\chi} \tilde{\chi}}}{4m^2_{\tilde{\chi}}-m^2_{a_2}+m_{a_2}\Gamma_{a_2}}+2\sum_{k=1}^{5}\frac{c_{a \tilde{\chi}\tilde{\chi}_k} c_{h\tilde{\chi}\tilde{\chi}_k}}{m_h+m_{\tilde{\chi}_k}-m_{\tilde{\chi}}}\right]^2 ,
\label{crosssectionHA}
\end{multline}
where the heaviest pseudoscalar is denoted by $a_2$ (we neglect the small width of the $a$).  

From Fig.~\ref{fig:fit-regions}, we see that the best-fit regions have $m_{\tilde{\chi}} \sim 100 \GeV$ and $\delta \sim 0.1$. The best-fit cross section, on the other hand, depends on the value of the astrophysical uncertainty factor $\mathcal{A}$ but is consistent with the freeze-out cross section $\sim 2 \times 10^{-26} \text{cm}^3 / \text{s}$ required for the WIMP miracle. In the absence of resonance enhancement, additional important annihilation channels (e.g.~mediated by the $Z$-boson) and coannihilation, the cross section at freeze-out is of the same size as \eqref{crosssectionHA} (deviations from these conditions are tolerable to some extent given the relatively large range for the uncertainty factor $\mathcal{A}$). Assuming the values for $\delta$, $m_{\tilde{\chi}}$ and the target cross section above, we will now show that \eqref{crosssectionHA} can be large enough to explain both the GCE and the observed DM abundance without relying on special enhancements such as a resonance. Considering only the $t$-channel exchange of the second neutralino $\tilde{\chi}_2$, the cross-section may be approximated as
\begin{multline}
 \langle \sigma_{ha} v \rangle_0 \, \approx \, 2\times 10^{-26}\, \frac{\text{cm}^3}{\text{s}} \left(\frac{c_{a \tilde{\chi}\tilde{\chi}_2} \, c_{h\tilde{\chi}\tilde{\chi}_2}}{0.2\times 0.2}\right)^2
 \,\left(\frac{100 \GeV}{m_{\tilde{\chi}}}\right)
 \sqrt{\frac{2m_{\tilde{\chi}}-125 \GeV}{2\cdot 100 \GeV-125 \GeV}}
\,\sqrt{\frac{\delta}{0.1}}\\ \times  \,\left(\frac{125 \GeV - 100 \GeV+150 \GeV}{125  \GeV-m_{\tilde{\chi}}+m_{\tilde{\chi}_2}}\right)^2 \, ,
\label{subcrosssectionHA1}
\end{multline}
where we set $m_h=125 \GeV$.
We have focused on the second neutralino because the LSP necessarily has a small Higgsino-fraction as discussed in Sec.~\ref{CGENMSSM:Composition}. This limits the size of the couplings $c_{a \tilde{\chi}\tilde{\chi}}$ and $c_{h\tilde{\chi}\tilde{\chi}}$ that would appear in \eqref{subcrosssectionHA1} for $t$-channel exchange of the LSP and thereby the size of the cross section. Since there is no limit on the Higgsino fraction of the second neutralino, the couplings $c_{a \tilde{\chi}\tilde{\chi}_2}$ and $c_{h\tilde{\chi}\tilde{\chi}_2}$ can be much larger. Thus, if the second neutralino is relatively light, we see that a sufficiently large cross section to explain the GCE and the observed DM abundance can be obtained.

Similarly, considering only the $s$-channel exchange of the pseudoscalar $a$, the cross section is given by
\begin{multline}
 \langle \sigma_{ha} v \rangle_0 \, \approx \,  2 \times 10^{-26} \, \frac{\text{cm}^3}{\text{s}}  
 \left(\frac{c_{a a h}}{v}\right)^2 \left|\frac{c_{a \tilde{\chi} \tilde{\chi}}}{0.08}\right|^2
\left(\frac{100 \GeV}{m_{\tilde{\chi}}}\right) \sqrt{\frac{2 \, m_{\tilde{\chi}}-125 \GeV}{2 \cdot 100 \GeV - 125 \GeV}} \sqrt{\frac{\delta}{0.1}} \\ 
\times  \left(\frac{4 \cdot 100 \GeV - 125 \GeV}{4 \, m_{\tilde{\chi}}-125 \GeV}\right)^2  ,
\label{subcrosssectionHA2}
\end{multline}
where we set $m_h=125 \GeV$. 
The coupling $c_{a a h}$ between the Higgs and two lightest pseudoscalars is dimensionful and determined by $v$, $v_s$ and $A_\lambda$ (for brevity, we do not give the expression for this coupling which can be found in~\cite{Ellwanger:2009dp}). Since $v_s$ and $A_\lambda$ can be much larger than $v$, the suppression from the coupling $c_{a \tilde{\chi} \tilde{\chi}}$ for a small Higgsino fraction may be overcome by the coupling $c_{aah}$. Thus, we again see that a sufficiently large cross section to explain the GCE and the observed DM abundance can be obtained.

However, we find in our scan below that generally neither of the terms in \eqref{crosssectionHA} dominate the others, but that they must all be considered and can interfere with each other. When all the contributions are combined they allow for a sufficiently large cross section without the need for a resonant enhancement.

\section{Parameter scans}
\label{ParameterScans}

In order to find parameter sets for which the NMSSM can explain the GCE via the $b \bar b$- or $ha$-channel, we have performed numerical scans using the programs \texttt{NMSSMTools\;4.4.0}~\cite{Ellwanger:2004xm,Ellwanger:2005dv,Belanger:2005kh} and \texttt{micrOMEGAs\;3.0}~\cite{Belanger:2005kh}.  To simplify the parameter space, we decouple the first and second generation of squarks and all generations of sleptons, fixing their masses at $2 \TeV$ (similar to the scan performed in \cite{Gherghetta:2012gb}). 
We similarly decouple the wino, choosing $2 \TeV$ for its mass. Due to the small Yukawa couplings of these sparticles and the small gauge couplings, these masses are compatible with naturalness. 
We fix the soft masses of the stops and sbottoms at $1 \TeV$ with vanishing trilinear terms $A_{t,b}$ and the gluino mass at $2 \TeV$. These values are chosen such that the collider constraints on their masses are fulfilled but the fine-tuning still remains approximately optimal. Note that the physics of the DM is relatively independent of the properties of these particles, so as long as collider constraints on these masses are satisfied, we do not expect our results to qualitatively change if these are allowed to take different values.
We have performed three different scans, for (i) a bino-like LSP with dominant annihilation to $ha$, (ii) a singlino-like LSP with dominant annihilation to $ha$ and (iii) a bino-like LSP with dominant annihilation to $b\bar b$.

We impose all relevant collider and flavour constraints which are implemented in \texttt{NMSSMTools \, 4.4.0} (without trying to explain the discrepancy from the SM in the anomalous magnetic dipole moment of the muon), which in particular require a Higgs at 125 GeV with SM-like couplings consistent with the signal strengths measured at the LHC.
Most points that pass all experimental constraints have $m_{a}> m_h/2$ and thus decays $h\to a a$ are kinematically forbidden. Indeed, a significant decay rate $h\to a a$ would induce large deviations from LHC measurements in the branching fractions of the Higgs decays into photons, $b$-quarks, $\tau$-leptons, $W$- and $Z$-bosons. 
We only find a few points with $m_{a}< m_h/2$ and these are all close to the threshold. Since the pseudoscalars can have a significant singlet-admixture, LHC searches do  not significantly constrain the range of pseudoscalar masses that we consider~\cite{Kozaczuk:2015bea}.
For all points in the scans, the couplings $\lambda$ and $\kappa$ do not evolve toward a Landau pole below $20 \TeV$. 
We use \texttt{micrOMEGAs\;3.0} to calculate the relic abundance of the LSP and the cross-sections relevant for direct and indirect detection. We impose the range $0.107<\Omega_{\tilde{\chi}}h^2<0.131$ when we require that the LSP abundance be consistent with the observed DM abundance. This range is obtained from the best-fit value for the DM abundance from Planck \cite{Ade:2013lta} after including a $\pm 10\%$ theory error in order to account for uncertainties in our calculations (the measurement error quoted by Planck is subsumed by this).

In addition, we check constraints from electroweak precision measurements which are not implemented in \texttt{NMSSMTools\;4.4.0}.
These can be important for large $\lambda \sim 1$ since the mixing of the Higgsinos with the other neutralinos then induces sizeable deviations from the custodial symmetry. We calculate the Peskin-Takeuchi parameters $S$ and $T$ using the formulas from Refs.~\cite{Cho:1999km,Hagiwara:1994pw,Martin:2004id} for the neutralino-chargino and stop-sbottom sector and those from Refs.~\cite{Barbieri:2006bg,Franceschini:2010qz} for the Higgs and singlet scalars.
We compare the results with the $95\%$ exclusion region determined in \cite{Baak:2014ora}. As was found previously, for $\lambda \sim 1$ the neutralino/chargino-sector tends to provide the largest contribution to $T$ and $\tan\beta \gg 1$ is ruled out due to custodial-symmetry breaking.

\subsection{Parameter scan for the $ha$-channel}\label{Higgs-pseudoscalar-channel}
We have performed two random scans using \texttt{NMSSMTools\;4.4.0}~\cite{Ellwanger:2004xm,Ellwanger:2005dv,Belanger:2005kh} for the Higgs-pseudoscalar channel, one optimised for bino-like LSPs and the other for singlino-like LSPs. 
The masses of the squarks, sleptons, gluino and wino are fixed as mentioned previously and the remaining free parameters are scanned over the ranges shown in Table~\ref{tab:HAscan}. Here $A_\lambda$ is partly determined by the requirement that the singlet admixture to the Higgs is small.
To discuss this in more detail, let us rotate from the $(h_d^0, h_u^0)$ basis into the $(h_{_{\rm SM}},H)$ basis, where only $h_{_{\rm SM}}$ obtains a VEV and thus couples to SM particles precisely like the SM Higgs. The admixture of the orthogonal state $H$ and the singlet $s$ to the lightest $CP$-even Higgs then drives its couplings away from SM values. The size of the admixture with $s$ is determined to leading order by the mass term, $m_s$ of $s$ and its mixing mass term, $m_{hs}$ with $h_{_{\rm SM}}$:
\begin{equation}
\mathcal{L} \, \supset \, \frac{1}{2} \, m_s^2 \, s^2 \, + \, m_{hs}^2 \, h_{_{\rm SM}} s \,,
\label{Higgs-singlet-mixing}
\end{equation}
where
\begin{eqnarray}
m_{s}^2  &= &\kappa \, A_\kappa v_s \, + \, 4 \, \kappa^2 v_s^2\, + \, \frac{\lambda}{4 v_s}  A_\lambda v^2 \sin 2 \beta\,, \\ 
m_{hs}^2 & = & 2 \lambda^2 v_s v \, - \sin 2 \beta \,\lambda\, v (A_\lambda + 2 \kappa v_s)\,.
\end{eqnarray}
The sizes of $A_\lambda$ and $A_\kappa$ are limited by the requirement that the Higgs sector is non-tachyonic.
Furthermore, as we have discussed in Sec.~\ref{CGENMSSM:Composition}, small $\kappa \sim 0.1$ is required for a singlino-dominated LSP with a small Higgsino admixture. 
We find in our scan that this then implies that $m_s$ is not much larger than the Higgs mass. In order to suppress the singlet mixing with the Higgs and the resulting deviations in the Higgs couplings, we therefore fix $A_{\lambda}$ to minimise the mixing matrix element $m_{hs}^2$. More precisely, we set $A_\lambda=A_\lambda^0+\Delta A_\lambda$, where $A_\lambda^0$ is determined by $m_{hs}^2=0$ and $\Delta A_\lambda$ is randomly chosen in the range $[-50,50] \GeV$.
Although a bino-dominated LSP allows for larger values of $\kappa$, we find that $\mu_{\rm eff}$ must be relatively small for a sufficiently large cross section (as suggested in \eqref{subcrosssectionHA1}, where a Higgsino-like $\tilde{\chi}_2$ of mass $m_{\tilde{\chi}_2}\sim 150\,\GeV$ is preferred). This again results in a relatively light singlet-like state of several hundred GeV.
In order to obtain a Higgs with SM-like couplings, we therefore also impose this condition on $A_{\lambda}$ for the bino-dominated case. We have checked that our findings are qualitatively unchanged after relaxing this condition.

\begin{table}[bt]\centering
\begin{tabular}{ l  c  c  c ccccc}
\toprule
& $\Delta A_\lambda$ [GeV] & $A_\kappa$  [GeV] & $\mu_{\rm eff}$  [GeV] &  $M_1$  [GeV] &$\lambda$ & $\kappa$ & $\tan \beta$ \\
\midrule
bino-like & [-50,50] & [-100,100] & [-300,-100]  & [60,170] & [0.6,1.4] & [0.1,1.6] & [2,5]  \\
singlino-like &  [-50,50] & [-100,100] & [-600,-200] & 2000 & [0.6,1.4] & [0.05,0.5] & [2,5] \\
\bottomrule
\end{tabular}
\caption{Parameter ranges for the two random scans for the $ha$-channel. The first and second line are for the scan optimised for bino-like and singlino-like LSPs, respectively.
}
\label{tab:HAscan}
\end{table}

\begin{figure}[bt]\centering
\begin{subfigure}{0.49\linewidth}
\includegraphics[width=\linewidth]{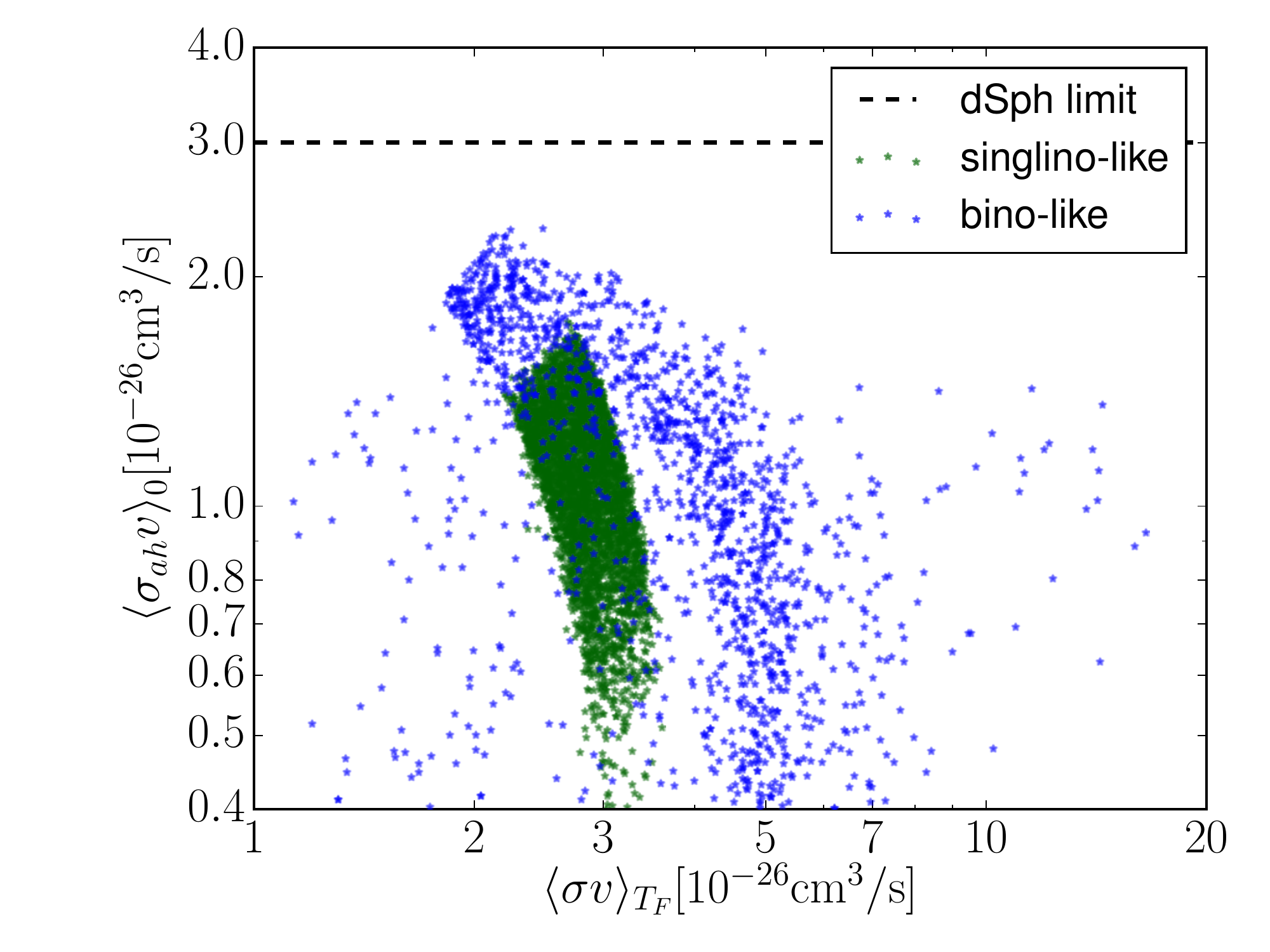}
\caption{\label{fig:HAscanLSP1}}
\end{subfigure}
\hfill
\begin{subfigure}{0.49\linewidth}
\includegraphics[width=\linewidth]{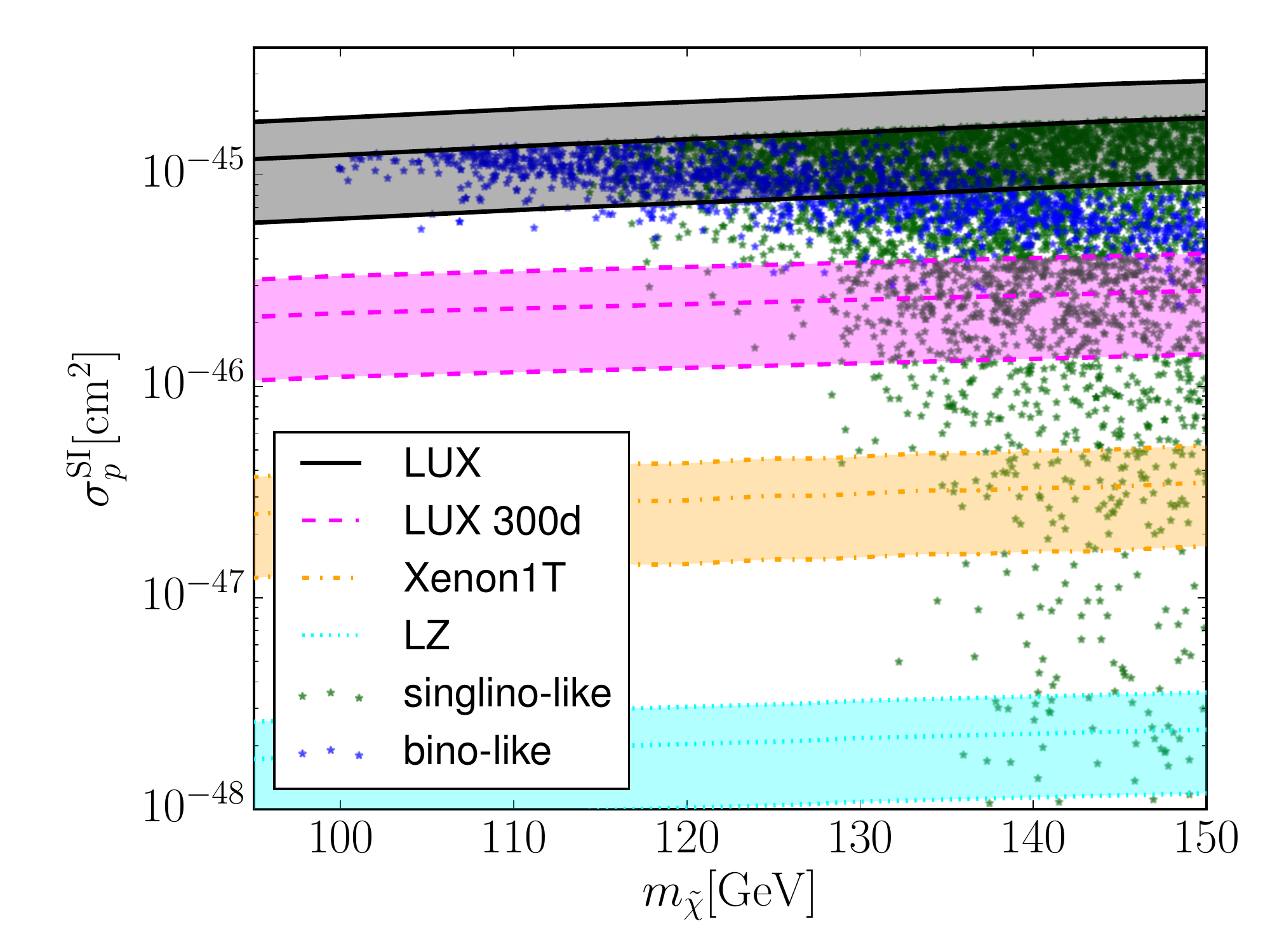}
\caption{\label{fig:HAscanLSP2}}
\end{subfigure}
\caption{Results from the two random scans with parameter ranges given in Table \ref{tab:HAscan} optimised for the $ha$-channel. We require a branching fraction of annihilations into $ha$ of at least 85\%. The left panel (a) shows the annihilation cross section at the present time, $\langle \sigma_{ha} v\rangle_0$, and at freeze-out, $\langle\sigma v \rangle_{T_F}$. The right panel (b) shows the spin-independent cross section for scattering of LSPs off nucleons versus the LSP mass.}
\label{fig:HAscanLSP}
\end{figure}

Results from the scans are presented in Fig.~\ref{fig:HAscanLSP}. All points in the scatter plots have an LSP abundance in the range $0.107<\Omega_{\tilde{\chi}}h^2<0.131$, consistent with the observed DM abundance as discussed above. Furthermore, all points have masses $m_{\tilde{\chi}}$ and $m_a$ and an annihilation cross section $\langle \sigma_{ha} v \rangle_0$ into $ha$ in our galaxy within the best-fit ranges as discussed at the end of  Sec.~\ref{GCE-section}. We also demand a branching ratio to $ha$ of at least 85\% in order to have a relatively pure annihilation into these states.
In Fig.~\ref{fig:HAscanLSP1}, we show a scatter plot of the annihilation cross-sections during freeze-out and at late times. 
The approximate degeneracy of these cross sections when $\langle\sigma_{ha} v\rangle_0\sim (1-3)\times 10^{-26}\,\text{cm}^3/\text{s}$ reflects the dominance of the $\tilde{\chi}\tilde{\chi}\rightarrow ha$ process and the absence of a large special enhancement such as a resonance or coannihilation.
The late-time cross section can reach values of up to $\sim 2\times 10^{-26}\,\text{cm}^3/\text{s}$ for the bino-dominated LSPs, 
whereas it is smaller for the singlino-dominated LSPs. These cross sections are nevertheless consistent with the best-fit regions prescribed in Sec.~\ref{CGESection} once uncertainties in the DM distribution are accounted for. 
Larger late-time cross sections are too big for the LSP to freeze-out with the required relic abundance. This is because additional processes that are $p$-wave suppressed in our galaxy (most notably those mediated by a singlet-like $CP$-even scalar) activate and contribute significantly in the hotter early universe. Furthermore, we find that the points with bino-dominated LSPs that have significantly enhanced cross-sections at freeze-out, notably most of the ``column'' clustering at $\langle\sigma v\rangle_{T_F}\sim 5\times 10^{-26}\,\text{cm}^3/\text{s}$ in Fig.~\ref{fig:HAscanLSP1}, as well as the diffuse region up to $\langle\sigma v\rangle_{T_F}\sim 2 \times 10^{-25}\,\text{cm}^3/\text{s}$, 
undergo coannihilation with a Higgsino-like second-lightest neutralino. The larger effective cross section at freeze-out is partly because the $aa$ and $hh$ channels are not $p$-wave suppressed for coannihilation since the Pauli exclusion principle no longer applies to the initial coannihilating particles. The few points with $\langle\sigma v\rangle_{T_F}\sim 10^{-25}\,\text{cm}^3/\text{s}$ have this coannihilation resonantly enhanced by a singlet-like $CP$-even scalar. Because these enhancements have a strong temperature dependence, they diminish as the universe cools such that the LSP still freezes-out with the observed DM relic density (the cross sections plotted in Fig.~\ref{fig:HAscanLSP1} are calculated for the instantaneous temperature at which the LSP begins to freeze-out). We emphasise, however, that neither a resonance nor coannihilation is necessary for a bino-like LSP to explain the GCE as can be appreciated by the fact that there are many points with $\langle\sigma_{ha} v\rangle_0\approx \langle\sigma v\rangle_{T_F}$ in Fig.~\ref{fig:HAscanLSP1}.
We have also plotted, as a dashed line in Fig.~\ref{fig:HAscanLSP1}, our estimate from Sec.~\ref{GCE-section} of the limit on $\langle \sigma_{ha}v \rangle_0$ arising from the latest Fermi-LAT study of dwarf spheroidals. All our points satisfy this limit.

The characteristics of the two types of LSPs are quite distinct. The points with a singlino-like LSP are dependent upon a large $|A_{\lambda}|\sim 1-2.5$ TeV. 
This is because the neutralinos $\tilde{\chi}_2$ and $\tilde{\chi}_3$ are typically relatively light and Higgsino-dominated, leading to
a large contribution to the cross section. 
Although the significance of the $t$-channel exchanges of $\tilde{\chi}_2$ and $\tilde{\chi}_3$ in \eqref{subcrosssectionHA1} diminishes as $|\mu_{\rm eff}|$ increases (as this determines the Higgsino masses), we only scan over relatively small values $|\mu_{\rm eff}| \leq 600 \GeV$ to try to maintain naturalness.
The contribution to the cross section from the $s$-channel exchange of the $a$ is then required to be subdominant. To achieve this, $A_{\lambda}$ is selected to cancel against the other dominant terms in the coupling $c_{aah}$ to keep it small (the term proportional to $A_{\lambda}$ is multiplied by the singlet component of $h$ or the doublet component of $a$, both of which are small in this region and hence $A_{\lambda}$ must be several TeV for this term to be significant). Exacerbated by the large $\lambda$, these large values for $A_{\lambda}$ have a similar impact as large stop masses on the fine-tuning of the Higgs VEV and are therefore undesirable from the perspective of naturalness. However, the lower end of this range with $|A_{\lambda}|\sim 1 \TeV$ suggests that this scenario may be still compatible with the optimally natural region.

The points with a bino-dominated LSP do not require large $A_\lambda$ and lie directly in the region of the NMSSM parameter space with minimal fine-tuning. Note that we find points with $\lambda$ spanning the full range over which we scan (see Table~\ref{tab:HAscan}), which includes points both below and within the large-$\lambda$ regime (where the renormalization group running of $\lambda$ reaches a Landau pole before the GUT scale). In particular, all of these have small $|\mu_{\rm eff}|\sim 200 \GeV$ that both maximises the couplings $c_{a\tilde{\chi}\tilde{\chi}_2}$ and $c_{h\tilde{\chi}\tilde{\chi}_2}$ (through maximising the Higgsino fraction of $\tilde{\chi}_2$ within experimental constraints) and keeps the $\tilde{\chi}_2$ light. 
Note that in the bino-dominated case, $c_{a\tilde{\chi}\tilde{\chi}_2}$ and $c_{h\tilde{\chi}\tilde{\chi}_2}$ cannot be much larger than $0.2$ because, unlike the singlino-dominated case above, the dominant terms in these couplings are proportional to $g_1$ rather $\lambda$. 
It is thus preferred from \eqref{subcrosssectionHA1} that $\tilde{\chi}_2$ be as light as possible. Substantial contributions to the annihilation cross-section also arise from the $s$-channel annihilation via the second pseudoscalar $a_2$, which has mass typically in the range $\sim 300-500\,\GeV$ because of the small values of $\mu_{\rm eff}$. The consequences of direct detection on the coupling $c_{a\tilde{\chi}\tilde{\chi}}$ for a singlet-like $a$ described in Sec.~\ref{CGENMSSM:Mediators} prevent the $s$-channel annihilation via the $a$ from dominantly contributing to the cross-section (in contrast to the singlino-dominated case, where the $\kappa$-dependent term in \eqref{CaChiChiLimitSinglino} allows $c_{a\tilde{\chi}\tilde{\chi}}$ to be much larger), although we do not claim that our scan is exhaustive enough to discount this process altogether. Despite being more appealing, the bino-case is also more severely constrained, particularly by electroweak precision measurements and direct detection.

In Fig.~\ref{fig:HAscanLSP2}, we show a scatter plot of the cross section for spin-independent DM-nucleon scattering versus the DM mass. We also plot the current limits from LUX and the projected limits from LUX with 300 live days of measurements~\cite{LUXtalk}, XENON1T~\cite{XENON:2013SnowMass} and LZ~\cite{LZ:2013SnowMass} (XENON10T has a similar projected sensitivity~\cite{XENON:2013SnowMass}) for a local DM density $\rho_\odot = 0.3 \GeV/\text{cm}^3$.
Note that the uncertainties in the DM distribution invoked above may be correlated with the bounds from direct detection. Indeed, the range $\mathcal{A} \in [0.17,5.3]$ that we consider for the astrophysical uncertainty factor is determined by the uncertainties in the slope parameter $\gamma$, the scale radius $R_s$ and the local DM density $\rho_\odot$~\cite{Agrawal:2014oha,Calore:2014nla}. 
To illustrate how the direct-detection bounds are affected by the latter uncertainty, we plot them as shaded bands for local DM densities between $\rho_\odot=0.2 \GeV/\text{cm}^3$ and $\rho_\odot=0.6 \GeV/\text{cm}^3$. The LUX collaboration plans to update their analysis to 300 live days of measurements in the near future, extending the sensitivity by a factor of 5~\cite{LUXtalk}. This would 
test a considerable portion of the singlino-like points and almost the entirety of the bino-like points from our scan of the $ha$-channel. The vast majority of points with a singlino-like LSP would then be probed by XENON1T, although we find a few such points that would evade even the projected LZ limits.

\subsection{Parameter scan for the $b \bar b$-channel}\label{BinoResonance}
\begin{figure}[bt]\centering
\begin{subfigure}{0.49\linewidth}
\includegraphics[width=\linewidth]{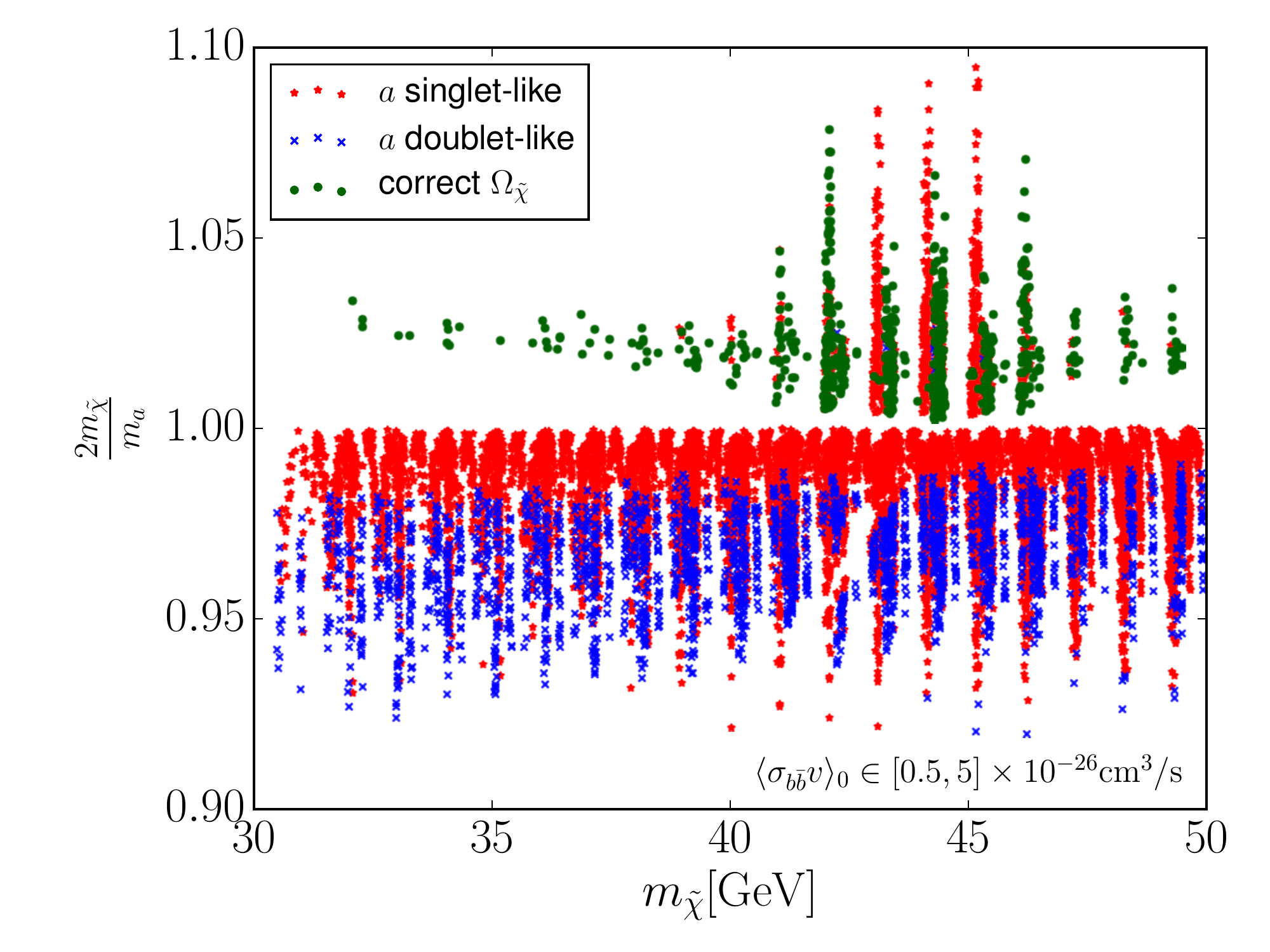}
\caption{
\label{fig:resonance}}
\end{subfigure}
\hfill
\begin{subfigure}{0.49\linewidth}
\includegraphics[width=\linewidth]{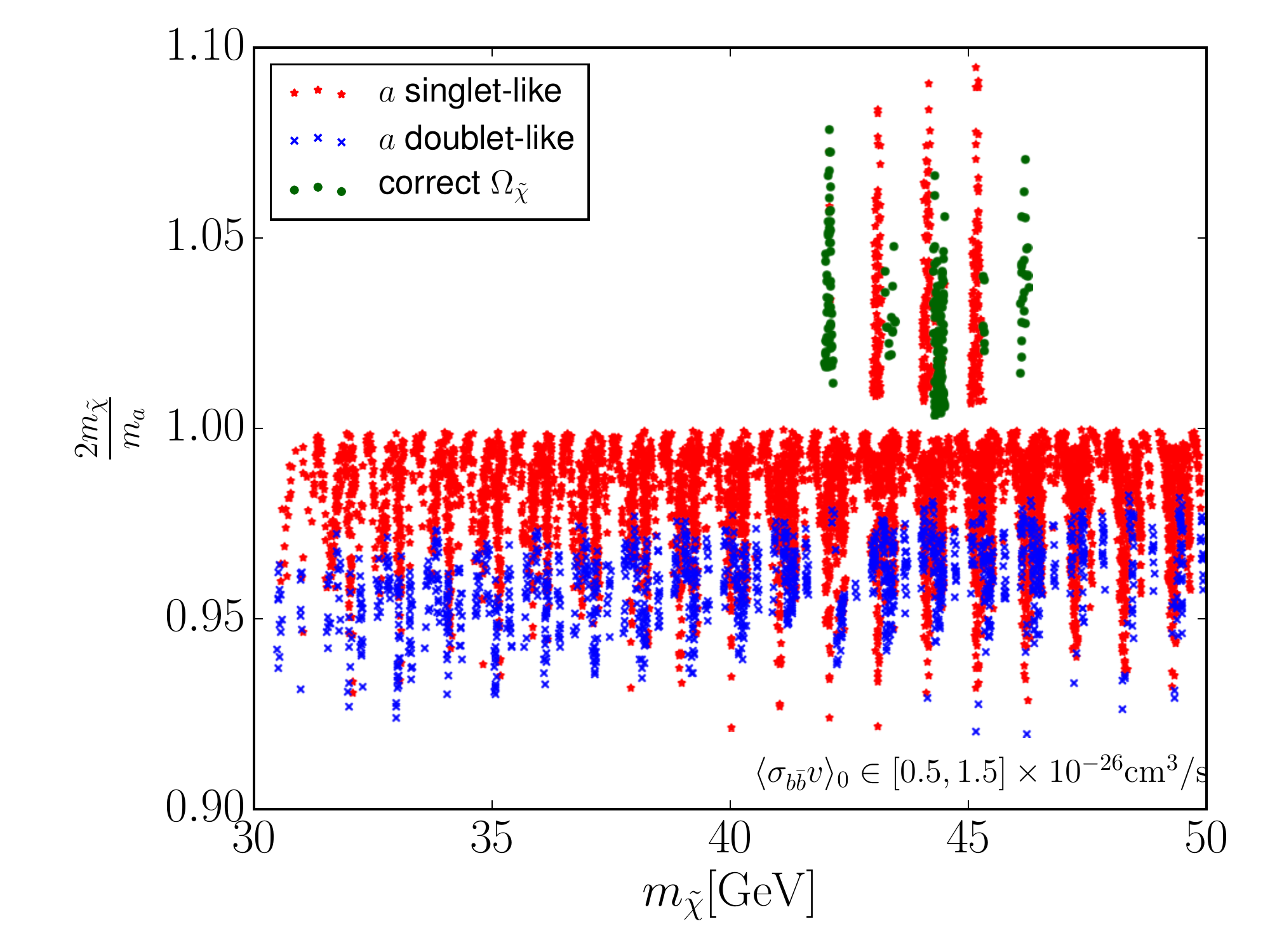}
\caption{
\label{fig:resonanceD}}
\end{subfigure}
\caption{ 
Scatter plots of the LSP mass $m_{\tilde{\chi}}$ versus the ratio $2 m_{\tilde{\chi}}/m_{a}$. Green points have an LSP relic abundance within $10\%$ of the DM density determined by Planck, whereas for red and blue points it is smaller.
Red and blue points have a pseudoscalar $a$ which consists dominantly of $s_I$ and $H_I$, respectively. For the left panel (a), the late-time cross section is within the range $\langle\sigma_{b\bar{b}} v\rangle_0\in [0.5,5] \times 10^{-26}  \, \text{cm}^3/\text{s}$, where the upper limit satisfies the published dwarf-spheroidal bounds from Fermi-LAT~\cite{Ackermann:2013yva}. For the right panel (b), the upper limit satisfies the latest, preliminary dwarf-spheroidal bounds from Fermi-LAT~\cite{AndersonTalk:2014FermiSymposium}, $\langle\sigma_{b\bar{b}} v\rangle_0 \in [0.5,1.5] \times 10^{-26} \, \text{cm}^3/\text{s}$.}
\label{fig:resonanceComb}
\end{figure}

\begin{figure}[bt]\centering
\includegraphics[width=0.65\linewidth]{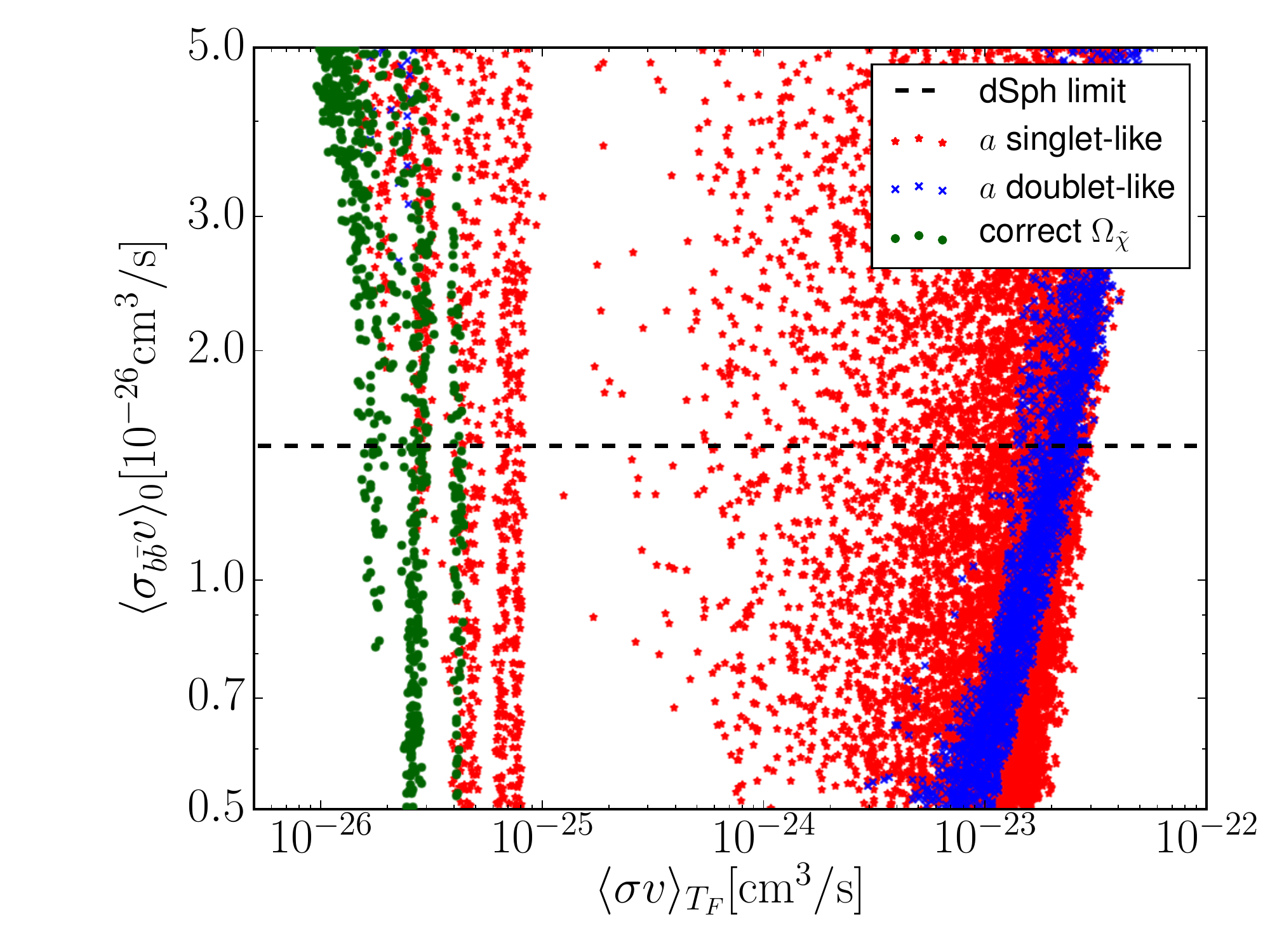}
\caption{Scatter plot of the freeze-out cross section $\langle\sigma_{b\bar{b}} v \rangle_{T_{\rm F}}$ versus the late-time cross section $\langle \sigma_{b\bar{b}} v \rangle_0$. Green points have an LSP relic abundance within $10\%$ of the DM density determined by Planck, whereas for red and blue points it is smaller. Red and blue points have a pseudoscalar $a$ which consists dominantly of $s_I$ and $H_I$, respectively.}
\label{fig:crosssections}
\end{figure}

\begin{figure}[bt]\centering
\begin{subfigure}{0.49\linewidth}
\includegraphics[width=\linewidth]{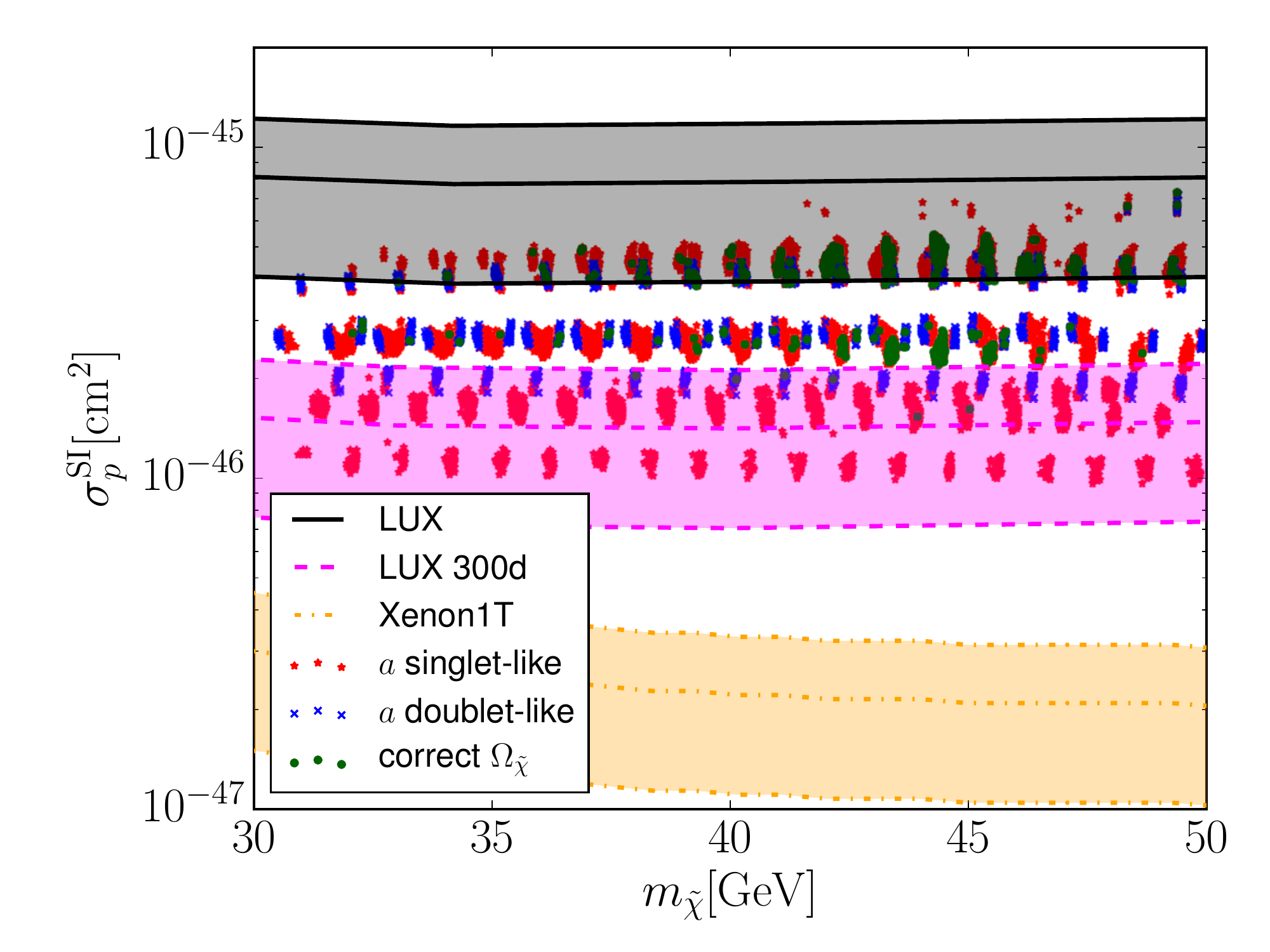}
\caption{\label{fig:bbDD}}
\end{subfigure}
\hfill
\begin{subfigure}{0.49\linewidth}
\includegraphics[width=\linewidth]{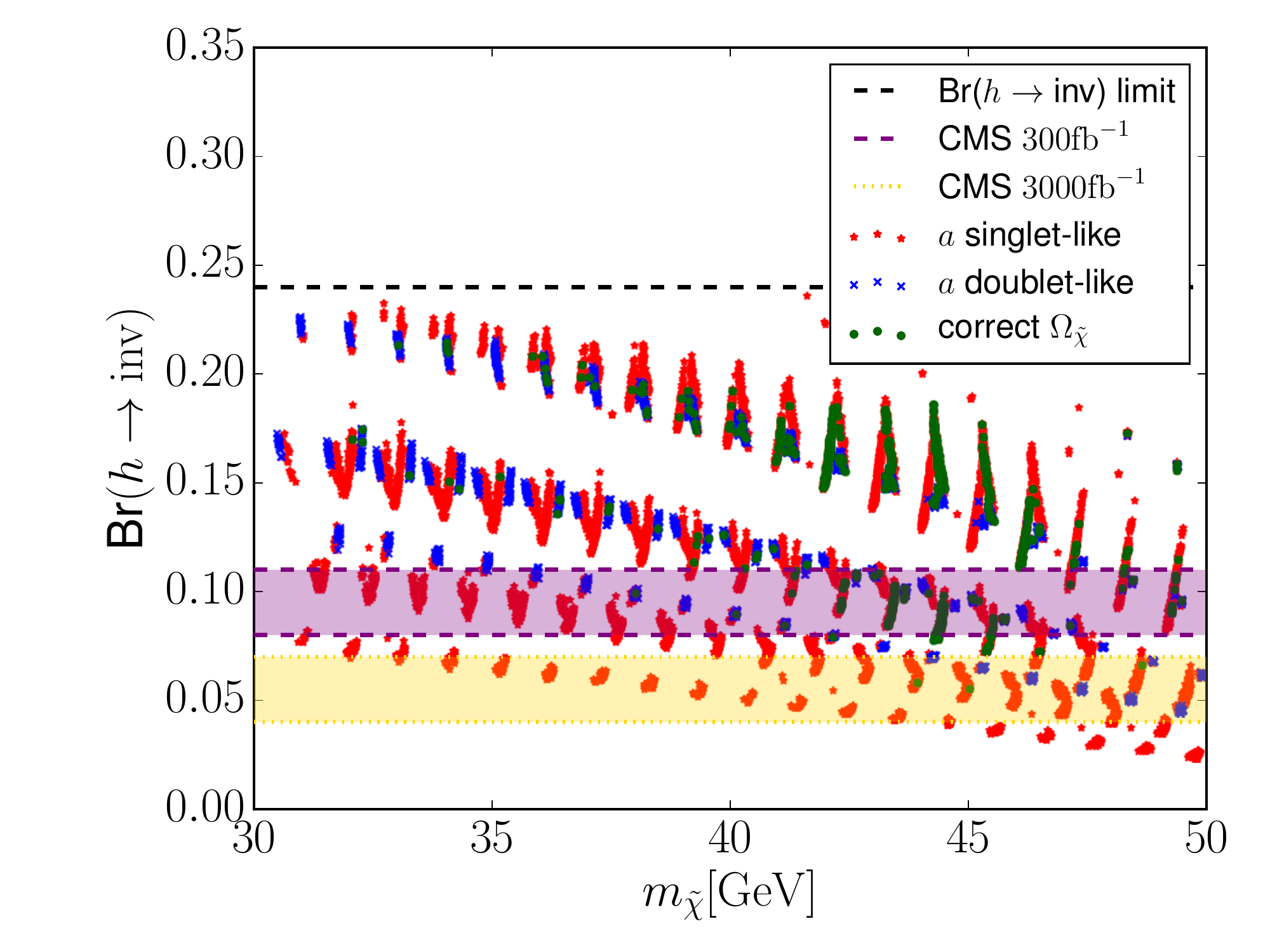}
\caption{\label{fig:bbHinv}}
\end{subfigure}
\caption{The left panel (a) shows the spin-independent cross section for scattering of LSPs on nucelons versus the LSP mass, while the right panel (b) shows the branching ratio of invisible Higgs decays versus the LSP mass. 
Green points have an LSP relic abundance within $10\%$ of the DM density determined by Planck, whereas for red and blue points it is smaller.
Red and blue points have a pseudoscalar $a$ which consists dominantly of $s_I$ and $H_I$, respectively. 
}
\label{fig:DDHinv}
\end{figure}

We have performed a grid scan for the $b \bar b$-channel and a bino-dominated LSP using \texttt{NMSSMTools\, 4.4.0} \cite{Ellwanger:2004xm,Ellwanger:2005dv,Belanger:2005kh}. We have fixed the masses of the squarks, sleptons, gluino and wino to the values mentioned previously and have scanned over the remaining free parameters of the scale-invariant NMSSM with ranges and step sizes as summarised in Table~\ref{ScanSumTable}. The range for $M_1$ is chosen such that the LSP is dominantly bino-like. 
We show scatter plots of the results from this scan in Figs.~\ref{fig:resonanceComb} to~\ref{fig:DDHinv}. As expected, we find that annihilation into $b \bar b$ is the dominant channel. 
We demand that at least $80\%$ of the annihilations are into $b\bar b$ states to remove the very few points with a slightly smaller branching ratio to $b \bar b$.
The cross sections are restricted to lie in the range $\langle \sigma_{b\bar{b}} v \rangle_{0}\in [0.5,5] \times 10^{-26} \, \text{cm}^3/\text{s}$ required to explain the GCE discussed in Sec.~\ref{GCE-section}. Green points have an LSP abundance in the range $0.107<\Omega_{\tilde{\chi}}h^2<0.131$ consistent with the observed DM abundance as discussed above.
Red and blue points, on the other hand, have smaller relic densities. We keep the latter points because, in  Sec.~\ref{NLSPdecays}, we present a model in which NLSP decays can regenerate an under-produced LSP. We do not include points with abundances above the aforementioned range (which could become viable e.g.~if there is significant entropy production after freeze-out). Blue points have a pseudoscalar mediator with a greater doublet fraction than singlet fraction, while for the red points the singlet fraction is larger.

\begin{table}[bt]
\begin{center}
\begin{tabular}{l c  c  c  c c c  c c}
\toprule
& $A_\lambda$ [GeV] & $A_\kappa$  [GeV] & $|\mu_{\rm eff}|$  [GeV] & $\frac{\mu_{\rm eff}}{|\mu_{\rm eff}|}$ &  $M_1$  [GeV] &$\lambda$ & $\kappa$ & $\tan \beta$\\
\midrule
 range & [-1000,1000] & [-300,300] & [300,600] & $\pm1 $ & [30,50] & [0.8,1.4] & [0.8,1.4] & [1,4]\\
 step & 10 & 1 & 100 & . & 1 & 0.05 & 0.05 & 1\\
\bottomrule
\end{tabular}
\end{center}
\caption{Parameter ranges and step sizes used in the grid scan for the bino-dominated LSP annihilating into $b \bar b$. For $\mu_{\rm eff}$, the scan is over both possible signs. \label{ScanSumTable}}
\end{table}

We have seen in  Sec.~\ref{AnalyticalArgumentBB} that constraints on the Higgsino fraction make it difficult to obtain the annihilation cross section required for the GCE without an additional enhancement. Such an enhancement arises if $m_{a} \simeq 2 m_{\tilde{\chi}}$ so that the annihilation happens close to resonance. 
Focusing on this enhancement, we have scanned over the region $1/1.1< 2m_{\tilde{\chi}} / m_a<1/0.9$, where the ratio $2m_{\tilde{\chi}} / m_a$ measures the proximity to resonance. 
We show a scatter plot of this ratio versus the LSP mass in Fig.~\ref{fig:resonanceComb}. 
Notice that only very few points come close to the upper or lower limit of our scan region (which approximately coincides with the plotted range) and that the vast majority of points lie closer to resonance, illustrating the necessity of this enhancement.\footnote{It should be noted that higher-order quantum corrections to the pseudoscalar mass than those included in \texttt{NMSSMTools\;4.4.0} may perturb it sufficiently to significantly change the cross-section in the vicinity of the resonance. However, we do not expect that these will qualitatively change our results, but would rather exchange points with otherwise similar characteristics in and out of the region satisfying the fits to the GCE.}
Points in Fig.~\ref{fig:resonance} have cross sections in the range $\langle\sigma_{b\bar{b}} v\rangle_0\in [0.5,5]\times 10^{-26} \, \text{cm}^3/\text{s}$,
where the upper limit is determined by the published dwarf-spheroidal bounds from Fermi-LAT~\cite{Ackermann:2013yva}.
Points in Fig.~\ref{fig:resonanceD} are restricted to lie in the smaller range $\langle\sigma_{b\bar{b}} v\rangle_0 \in [0.5,1.5]\times 10^{-26} \, \text{cm}^3/\text{s}$,
where the upper limit satisfies the latest, preliminary dwarf-spheroidal bounds from Fermi-LAT~\cite{AndersonTalk:2014FermiSymposium}. Note that we do not apply this second range to other plots in this section.

A scatter plot of the cross section at freeze-out versus that at the present time is shown in Fig.~\ref{fig:crosssections}.  Notice that for many points, the former is orders of magnitude larger than the latter. This can be understood by noting that when
$m_a$ is close to $2 m_{\tilde{\chi}}$, the annihilation is enhanced by the $s$-channel propagator at small velocities.
Then if $m_{a} > 2 m_{\tilde{\chi}}$, the annihilation can happen even closer to resonance during freeze-out due to the larger kinetic energy of the LSP in the thermal plasma. This enhances the cross section at freeze-out relative to that at the present time~\cite{Griest:1990kh} 
and thus reduces the relic abundance (as follows from Fig.~\ref{fig:crosssections}, these points typically have $\langle\sigma_{b\bar{b}} v \rangle_{T_{\rm F}}\approx \mathcal{O}(10^3)\times \langle \sigma_{b\bar{b}} v \rangle_{0}$, leading to LSP relic densities $\Omega_{\tilde{\chi}} h^2 \lesssim 10^{-3}$). Indeed, the majority of points in Fig.~\ref{fig:resonanceComb} that correspond to underabundant LSPs (in red and blue) have $m_{a} > 2 m_{\tilde{\chi}}$.
Notice also that points with a singlet-dominated pseudoscalar (in red) tend to be closer to resonance than those for which it is dominantly doublet-like (in blue). This can be understood from the fact that the singlet does not couple to the SM and the doublet couples more strongly to neutralinos, so the product of couplings $c_{a \tilde{\chi} \tilde{\chi}} \cdot c_{a b \bar b}$ becomes smaller with growing singlet-admixture to the pseudoscalar (cf.~Eqs.~(\ref{ga1neut}) and (\ref{ga1bb})).
On the other hand, notice from Fig.~\ref{fig:resonanceComb} that all points for which the LSP abundance accounts for the DM (in green) have $m_{a} < 2 m_{\tilde{\chi}}$. The larger kinetic energies at freeze-out then reduce the cross section for the process $\tilde{\chi} \tilde{\chi} \rightarrow a^* \rightarrow b \bar b$ compared to the corresponding cross section in our galaxy. 
For the green points with LSP masses close to $m_Z/2$,
this is counterbalanced by the resonant $Z$-boson mediated annihilation. This process does not contribute significantly to the cross section in our galaxy as it is helicity-suppressed in the $s$-wave and the width of the $Z$-boson is too broad, but becomes important in the early universe (where the $p$-wave becomes important) and raises the cross section to the required value to account for the DM.  
This effect was discussed previously in Refs.~\cite{Cheung:2014lqa,Guo:2014gra,Cao:2014efa}. 

Away from the $Z$ resonance, both the cross section at freeze-out and today are simply dominated by $a$-mediation to $b \bar b$-pairs, with the former being smaller than the latter. The freeze-out cross section is determined by the requirement that the LSP accounts for the DM, $\langle \sigma v \rangle_{T_F} \sim 2\times 10^{-26} \text{cm}^3/\text{s}$. Correspondingly, these points tend to have late-time cross sections that are larger than this, $\langle \sigma_{b\bar{b}} v \rangle_{0} \gtrsim 3\times 10^{-26} \text{cm}^3/\text{s}$. These large cross sections are still compatible with the GCE because of the relatively large uncertainties in the fits in Tables~\ref{tab:data} and \ref{tab:bestfit} and the large astrophysical uncertainty factor $\mathcal{A}$.
Indeed, if we impose the preliminary dwarf-spheroidal bounds as in Fig.~\ref{fig:resonanceD}, $\langle \sigma_{b \bar b} v \rangle_0 \lesssim 1.5\times 10^{-26}\,\mathrm{cm^3}/\mathrm{s}$ 
for $m_{_{\rm DM}} \sim 30 - 50 \GeV$~\cite{AndersonTalk:2014FermiSymposium}, only green points with masses close to $m_Z/2$ survive. 
Similarly, points for which the LSP is underabundant (in red and blue) with $ m_{a} < 2m_{\tilde{\chi}} $ in Fig.~\ref{fig:resonanceComb} have relatively large late-time cross sections. For these points, however, the thermal broadening of the resonance does not sufficiently reduce the freeze-out cross section to account for the DM.

In Fig.~\ref{fig:bbDD}, we show a scatter plot of the cross-section for spin-independent LSP-nucleon scattering versus the LSP mass.  As discussed in Sec.~\ref{Higgs-pseudoscalar-channel} for Fig.~\ref{fig:HAscanLSP2}, we plot the current limits from LUX and the projected limits from LUX with 300 live days of measurements~\cite{LUXtalk} and XENON1T~\cite{XENON:2013SnowMass} as shaded bands covering local DM densities between $\rho_\odot=0.2 \GeV/\text{cm}^3$ and $\rho_\odot=0.6 \GeV/\text{cm}^3$. The LUX collaboration 
will test a considerable portion of the points from our scan for the $b \bar b$-channel in the near future. The remaining points will be tested by XENON1T.

In Fig.~\ref{fig:bbHinv}, we show a scatter plot of the branching ratio for Higgs-to-invisible decays versus the LSP mass. The better precision in the measurement of the Higgs properties projected for Run 2 of the LHC and 
after the high-luminosity upgrade will allow for another way of probing our points. Indeed, the CMS collaboration expects to probe branching ratios for invisible Higgs decays at the level of ${\rm Br}(h\rightarrow {\rm inv})\in [8,11]\%$ at $1\sigma$ for an integrated luminosity of 300 fb$^{-1}$ and ${\rm Br}(h\rightarrow {\rm inv})\in [4,7]\%$ at $1\sigma$ for 3000 fb$^{-1}$~\cite{CMS:2013xfa}. We have plotted these ranges as colored bands in Fig.~\ref{fig:bbHinv}. As one can see, if no significant deviations from a SM-like Higgs are found at the LHC during the high-luminosity run, only a small slice of parameter space with DM masses in the range $46$ GeV $\lesssim m_{\tilde{\chi}} \lesssim m_h/2$ would survive. 

To conclude, we find a small number of points (around 5 $\%$ of all points which do not overclose the universe) that can simultaneously accommodate the GCE via the $b \bar b$-channel and the observed DM abundance from thermal freeze-out. The majority of our points, however, have an underabundant LSP. In the next section, we discuss a model where the right abundance of LSPs is obtained from late NLSP decays. This introduces a new degree of freedom that would allow for the observed DM abundance to be separately accounted for from the GCE. The majority of the viable points will be tested by LUX~\cite{LZ:2013SnowMass} this year, while the remainder will be probed by XENON1T~\cite{XENON:2013SnowMass}.

\section{Seesaw extension with sneutrino LSPs or NLSPs}
\label{NLSPdecays}

\subsection{Seesaw in the scale-invariant NMSSM}\label{NeuIntro}
We consider a simple extension of the NMSSM, the seesaw~\cite{Minkowski:1977sc, Yanagida:1980, Glashow:1979vf, Gell-Mann:1980vs, Mohapatra:1980ia}, that provides new possibilities for the dark sector and simultaneously addresses the generation of neutrino masses absent in the original model. 
We introduce three right-handed neutrinos and their scalar superpartners, i.e. gauge-singlet superfields $N_i$ (the $i$ denotes the flavour index). The NMSSM superpotential is extended to include the terms
\be
W_{\rm seesaw} \, = \, \lambda_{N} \, SNN \, + \, y_{\!_N} L\cdot H_u \, N \, , \label{NeuSuper}
\ee
and the soft supersymmetry breaking terms are
\be
\mathcal{L_{\text{soft-seesaw}}} \, = \, -\frac{1}{2}m^2_{\tilde{N}}\tilde{N}\tilde{N}^* \, - \, (\lambda_NA_{\lambda_N}S\tilde{N} \tilde{N} \, + \, y_{\!_N} A_{y_{\!_N}} \tilde{L} \cdot H_u \, \tilde{N} \, + \, \text{h.c.}) \, ,
\label{NeuSoft}
\ee
where we have suppressed the flavour indices on the supermultiplets $N$ and $L$, along with those in the parameters 
$\lambda_N,y_{\!_N},m^2_{\tilde{N}},A_{\lambda_N}$ and $A_{y_{\!_N}}$.

The Majorana masses $m_N=2\lambda_N v_s$ of the fermionic components are generated by the VEV of the scalar component of the singlet $S$ and because of this link with the Higgs sector are naturally at most of order TeV.
Provided that the Majorana masses are hierarchically larger than the Dirac masses, the seesaw mechanism yields active neutrino masses
\begin{equation}
m_{\nu} \, \approx \, - \frac{ v^2\sin^2 \hspace{-.05cm} \beta}{2 \, v_s} \,y_{\!_N}\lambda_N^{-1}y_{\!_N}^T \, .
\end{equation}
In order for the active neutrino masses to be smaller than $\sim \eV$, the Yukawa couplings then need to be $|y_{\!_N}|\lesssim 10^{-6}$.

As $y_{\!_N}$ is much smaller than the other couplings, the model is approximately invariant under a $\mathbb{Z}_2$ parity under which only the right-handed neutrino supermultiplet is odd (cf.~Eqs.~\eqref{NeuSuper} and \eqref{NeuSoft}). Processes that involve odd numbers of these particles are therefore suppressed by the small coupling $y_{\!_N}$, whether mediated by the Yukawa coupling or mass mixing. 
If the lightest of the right-handed sneutrinos is not the LSP, this approximate symmetry can have the interesting implication that this particle is relatively long-lived. We will later exploit this property in connection with the non-thermal generation of DM.

In order to simplify the discussion, we restrict to just one flavour of right-handed neutrinos in the following.
We assume that $CP$ is conserved in the sneutrino sector and take all couplings and masses to be real. The real and imaginary parts of the sneutrino  $\tilde{N}=(\tilde{N}_R+i\tilde{N}_I)/\sqrt{2}$ then do not mix with each other. Mass mixing with the left-handed sneutrino sector is proportional to the small coupling $y_{\!_N}$ and can be ignored. The masses of the right-handed sneutrinos are then given by
\bea
m_{\tilde{N}_{I,R}}^2 \, = \, m_{\tilde{N}}^2 \, + \, 2\lambda_N v_s^2 \, \pm \, \lambda_N \, (\lambda \, v^2\sin2\beta - 2A_{\lambda_N}v_s - 2\kappa \, v_s^2) \, ,
\eea
where the $+$ and $-$ correspond to $\tilde{N}_I$ and $\tilde{N}_R$ respectively. 
These masses can be regarded as independent free parameters, simply exchanged for the parameters $m_{\tilde{N}}^2$ and $A_{\lambda_N}$ in the Lagrangian. This in particular fixes 
\bea
A_{\lambda_N} \, = \, \frac{1}{2v_s}\left(\lambda \, v^2\sin 2\beta-2\kappa \, v_s^2+\frac{1}{2\lambda_N}(m_{\tilde{N}_R}^2-m_{\tilde{N}_I}^2)\right) \, .\label{AlambdaN}
\eea
The size of $A_{\lambda_N}$ is then partly controlled by the mass-splitting between the $CP$-even and $CP$-odd sneutrino.

The possibility that the sneutrino is the LSP has been discussed previously in \cite{Cerdeno:2009dv} and \cite{Cerdeno:2014cda}, the latter of which also provides a cursory illustration of compatibility with the GCE. Because scalar DM respects Bose-Einstein statistics and its annihilation is not $p$-wave suppressed, it is easier to find amenable regions of parameter space 
than for the neutralino, although direct-detection constraints are still strong. Fits to the spectrum of the GCE for annihilating sneutrino DM (including scalar annihilation channels) were examined more closely in \cite{Cerdeno:2015ega}. Collider signatures of this scenario were discussed in \cite{Cerdeno:2013oya}. 

In Sec.~\ref{SneutrinoLSP}, we begin by exhibiting the case of sneutrino DM annihilating dominantly into Higgs scalars and pseudoscalars. Then in Sec.~\ref{SneutrinoNLSP}, under the assumption of an especially small coupling $y_{\!_N}$, we make use of the approximate $\mathbb{Z}_2$ invariance in the right-handed sneutrino sector to point-out the distinct possibility of metastable sneutrino NLSPs that decay into neutralino LSPs. This is applicable to the parameter regions in Sec.~\ref{CGENMSSM} where the annihilation cross-section of the neutralino LSP in the early universe is too large.

\subsection{Sneutrino LSP}
\label{SneutrinoLSP}
We now analyse the possibility of sneutrino DM to explain the GCE.  We focus on annihilation into Higgs scalars and pseudoscalars, with the mass and cross-sections prescribed in Sec.~\ref{CGESection}, complementing the recent analysis for smaller values of $\lambda$ and $\kappa$ in Ref.~\cite{Cerdeno:2015ega}.
The case of annihilation into light quarks has been discussed previously in Ref.~\cite{Cerdeno:2014cda}. For simplicity, we consider only one flavour of right-handed (s)neutrinos and neglect flavour mixing. Furthermore, we assume that the mass eigenstate $\tilde{N}_I$ is the LSP (henceforth denoted $\tilde{N}$ except where it is necessary to distinguish it from the other sneutrino), although the subsequent analysis would be similar if $\tilde{N}_R$ was the LSP.

The couplings of the right-handed (s)neutrino to the Higgs, the lightest pseudoscalar and the neutralinos are listed in Appendix \ref{app:couplings}. Note that these couplings are controlled by either $\lambda_N$ or $\lambda_N A_{\lambda_N}$. The coupling $\lambda_N$ is constrained in several ways, the first of which we discuss is by direct-detection experiments. Analogously to Sec.~\ref{CGENMSSM:Composition}, we assume isospin symmetry and obtain~\cite{Jungman:1995df,Andreas:2008xy} 
\begin{align}
a_u&=-\frac{m_u}{2\sqrt{2} v} \frac{\mathcal{S}_{12}}{\sin\beta} \frac{c_{h\tilde N\tilde N}}{2 m_h^2  m_{\tilde N}}\, ,&
a_d&=-\frac{m_d}{2\sqrt{2} v} \frac{\mathcal{S}_{11}}{\cos\beta} \frac{c_{h\tilde N\tilde N}}{2  m_h^2  m_{\tilde N}}
\end{align}
for the coefficients which enter in Eqs.~\eqref{eq:DD} and \eqref{eq:fp}. Assuming a Higgs with SM-like couplings and that no other scalars contribute to LSP-nucleon scattering, for a sneutrino mass $m_{\tilde N}\sim 100 \GeV$, the bound on the direct-detection cross section by LUX gives $\lambda_N\lesssim 0.1$.

\begin{figure}[btp]\centering
\begin{subfigure}{0.3\linewidth}
\FMDG{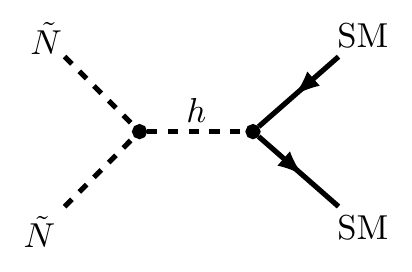}
\end{subfigure}
\hspace{1cm}
\begin{subfigure}{0.3\linewidth}
\FMDG{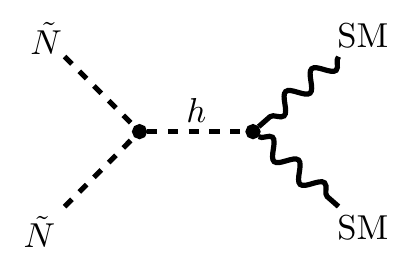}
\end{subfigure}
\caption{Feynman diagram contributing to sneutrino annihilation to SM fermions and gauge bosons.}
\label{FeynmanDiagramsSneutrinosSM}
\end{figure}

The annihilation of sneutrinos into SM fermions and gauge bosons is mediated via $s$-channel Higgs exchange, as shown in Fig.~\ref{FeynmanDiagramsSneutrinosSM}. The cross section is given by
\bea
\langle\sigma_{_{\rm SM}} v\rangle_0 \, \approx \, \frac{c_{h\tilde{N}\tilde{N}}^2}{m_{\tilde{N}}(4 \, m_{\tilde{N}}^2-m_h^2)^2} \, \Gamma_h(2m_{\tilde{N}}) \, ,\label{SneutoSMCS}
\eea
where $c_{h\tilde{N}\tilde{N}}$ is the coupling of two sneutrinos to the Higgs and $\Gamma_h(2m_{\tilde{N}})$ is the decay width of the SM Higgs evaluated for a Higgs mass of $2m_{\tilde{N}}$. For $m_{\tilde{N}}\gtrsim m_W$, the dominant process here is $\tilde N \tilde N \to h^* \to W^+ W^-$, due to the large gauge couplings. The cross section is controlled by the coupling $c_{h\tilde{N}\tilde{N}}$ which is given in terms of the couplings and soft masses in Eq.~\eqref{cNNh}. It scales with the dimensionful parameters $v$, $v_s$ and $A_{\lambda_N}$. The latter two are multiplied by the singlet fraction of the Higgs which needs to be small in order to be compatible with the measured Higgs couplings, typically $\mathcal{S}_{13}\sim 0.1$. Since we restrict ourselves to $v_s$ and $A_{\lambda_N}$ below TeV, the contribution proportional to $v$ dominates the coupling $c_{h\tilde{N}\tilde{N}}$ and we can approximate it as 
\begin{equation}
|c_{h\tilde N \tilde N}| \, \approx \,  \sqrt{2} \, \lambda_N \lambda \, v \sin 2\beta \, \approx \, 15  \GeV \,\frac{ \lambda_N \lambda}{0.1} \, \frac{\sin2\beta}{0.6}
\label{eqn:SneutoSM}
\end{equation}
for a SM-like Higgs with $S_{11}\approx \cos\beta$ and $S_{12}\approx \sin\beta$. Demanding that the cross section for annihilation into SM particles be much smaller than the one required to explain the GCE,
$\langle\sigma_{_{\rm SM}} v\rangle_0 \ll  2\cdot 10^{-26}\text{cm}^3/\text{s}$, gives the bound $|c_{h\tilde{N}\tilde{N}}|< 20\,\text{GeV}$ for an LSP mass above the Higgs resonance and up to $\sim 10\,\GeV$ below the $W$ threshold. The bound is even stronger above the $W$ threshold, with $ |c_{h\tilde N\tilde N}|\lesssim 5\,-\,13\, \GeV$ for masses $m_{\tilde N}=100\,-\,150\,\GeV$. This translates into the bound $\lambda_N \lesssim 0.1$. Note that this bound becomes stronger if the astrophysical-uncertainty factor $\mathcal{A}$ is large and smaller cross sections are sufficient to generate the GCE (see Sec.~\ref{GCE-section}). Also note that, for $m_{\tilde{N}}\lesssim m_W$, the largest branching fraction is into $b \bar b$-pairs, although the branching fraction into $WW^*$-pairs is not much less important except for a small range of masses above the Higgs resonance that we do not consider further (below which constraints from invisible Higgs decays into sneutrinos give a bound on $c_{h\tilde N \tilde N}$ that is too strong for \eqref{SneutoSMCS} to give a large enough cross section for the GCE).

\begin{figure}[btp]\centering
\begin{subfigure}{0.3\linewidth}\centering
\FMDG{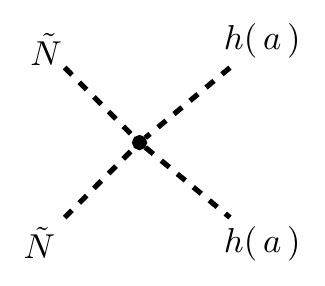}
\end{subfigure}
\hspace{1ex}
\begin{subfigure}{0.3\linewidth}\centering
\FMDG{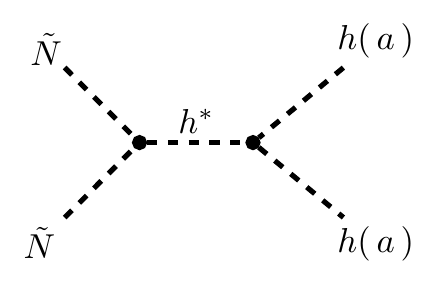}
\end{subfigure}
\hspace{1ex}
\begin{subfigure}{0.3\linewidth}\centering
\FMDG{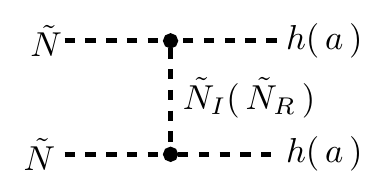}
\end{subfigure}
\caption{Feynman diagrams contributing to sneutrino annihilation to $hh$- or $aa$-pairs. Note that there is a $u$-channel diagram in addition to the $t$-channel one.}
\label{FeynmanDiagramsSneutrinos}
\end{figure}

The dominant processes that allow for the annihilation of sneutrinos into scalars or pseudoscalars from the Higgs and singlet sector are 1) via the four-point scalar coupling, 2) via $s$-channel mediation of the Higgs and 3) via $t$- and $u$-exchange of a sneutrino. We show the Feynman diagrams for these processes in Fig.~\ref{FeynmanDiagramsSneutrinos}. 
The cross section for the annihilation of sneutrinos into Higgs pairs is given by
\begin{equation}
 \langle\sigma_{hh} v\rangle_0\, \approx \, \frac{\sqrt{m_{\tilde N}^2-m_h^2}}{64 \pi \, m_{\tilde N}^3} \left| c_{hh\tilde N \tilde N} - \frac{ c_{h\tilde N\tilde N} \, c_{hhh}}{4 m_{\tilde N}^2-m_h^2} -\frac{2 \, c_{h\tilde N \tilde N}^2}{m_h^2-2m_{\tilde N}^2} \right|^2\,,
\end{equation}
where $c_{hh\tilde N \tilde N}$ is the coupling between two Higgs and two sneutrinos and $c_{hhh}$ is the triple Higgs coupling.
We find that it is difficult to obtain the required cross section to generate the GCE via the $hh$-channel while keeping annihilation into  SM fermions and gauge bosons subdominant. This is due to the fact that the $s$-, $t$- and $u$-channel processes are determined by the same coupling $c_{h\tilde N\tilde N}$ as the annihilation into SM particles, while the coupling $c_{hh\tilde N \tilde N}$ relevant for the direct annihilation is also controlled by $\lambda_N \lesssim 0.1$ (see Eq.~\eqref{chhNN}).
In addition, the annihilation into a Higgs pair is phase-space suppressed because it needs to happen close to threshold for a good fit to the GCE. This makes it difficult to enhance the $hh$-channel over annihilation to SM particles.
We have performed a short uniformly random scan of the parameter space in the natural region of the NMSSM with the LSP mass constrained to lie within $30$ GeV of the Higgs mass threshold. 
However, the branching fraction into $hh$ was never found to exceed $\sim 40 \%$ (with the remaining annihilations dominantly into $W^+ W^-$).
Although we did not search more extensively and cannot conclusively rule this possibility out, we will not consider it further.

Annihilations into singlet-like scalars or pseudoscalars are more promising for generating the GCE because the sneutrino can couple more strongly to relatively pure singlets than to doublet-like states. For a SM-like Higgs with small singlet fraction, the largest terms in the couplings $c_{h\tilde{N} \tilde{N}}$ and $c_{hh\tilde{N} \tilde{N}}$ (see Eqs.~\eqref{cNNh} and \eqref{chhNN}) are proportional to $\lambda \sin 2\beta$, whereas the terms proportional to the singlet fraction can make the corresponding couplings for a singlet-like state larger without affecting the annihilation cross-section into other SM particles such as $W^+W^-$. Scenarios with a light, $CP$-even singlet-like state (possibly lighter than the Higgs) were studied in \cite{Cerdeno:2015ega} for small $\lambda$ and $\kappa<0.1$.
However, such a light singlet is difficult to arrange in the more natural region of the NMSSM with large $\lambda$ and $\kappa$ as this state typically mixes with the Higgs and the mixing is enhanced with $\lambda$, leading to large deviations in the Higgs couplings from SM values. The mixing must then be tuned small in order for the Higgs couplings to be consistent with experiment (see Eq.~\eqref{Higgs-singlet-mixing}).

We will instead focus on the annihilation channel of light pseudoscalars.
These can arise naturally as the pseudo-Nambu-Goldstone boson of an approximate, spontaneously broken $R$-symmetry in the Higgs sector if $A_{\lambda}$ and $A_{\kappa}$ are relatively small.
The thermally-averaged annihilation cross section for this process is 
\bea
\langle\sigma_{aa} v\rangle_{T_F} \, \approx \, \frac{\sqrt{m_{\tilde{N}}^2-m_a^2}}{64\, \pi \, m_{\tilde{N}}^3} \, \, \, \left|c_{aa \tilde{N}\tilde{N}}-\frac{ c_{h \tilde{N}\tilde{N}} \, c_{haa}}{4m_{\tilde{N}}^2-m_h^2} - \frac{2 \, c_{a\tilde N_I \tilde N_R}^2}{m_a^2-m_{\tilde N_I}^2-m_{\tilde N_R}^2}\right|^2\label{Sneutoa1} \, ,
\eea
where $c_{h a a}$ is the Higgs coupling to two pseudoscalars and $c_{a\tilde N_I \tilde N_R}$, $c_{aa \tilde{N}\tilde{N}}$ and $c_{h \tilde{N}\tilde{N}}$ are the couplings of two sneutrinos to the pseudoscalar, two pseudoscalars and the Higgs, respectively. We have assumed that the cross-section is well-approximated by its velocity-independent component, although care must be taken in performing the thermal-average close to the annihilation threshold \cite{Gondolo:1990dk,Griest:1990kh} (we also ignore the Higgs resonance).

We will now determine regions for the couplings $\lambda_N$ and $A_{\lambda_N}$ (or, equivalently, $m_{\tilde{N}_R}$) where $\langle\sigma_{aa} v\rangle_{T_F}\sim 3\times 10^{-26}\text{cm}^3/\text{s}$ can be achieved so that the observed DM abundance is obtained from thermal freeze-out. We will use these regions for our scan below. To this end, we determine parameter values for which each term in \eqref{Sneutoa1} individually gives the required freeze-out cross-section. Since generically all three terms contribute, this gives approximate upper bounds for the corresponding parameters where this annihilation channel can be viable. The first two terms provide suitable regions for $\lambda_N$, while the third points to an appropriate range for $m_{\tilde{N}_R}$. Let us first consider the case that the four-point scalar coupling dominates the cross section.
The expression for $c_{aa\tilde{N}\tilde{N}}$ in terms of the couplings and soft masses is given in~(\ref{cNNaa}), where $\mathcal{P}$ is a diagonalisation matrix that relates the mass eigenstates $a_i$ of the pseudoscalars to the interaction eigenstates according to
\begin{equation}
\left( \begin{matrix}
  a_1 \\  a_2
 \end{matrix}
\right) \, = \, \mathcal{P} \, 
\left( \begin{matrix}
  h_{d,I}^0 \\  h_{u,I}^0 \\  s_I   
 \end{matrix} \right)\, .
\end{equation}
Assuming a very singlet-dominated pseudoscalar with $\mathcal{P}_{13}\approx 1$ and $\mathcal{P}_{11},\mathcal{P}_{12}\ll 1$, we find $c_{aa\tilde{N}\tilde{N}}\approx -2 \lambda_N(2\lambda_N+\kappa)$. The cross section does not exceed that required to obtain the observed DM adundance when $\lambda_N \lesssim 0.05$ for $\kappa\approx 1$, $m_{\tilde N}\approx100 \GeV$ and $m_a\approx 90 \GeV$. This bound is relatively independent of the values of $m_{\tilde N}$ and $m_a$ in the region of interest or becomes stronger (provided that the masses are not too close to the threshold, in which case \eqref{Sneutoa1} becomes invalid). This satisfies the bound above required for SM channels to be subdominant.
A similar argument for the case where the $s$-channel process dominates the cross section gives a comparable limit on $\lambda_N$.

Via $c_{a\tilde N_I \tilde N_R}$, the third term in \eqref{Sneutoa1} depends upon $\lambda_N A_{\lambda_N}$ multiplied by the large singlet component of the $a$ (see Eq.~\eqref{cNNa}).  
Since we fix $\lambda_N A_{\lambda_N}$ in terms of the sneutrino mass $m_{\tilde{N}_R}$ (see Eq.~\eqref{AlambdaN}), we can obtain a constraint on this mass from the requirement that the third term in \eqref{Sneutoa1} does not exceed the required freeze-out cross section. 
To this end, we take the limit $\lambda_N\rightarrow 0$ keeping $\lambda_N A_{\lambda_N}$ fixed (in which case the interference from the other amplitudes is negligible) and assume that 
$m_{\tilde{N}_R}^2\gg m_{\tilde N_I}^2,m_a^2$
so that the terms proportional to $m_{\tilde{N}_R}$ dominate.
In this limit, 
\begin{equation}
|c_{a\tilde{N}_I\tilde{N}_R}| \, \approx \, \mathcal{P}_{13} \frac{m_{\tilde N_R}^2}{\sqrt{8}\,  v_s}
\end{equation}
and the sneutrino annihilation cross-section is
\bea
\langle\sigma_{aa} v\rangle_{T_F} \, \approx \, \frac{\sqrt{1-m_a^2/m_{\tilde{N}}^2}}{4\, \pi \, m_{\tilde{N}}^2} \, \, \, \left(\frac{ \mathcal{P}_{13} m_{\tilde N_R}}{4\, v_s}\right)^4\, .
\eea
The cross section does not exceed that required to obtain the observed DM abundance when 
$m_{\tilde{N}_R}\lesssim 0.75 v_s/|\mathcal{P}_{13}|$ for $m_{\tilde{N}_I}\approx 150\,\GeV$ and $m_a \approx 0.9\, m_{\tilde{N}_I}$. This is roughly independent of $m_{\tilde{N}_I}$ and $m_a$ within the best-fit range.  
We have $v_s\lesssim 500 \GeV$ in the natural region of parameter space, so if $|\mathcal{P}_{13}|\gtrsim 0.9$, then $m_{\tilde{N}_R}\lesssim 400 \GeV$. We caution the reader again that this analysis is only approximate, but reasonably describes our target parameter space.

We have performed a random scan of the parameter space of the scale-invariant NMSSM extended by right-handed neutrinos. \texttt{NMSSMTools\;4.3.0} and \texttt{micrOMEGAs 3.0} were used to calculate the spectrum, phenomenological constraints, DM signals and relic abundance. The code was modified to calculate the relic abundance of the sneutrino LSP, using modified model files for \texttt{CalcHEP} \cite{Belyaev:2012qa} generated with \texttt{LanHEP} \cite{Semenov:2008jy}. All calculations of neutrino/sneutrino cross sections were performed at tree-level. The interactions mediated by the coupling $y_{_N}$ were omitted as they contribute negligibly for all processes of interest here. The ranges for the parameters used in the scan are given in Table \ref{SneuTab}. 
\begin{table}[btp]\centering
\setlength{\tabcolsep}{3.2pt}
\begin{tabular}{ l  c  c  c cccccc}
\toprule
&$A_\lambda$ [GeV] &$A_\kappa$  [GeV] &$|\mu_{\rm eff}|$  [GeV]  &$\lambda$ &$\kappa$ &$\tan \beta$ &$\Delta m_{\tilde{N}_I}$  [GeV] &$\lambda_N$ &$m_{\tilde{N}_R}$  [GeV] \\
\midrule
 &  [-100,100] & [-10,10] & [200,500]  & [0.5,1.5] & [0.2,1.4]  & [1.2,4] & [0,30] & [-0.1,0.1] & [$m_{\tilde{N}_I}$,400] \\
\bottomrule
\end{tabular}
\caption{
Parameter ranges for the random scans for the $aa$-channel, where $\Delta m_{\tilde{N}_I}\equiv m_{\tilde{N}_I} - m_a$.
Both signs of $\mu_{\rm eff}$ are scanned over.}
\label{SneuTab}
\end{table}
To concentrate on a light pseudoscalar, we have only scanned over relatively small values for $A_\lambda$ and $A_\kappa$. According to the heuristic analysis above, we have restricted $|\lambda_N|<0.1$ and $m_{\tilde{N}_R}<400 \GeV$. To ensure that the LSP mass lies in the best-fit region of the $aa$-channel as in Fig.~\ref{fig:fit-regions-aa}, $m_{\tilde{N}}$ was restricted to uniformly random values between $[m_a,m_a+30\,\text{GeV}]$, up to a maximum of $150$ GeV. The remaining NMSSM parameters were fixed as in Sec.~\ref{ParameterScans} with the addition that $M_1=2$ TeV. 
Likewise, collider and flavour constraints were also imposed as in Sec.~\ref{ParameterScans}. We also demand, as before, that the LSP has the necessary relic abundance to entirely account for DM, satisfies the constraints imposed by the LUX experiment and has an annihilation cross section consistent with the GCE.  
As we find $\lambda\lesssim 1.1$ for all points presented and no clear preference for $\tan\beta>3$, we expect all points to be consistent with electroweak precision tests and have refrained from explicitly checking this (see also Ref.~\cite{Gherghetta:2012gb}). We emphasise that our approach is not an exhaustive scan of the parameter space, but rather an illustration of how this model provides a natural, supersymmetric account of the GCE.

\begin{figure}[btp]\centering
\begin{subfigure}{0.49\linewidth}
\includegraphics[width=\linewidth]{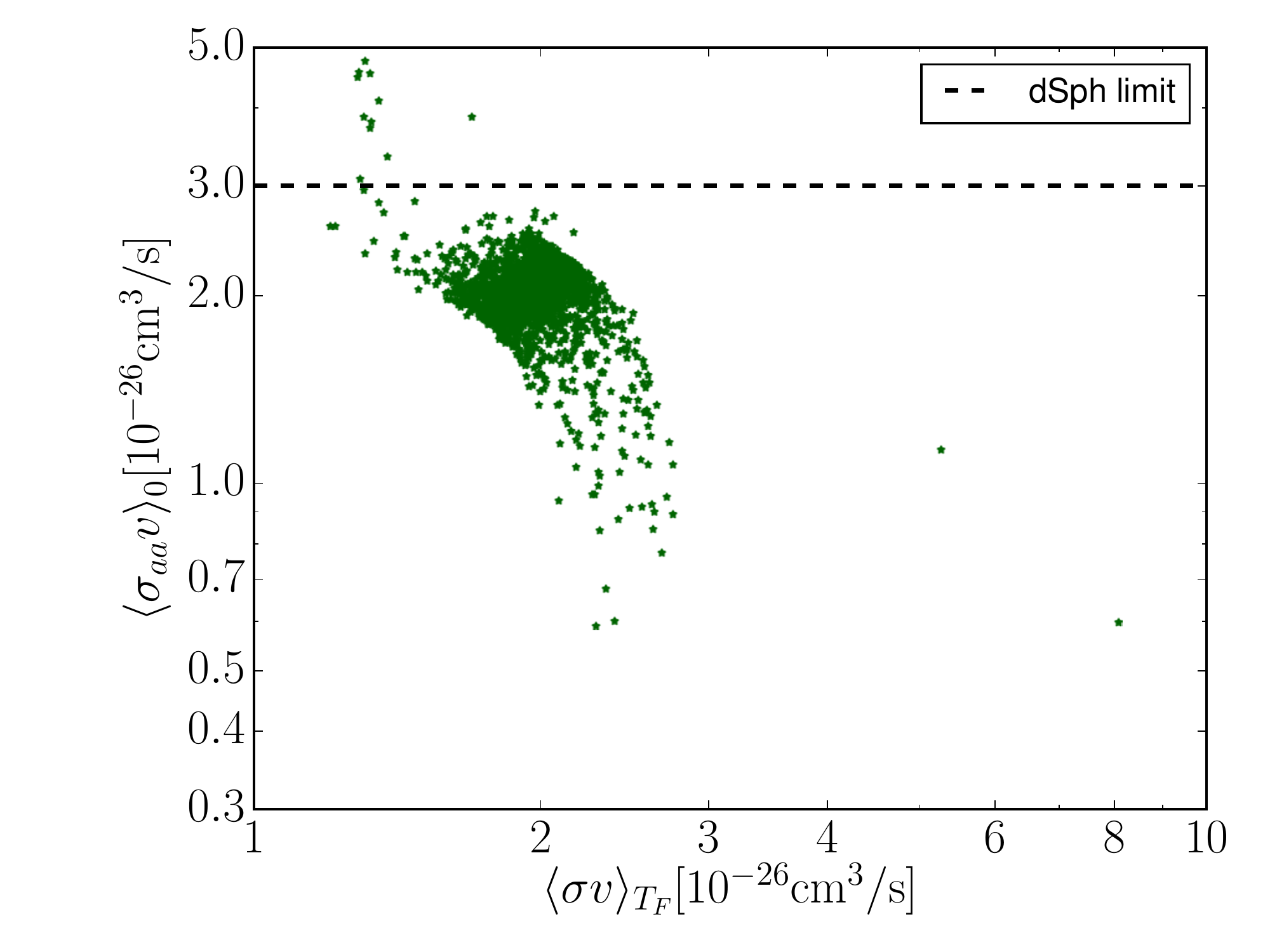}
\caption{\label{fig:Sneu60a}}
\end{subfigure}
\hfill
\begin{subfigure}{0.49\linewidth}
\includegraphics[width=\linewidth]{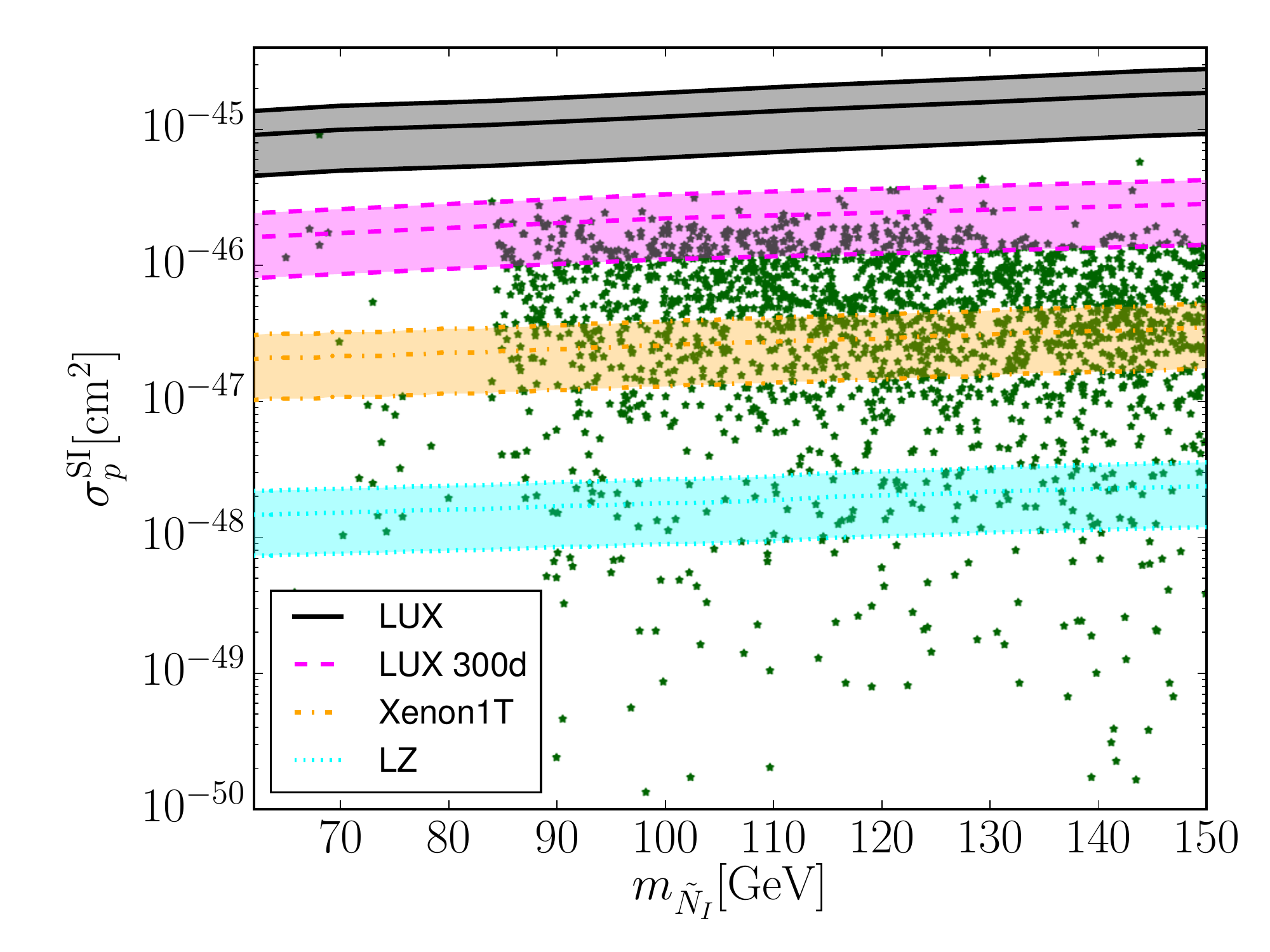}
\caption{\label{fig:Sneu60b}}
\end{subfigure}
\caption{Results from the random scan with parameters given in Table \ref{SneuTab}. The left panel (a) shows the annihilation cross section at the present time, $\langle \sigma_{aa} v\rangle_0$, and at freeze-out, $\langle\sigma v \rangle_{T_F}$. The right panel (b) shows the spin-independent cross section for scattering of LSPs off nucleons versus the LSP mass.}
\label{fig:Sneu60}
\end{figure}

The results from our scan are presented in Fig.~\ref{fig:Sneu60}. All points have a branching fraction for annihilations into $aa$ of at least 80\%. As suggested from the above analysis, most viable points lie in the range $|\lambda_N|\lesssim 0.04$. The appeal of this scenario is that the cross-sections are not severely constrained by experiments and are not dependent on special effects, like a resonance. The latter point becomes clear in Fig.~\ref{fig:Sneu60a}, where we plot the cross-sections at freeze-out and late-times. These are of comparable size for the vast majority of points. 
We find that the few outlying points with large $\langle \sigma_{aa} v\rangle_0\gtrsim 3\times 10^{-26}\text{cm}^3/\text{s}$ have smaller cross-sections in the early universe because the total freeze-out cross section has additional contributions from coannihilation with $\tilde{N}_R$,
while the points with larger $\langle \sigma v\rangle_{T_F}$ are resonantly enhanced by a singlet-like Higgs scalar. Just as for the case of the neutralino annihilating into $ha$, the typical cross sections for most of the points are close to half of that prescribed for the best-fit regions in Fig.~\ref{fig:fit-regions-aa} and therefore require an astrophysical-uncertainty factor $\mathcal{A}\sim 2$ to explain the GCE. We have plotted our estimate for the limit on $\langle \sigma_{aa}v \rangle_0$ from the latest Fermi-LAT study of dwarf spheroidals, that we have discussed in Sec.~\ref{GCE-section}, as a dashed line in Fig.~\ref{fig:Sneu60a}. The vast majority of points satisfies this limit. We show a scatter plot of the cross section for spin-independent scattering of LSPs off nucleons versus the LSP mass in Fig.~\ref{fig:Sneu60b}. As discussed in Sec.~\ref{Higgs-pseudoscalar-channel} for Fig.~\ref{fig:HAscanLSP2}, we plot the current limits from LUX and the projected limits from LUX with 300 live days of measurements~\cite{LUXtalk}, XENON1T~\cite{XENON:2013SnowMass} and LZ~\cite{LZ:2013SnowMass} as shaded bands for local DM densities between $\rho_\odot=0.2 \GeV/\text{cm}^3$ and $\rho_\odot=0.6 \GeV/\text{cm}^3$. Notice that the vast majority of points would satisfy the projected, updated LUX bounds, whereas only very few would evade the LZ bounds.

\subsection{Metastable sneutrino NLSP}
\label{SneutrinoNLSP}

As described in Sec.~\ref{BinoResonance}, the pseudoscalar resonance can enhance or suppress the annihilation cross-section of the neutralino LSP in the early universe compared to that at late times. For regions of parameter space that explain the GCE via the $b \bar b$-channel and satisfy $m_a>2m_{\tilde{\chi}}$, the thermal relic abundance is then too small to account for the DM content of our universe. 
We now discuss the possibility that a metastable right-handed sneutrino NLSP freezes-out and decays into the neutralino LSP to produce the observed DM abundance.
This realises a similar simplified DM scenario proposed in \cite{Fairbairn:2008fb}, albeit in a different context.

We will again restrict our discussion to one flavour of right-handed neutrinos and will neglect flavour mixing. It will turn out that the suppression of flavour mixing is a necessary consequence of the requirement of metastability of the sneutrino.   
Furthermore, we will continue to consider only the $CP$-odd mass eigenstate of the sneutrino. Another qualitatively similar possibility that we will not explore is that the $CP$-even or both states are metastable.

The right-handed sneutrino NLSP freezes-out before the LSP does 
because of the smaller LSP mass.
In order for the NLSP to contribute to the LSP relic abundance, it needs to decay after the LSP freezes-out. 
In addition, the NLSP lifetime must be sufficiently long to ensure that LSP annihilations do not resume once these are repopulated (the ``freeze-out and decay'' scenario in \cite{Cheung:2010gj}). 
Neglecting changes in entropy due to the late sneutrino decays, the total density fraction of DM is then given by
\begin{equation}
\Omega_{\rm DM} \, = \, \Omega_{\tilde{\chi}} \, + \, \frac{m_{\tilde{\chi}}}{m_{\tilde{N}}} \, \Omega_{\tilde{N}} \, .
\label{RelicAbundancesRelation}
\end{equation}
Here $\Omega_{\rm DM}$ is the DM density fraction, $\Omega_{\tilde{\chi}}$ is the density fraction of LSPs produced as a thermal relic and $\Omega_{\tilde{N}}$ is that of NLSPs expected today had they not decayed. For the points of interest from Sec.~\ref{BinoResonance}, $\Omega_{\tilde{\chi}}\ll\Omega_{\tilde{N}}$.

The requirement that the sneutrino be long-lived places an upper bound on the coupling $y_{\!_N}$ that determines the non-conservation of 
the approximate $\mathbb{Z}_2$ in the right-handed neutrino sector. We will investigate this bound below. However, $\mathbb{Z}_2$-preserving decays into the right-handed neutrino, such as $\tilde{N} \rightarrow \tilde{\chi} + N$, are controlled by the coupling $\lambda_N$. This cannot be much smaller than the electroweak gauge couplings for a sneutrino of mass $\sim \GeV - \TeV$ to freeze-out with a relic abundance that reproduces the observed amount of DM.\footnote{We do not consider alternative non-thermal production possibilities that involve a very small $\lambda_N$ such that the sneutrino is never in thermal equilibrium with the plasma to begin with, see~\cite{Medina:2014bga} for an example in the MSSM.} 
These fast decays must be kinematically forbidden if the sneutrino is to be long-lived and therefore the sneutrino mass must satisfy
\be
m_{\tilde{N}} \, < \, m_{\tilde{\chi}} \, + \, m_N  \, = \,  m_{\tilde{\chi}} \, + \, \frac{2 \, \lambda_N  \mu_{\rm eff}}{\lambda} \, .\label{metastable}
\ee 
The neutralino mass is $m_{\tilde \chi}\sim 30-50$ GeV in the parameter region of interest for the $b \bar b$-channel (see Sec.~\ref{BinoResonance}).
As we focus on $\lambda \sim 1$ to improve naturalness, we therefore require that $m_{\tilde{N}} \lesssim \mu_{\rm eff}$, taking $\lambda_N\sim\mathcal{O}(1)$, where $\mu_{\rm eff}$ is typically of order a few hundred GeV. 
The dominant sneutrino decay mode then is $\tilde{N}\rightarrow \tilde{\chi} + \nu$, which has a rate
\be
\Gamma(\tilde{N}\rightarrow\tilde{\chi}+\nu) \, = \, \frac{|c_{\tilde{N}\tilde{\chi}{\nu}}|^2}{8\pi }m_{\tilde{N}}\left(1-\frac{m_{\tilde{\chi}}^2}{m_{\tilde{N}}^2}\right)^2 \, ,
\label{SneutrinoDecayRate}
\ee
where 
\be
c_{\tilde{N}\tilde{\chi}{\nu}}=\frac{y_{\!_N}}{\sqrt{2}} \mathcal{N}_{14}-\frac{1}{2}g_1 \mathcal{N}_{11} N^{\nu}.\label{fastcoup}
\ee
The lightest sneutrino $\tilde{N}$ has a left-handed sneutrino component given by
\be
N^{\nu}  \, \approx \, (2\lambda_N v \, v_s\sin\beta+ v(\lambda v_s\cos\beta-A_{{y_{\!_N}}}\sin\beta)) \, \frac{y_{\!_N}}{m^2_{\tilde{N}}-m_{\tilde{\nu}}^2} \, ,
\ee
where $m_{\tilde{\nu}}$ is the mass of the left-handed sneutrino. The mixing factor given by the second term in (\ref{fastcoup}) is therefore also directly proportional to $y_{\!_N}\ll 1$.

To ensure that the replenished LSPs remain frozen out of chemical equilibrium, the Hubble rate must be faster than the annihilation rate of the LSPs as the sneutrinos decay. Calling 
$T_d$ the temperature at which the sneutrino decays with its mean lifetime, this requires that
\be
\langle\sigma v\rangle_{T_d} \, Y_\infty \, s(T_d)  \, \lesssim \, H(T_d) \, .
\label{SneuBound}
\ee 
Here, $\langle\sigma v\rangle_{T_d}$ is the thermally-averaged annihilation cross section of the LSPs evaluated at temperature $T_d$, $Y_\infty\equiv n_{\tilde{\chi}}(T_0)/s_0$ where $n_{\tilde{\chi}}(T_0)$ is the present-day LSP number density and $s_0=2.22\times 10^{-38}$ GeV$^3$ the present-day entropy density of the universe and $H(T_d)=\pi \sqrt{g_*/90}\,T^2_d/M_{\rm Pl}$ and $s(T_d)=(2\pi^2/45) g_{*} T_d^3$ are respectively the Hubble rate and entropy density at the time of the decays with $g_{*}$ the active relativistic degrees of freedom.  
The time interval between sneutrino freeze-out and decay could encompass the QCD phase transition, over which the number of degrees of freedom in the plasma must be carefully treated. We include this temperature dependence in our analysis, but for simplicity will treat this as a constant $g_{*}$ in the analysis presented below, which is sufficient for an order of magnitude estimate.\footnote{Note that the case of equality in (\ref{SneuBound}) approximately describes the special possibility of ``reannihilation'' \cite{Cheung:2010gj}. In this scenario, as the sneutrinos decay, the neutralino number density grows sufficiently for annihilations to resume. These annihilations continue rapidly until the decays finish, after which the neutralinos quickly freeze-out again when their number density falls to the critical level given by the equality in (\ref{SneuBound}) (technically the temperature of LSP freeze-out is actually slightly lower than that corresponding to the sneutrino lifetime). The resulting neutralino relic abundance is approximately independent of the original sneutrino number density, provided that the latter is at least as large, because the surplus neutralinos rapidly reannihilate.}

The inequality \eqref{SneuBound} provides an upper bound on the decay temperature $T_d$ of the sneutrinos:
\be
T_d \, \lesssim \, \sqrt{\frac{45}{8 \pi^2 g_*}} \frac{1}{Y_\infty \, M_{\rm Pl} \, \langle\sigma v\rangle_{T_d} } \, .
\label{TdBound}
\ee
Note, however, that $\langle\sigma v\rangle_{T_d}$ is itself a function of the temperature $T_d$. Indeed, for the points in parameter space that we consider here, the pseudoscalar resonance results in an enhancement of the annihilation cross section during freeze-out compared to late-times. 
The cross section decreases at temperatures below the LSP freeze-out temperature. Using parallel analysis to that of \cite{Gondolo:1990dk}, the cross section may be well-approximated in the vicinity of the resonance as
\be
\langle\sigma v\rangle_{T} \, \simeq \, \frac{|c_{a\tilde{\chi}\tilde{\chi}}|^2|c_{ab\bar b}|^2}{4\sqrt{\pi} \, m_a\Gamma_a}\Big(\frac{m_{\tilde{\chi}}}{T}\Big)^{\frac{3}{2}}\,\text{Re}\Big(\sqrt{z_R} \, e^{-xz_R} \, \text{erfc}\big(-i\sqrt{m_{\tilde{\chi}}z_R/T}\big)\Big) \, .
\ee
Here $z_R=\epsilon_R+i\gamma_R$ with $\epsilon_R=(m_a^2-4m_{\tilde{\chi}}^2)/4m_{\tilde{\chi}}^2$ and $\gamma_R=m_a\Gamma_a/4m_{\tilde{\chi}}^2$, and $\Gamma_a$ is the decay width of the pseudoscalar.
Using this, comparison of (\ref{SneuBound}) at different temperatures reveals that it is the closing of the thermal-broadening of the resonance with the cooling of the plasma that determines the temperature after which the neutralinos will remain frozen-out. From Sec.~\ref{BinoResonance}, typical values are $\epsilon_R\sim 0.015$, $\Gamma_a\sim 1 \MeV$, $|c_{ab \bar b}|\sim 0.05$ and $|c_{a\tilde{\chi}\tilde{\chi}}|\sim 0.01$ (where the cross section is most sensitive to $\epsilon_R$ and $\Gamma_a$). Equation~\eqref{TdBound} then reduces to $T_d\lesssim 0.1 \GeV$. This bound can weaken by up to an order of magnitude if the masses are further from resonance, but is otherwise relatively insensitive to variations in the parameters over their ranges considered in our scan.

This bound on the decay temperature $T_d$ corresponds to a bound on the decay rate $\Gamma_{\tilde{N}}$ of the sneutrino. Denoting the time and temperature of sneutrino freeze-out respectively by $t_{\tilde{N},F}$ and $T_{\tilde{N},F}$ and the time of sneutrino decay by $t_d$, we have
\be
\frac{1}{\Gamma_{\tilde{N}}} \, = \, t_d-t_{\tilde{N},F} \, = \, \sqrt{\frac{90}{g_*}} \frac{M_{\rm Pl}}{\pi} \, \left(\frac{1}{T_d^{2}} -\frac{1}{T_{\tilde{N},F}^{2}}\right).\label{SneuDecRate}
\ee
With $T_{\tilde{N},F} \sim m_{\tilde{N}}/25\sim \text{few GeV}$, the bound $T_d\lesssim 0.1 \GeV$ gives $\Gamma_{\tilde{N}}\lesssim 10^{-20} \GeV$.
Using \eqref{SneutrinoDecayRate} and neglecting the second term, this translates into $y_{\!_N}\lesssim 10^{-8}$ for $m_{\tilde{N}}\approx 1.1\cdot m_{\tilde{\chi}}$ and $|N_{14}|\approx 0.01$ (as we typically find for a bino-like LSP, see Sec.~\ref{CGENMSSM:Composition}), with a stronger bound for larger sneutrino mass. 
To ensure that the decays occur before big bang nucleosynthesis, we require that $T_d\gtrsim 1\,\text{MeV}$. This corresponds to $\Gamma_{\tilde{N}} \gtrsim 10^{-25}\GeV$ and  $y_{\!_N}\gtrsim 10^{-12}$.\footnote{As this is peripheral to our analysis, we do not carefully consider the possibility that big bang nucleosynthesis may be compatible with an epoch of decaying sneutrinos.}

Recall that we have restricted our discussion to one flavour of right-handed sneutrinos. 
In general, the sneutrino may decay into any of the active neutrino flavours, in which case the decay width \eqref{SneutrinoDecayRate} and coupling \eqref{fastcoup} generalise and the bound on $\Gamma_{\tilde{N}}$ derived above applies to the sum of partial decay widths for each flavour. The bound on the coupling $c_{\tilde{N}\tilde{\chi}\nu}$ which follows from \eqref{SneuDecRate} then applies to the sum of each of these generalised couplings added in quadrature. The resulting bound on the coupling $y_{\!_N}$ derived above then applies to each of the Yukawa couplings of the superfield containing the NLSP to the active neutrino flavours. This implies that the neutrino superpartner of the NLSP cannot significantly contribute to the mass generation of the active neutrinos. The other two flavours then need to have corresponding couplings $y_{\!_N} \sim 10^{-6}$ in order to reproduce the known mass splittings among the active neutrinos. Furthermore, this implies that flavour mixing between the NLSP and the other two right-handed sneutrino flavours must also be sufficiently suppressed to prevent premature decays. Note that this includes the suppression of flavour mixing that originates from the couplings $\lambda_N$ and $A_{\lambda_N}$ and the masses $m_{\tilde{N}}^2$. This means that the right-handed neutrino superfield that contains the NLSP is effectively decoupled from the other two flavours and the lepton doublets, thereby justifying in hindsight our restriction to one flavour.
The $\mathbb{Z}_2$ parity mentioned in Sec.~\ref{NeuIntro} may be restricted to this decoupled flavour.

\begin{figure}[bt]
\centering
\includegraphics[width=.45\linewidth]{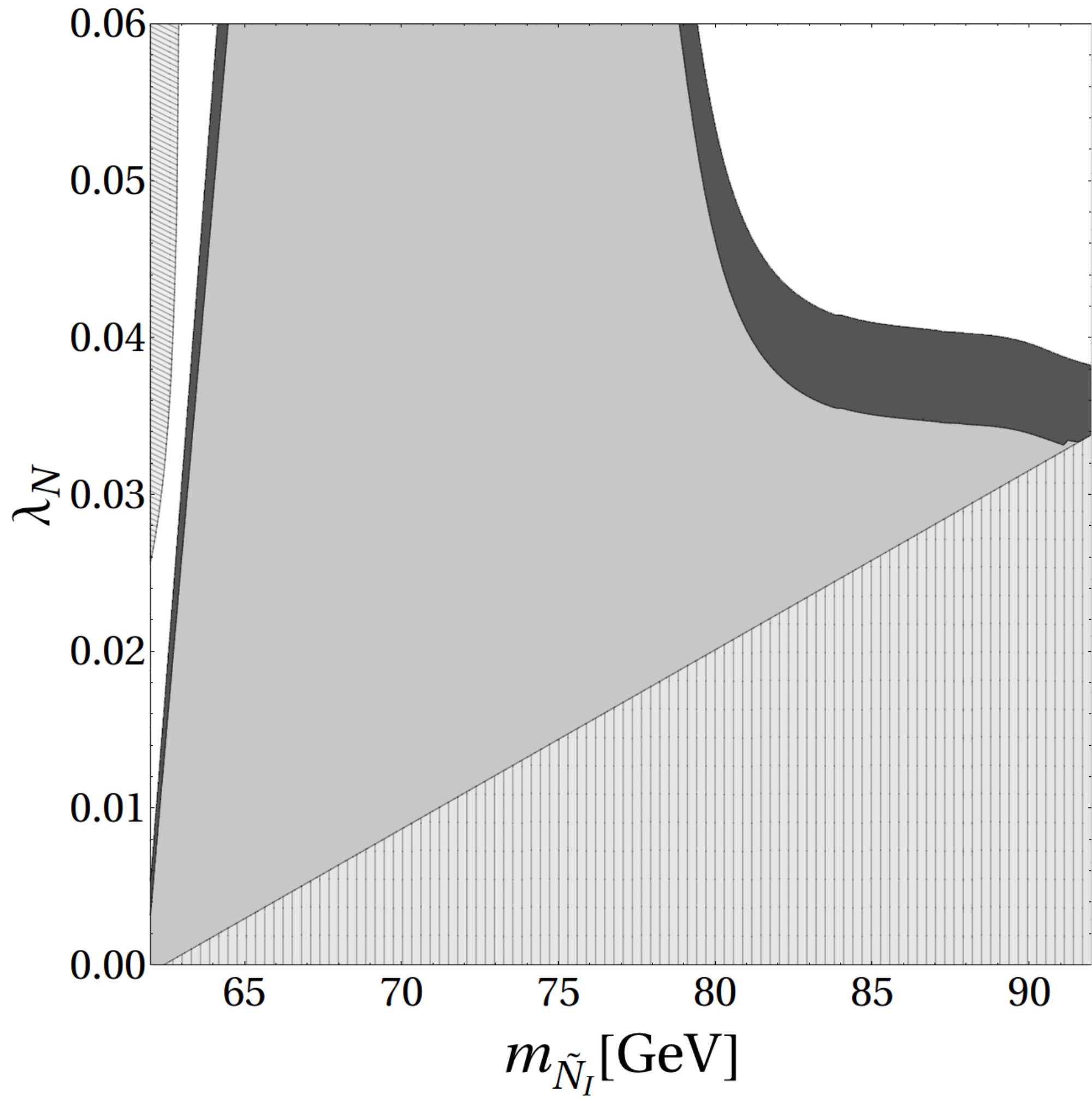}
\caption{Scan over $\lambda_N$ and $m_{\tilde{N}}$ for an LSP of mass $63 \GeV$. The NMSSM parameters are selected as $\tan\beta=3$, $\lambda=0.8$, $\kappa=1$, $\mu_{\rm eff}=-350 \GeV$, $A_{\lambda}=-10 \GeV$, $A_\kappa=12.86 \GeV$ and $M_1=60 \GeV$, which give $m_a=125.3 \GeV$. The remaining sparticle and gaugino masses were fixed as in Sec.~\ref{BinoResonance} and we set $m_{\tilde{N}_R}=300 \GeV$. 
This region in parameter space has a bino-like LSP which can explain the GCE but would be underabundant in the absence of the sneutrino NLSP. Dark grey denotes regions where the sneutrino freezes-out with a relic density that produces $\tilde{\chi}$ DM consistent with the density determined by Planck. The sneutrinos are underproduced in the white regions and overproduced in the inner light grey regions. The shaded wedge violates the metastability condition, while the invisible branching fraction of the Higgs exceeds the experimental bound in the shaded slice in the upper-left corner.
}\label{fig:MetaSneu}
\end{figure}

The freeze-out and decay of the $\tilde{N}$ may be complicated by its interactions with the $N$ and the $\tilde{\chi}$. Furthermore, the $N$ may also be metastable and freeze-out with the $\tilde{N}$ and $\tilde{\chi}$, making this an effectively multi-component DM model. Note that the $\tilde{N}_R$ can similarly complicate the analysis if it is metastable. However, here we will assume that its mass is sufficiently high so that it decays on short timescales and can be neglected. The behaviour of the $N$ must be carefully treated. If the mass of the $N$ satisfies $m_N>m_{\tilde{N}}+m_{\tilde{\chi}}$ then it decays rapidly into the $\tilde{N}$ and $\tilde{\chi}$ and plays no role in DM production. This occurs for large $\mu_{\rm eff}$ or $\lambda_N$
(for example, for the parameter values used in the scan in Fig.~\ref{fig:MetaSneu}, the right-handed neutrino is short-lived and the sneutrino freezes-out with the necessary abundance for $\lambda_N\approx-0.15$ and $m_{\tilde{N}}\approx 69$ GeV). However, if these fast decays are kinematically forbidden, the right-handed neutrino will also be metastable and have a lifetime similarly determined by the small coupling $y_{_N}$. 
This can complicate the production of DM in several ways. 

Firstly, the self-annihilation cross-section of the right-handed neutrino is typically smaller than that of the sneutrino by one or two orders of magnitude. The neutrino will therefore freeze-out with a number density greater than the sneutrinos by roughly the same factor (assuming that the $s$-wave approximation holds). If kinematically permitted, a non-negligible fraction of these can decay into pairs of neutralinos through, for example, the decay chain $N\rightarrow \nu a$ followed by $a\rightarrow\tilde{\chi} \tilde{\chi}$, thereby contributing considerably to the LSP density. A possibility that we leave for future work is that the LSPs are entirely produced by the decaying neutrinos while the sneutrinos freeze-out with negligible density.

Secondly, the $N$ and the $\tilde{N}$ could affect their respective relic densities through ``semi-annihilation'' processes such as $\tilde{N}N\rightarrow \tilde{\chi}a$ that could keep them in equilibrium for longer.\footnote{The destruction of a single $\tilde{N}$ must either consume or produce a right-handed neutrino or be suppressed by the small coupling $y_{\!_N}$. Due to $R$-parity conservation, the $\tilde{\chi}$ or a heavier $R$-odd particle must also be produced or consumed in such a reaction. Hence our use of terminology - the $\tilde{N}$, $N$ and $\tilde{\chi}$ may be regarded as particles in a $\mathbb{Z}_2\times\mathbb{Z}_2$ dark sector.} If these processes are significant, the departure of the $N$ from chemical equilibrium during the freeze-out of the $\tilde{N}$ would inextricably couple the dynamics of the thermal production of both species. In this case, a reliable calculation of the relic abundance of the sneutrino would require solving a pair of coupled Boltzmann equations. This is beyond the scope of this paper. Let us, however, point-out that the relic abundance of the  $\tilde{N}$ can be reliably calculated from the self-annihilation rate alone if the $N$ freezes-out at a lower temperature and the ``semi-annihilation'' rate of the $\tilde{N}$ with the $N$ (and, for that matter, with the $\tilde{\chi}$) is subdominant. Although the $N$ and $\tilde{\chi}$ are lighter and hence much more abundant than the $\tilde{N}$, the semi-annihilation rates may be nevertheless mildly suppressed compared to the self-annihilation rate of the sneutrinos. This is both because the cross sections are smaller (typically suppressed relative to the self-annihilation cross-section by the small singlino fraction of the neutralino) and the metastability condition \eqref{metastable} in conjunction with the condition $m_a\approx 2m_{\tilde{\chi}}$ implies that the reaction is often close to or below threshold.

We have performed a scan in the sneutrino parameter space for an example point in the NMSSM for which a bino-like LSP could be responsible for the GCE but would otherwise be thermally underproduced (see Sec.~\ref{BinoResonance}). 
We have used \texttt{NMSSMTools~4.3.0} and \texttt{micrOMEGAS~3.0} with the same set-up as in Sec.~\ref{SneutrinoLSP} to calculate the relic abundance of sneutrinos under the assumption that these particles are stable (again neglecting $y_{_N}$-mediated interactions and assuming that these are small enough to satisfy the criterion above). However, this calculation incorporates only self-annihilations in the Boltzmann equations (and coannihilation with heavier $R$-odd particles) and does not include the possibility of semi-annihilations. 
To check for the latter, we have compared the semi-annihilation rates for the reactions $\tilde{N}  N\rightarrow\tilde{\chi} a\,(\text{or}\,h)$ and $\tilde{N}\tilde{\chi}\rightarrow N  a\,(\text{or}\,h)$ (assuming that the $N$ is still in chemical equilibrium during the freeze-out of the $\tilde{N}$) to the self-annihilation rate of the $\tilde{N}$ at the freeze-out temperature computed by \texttt{micrOMEGAS~3.0}, using cross sections calculated by \texttt{CalcHEP} that were then thermally averaged. Likewise, the $N$ self-annihilation rate was computed and compared to verify the assumption that it was in equilibrium at the beginning of freeze-out of the $\tilde{N}$. Where this holds and the semi-annihilation rates are smaller than the sneutrino self-annihilation rate by at least a factor of $10$, we can safely neglect the impact of the $N$ and $\tilde{\chi}$ on the freeze-out of the $\tilde{N}$ (the number densities of the lighter $N$ and $\tilde{\chi}$ will decrease faster than that of the $\tilde{N}$ whether in equilibrium or freezing-out, so the semi-annihilation rates will also diminish faster as the temperature falls). We provide an example region where this occurs in Fig.~\ref{fig:MetaSneu} as a proof of principle. Over this region, the $N$ is metastable and freezes-out after the $\tilde{N}$, while the semi-annihilations are weak enough to be ignored. Furthermore, its mass satisfies $m_N<2m_{\tilde{\chi}}$ so that decays into the LSP are kinematically forbidden. The mass of the $\tilde{N}_R$, on the other hand, is large enough to ensure that it is short-lived by $\lambda_N$-mediated decays. 
Invisible Higgs decays into sneutrinos or right-handed neutrinos can provide a further constraint on the parameter space if these species are light enough. As before, we require that the branching fraction for invisible Higgs decays satisfies ${\rm Br}(h\rightarrow {\rm inv}) \lesssim 0.24$~\cite{Belanger:2013xza,Bechtle:2014ewa}, which we compute at tree-level.

We also check the possibility of conversion of the $\tilde{N}$ into a $\tilde{\chi}$ by $\mathbb{Z}_2$-violating annihilation with active neutrinos $\nu$ mediated by the small coupling $y_{\!_N}$. Despite this small coupling, the much greater abundance of the $\nu$ could enhance the rate for this process relative to the self-annihilation rate of the $\tilde{N}$. The largest of the processes mediated by the couplings $y_{\!_N}$ is $\tilde{N} \nu\rightarrow\tilde \chi a$, where the $a$ may be on- or off-shell depending upon the total centre-of-mass energy. However, a rough estimate indicates that this process can be safely neglected. Indeed, the sneutrino number density in the non-relativistic regime in equilibrium is $n_{\tilde{N}}\sim (m_{\tilde{N}} T)^{\frac{3}{2}} \exp(-\frac{m_{\tilde{N}}}{T})$ and the left-handed neutrino number density is $n_\nu\sim T^3$. Ignoring this process, freeze-out occurs at temperature $m_{\tilde{N}}/T\sim 20-30$, so $n_\nu/n_{\tilde{N}}\lesssim 10^{11}$. 
An order-of-magnitude estimate for the annihilation cross-section for the process $\tilde{N} \nu\rightarrow\tilde \chi a$ gives $\langle\sigma v\rangle\sim y_{\!_N}^2/m^2$, where $m\sim 10-100\,\GeV$ is a characteristic energy scale and $ y_{\!_N}^2\lesssim 10^{-16}$.
Noting that the sneutrino self-annihilation cross-section at freeze-out is $\sim  10^{-9}\,\GeV^{-2}$ (as required to obtain the right DM abundance), the ratio of the annihilation rate for $\tilde{N}$ with $\nu$ over that for $\tilde{N}$ with $\tilde{N}$ is $\sim y_{_N}^2 10^{20}/m^2\lesssim 10^4/m^2$. This shows that the process $\tilde{N} \nu\rightarrow\tilde \chi a$ may be significant only close to the upper bound on $y_{_N}$ determined above, but negligible for smaller values (i.e. $y_{_N}\lesssim 10^{-9}-10^{-10}$). However, if this reaction is below threshold, which is typically true given the metastability condition \eqref{metastable} and the comparable masses of the $N$ and the $\tilde\chi$ in the region where the $\tilde{N}$ freezes-out with the required abundance, the thermally-averaged cross-section will be much more suppressed. We do not seek to determine this bound on $y_{_N}$ more precisely as it is clear that, at worst, this is only an order of magnitude stronger than that derived above and there is always available parameter space where $y_{_N}$ can be made sufficiently small.

\section{Conclusions}\label{Conclusions}

An excess in the $\gamma$-ray spectrum originating from the galactic centre has been observed in the Fermi-LAT data~\cite{Hooper:2013rwa,Huang:2013pda,Gordon:2013vta,Abazajian:2014fta,Daylan:2014rsa,Calore:2014xka},  and was recently  confirmed by the Fermi-LAT collaboration~\cite{MurgiaTalk:2014FermiSymposium}. Notwithstanding a possible astrophysical origin for this excess, it is well-explained by DM annihilating into $b \bar b$-  or Higgs-pairs. Supersymmetric models provide a unifying framework to study the implications of DM annihilating via these channels since they contain a DM candidate and address many issues that arise beyond the SM. 

A suitable DM candidate in supersymmetric models is the neutralino. Since the annihilation of neutralinos into Higgs pairs is 
$p$-wave suppressed, we have considered the alternative channel of a Higgs and a Higgs-sector pseudoscalar.  The latter predominantly decays into $b \bar b$-pairs and thereby behaves similarly to the Higgs. We have performed a fit to the excess for this channel and found it to be just as good or even better than the $b \bar b$- and Higgs-channel for a relatively light pseudoscalar in the final state, with mass below about $120\,\GeV$. The quality of the fit prefers that the annihilation occurs as close to threshold as possible, but is typically acceptable to within $40\,\GeV$. The best-fit cross-section lies within $0.4- 30  \times 10^{-26}\,\text{cm}^3/\text{s}$, where the large range is mainly determined by uncertainties in the DM halo profile. 

In the MSSM the required light pseudoscalar is in strong tension with collider and flavour constraints.
We have therefore considered the NMSSM, which has an additional singlet superfield. Mixing of the singlet pseudoscalar with the MSSM pseudoscalar then eases the experimental constraints.
In addition, in the regime of a large Higgs-singlet coupling $\lambda$, the naturalness of the model is considerably improved. 
Direct-detection experiments limit the Higgsino fraction of the neutralino. Focusing instead on bino- and singlino-dominated neutralinos, we have performed a numerical scan over the NMSSM parameter space and found interesting regions that explain the excess via the Higgs-pseudoscalar channel. 
The points that we found do not rely on resonant enhancement of the annihilation cross-section, which was typically close to the canonical WIMP cross-section $\sim 2\times 10^{-26}\text{cm}^3/\text{s}$. However,  the masses $m_a$ and $m_{\tilde{\chi}}$ of the pseudoscalar and the neutralino have to be chosen such that the annihilation is close to threshold ($2 m_{\tilde{\chi}} \approx m_a +m_h$). Furthermore, to avoid large deviations in the Higgs couplings from SM values, we found it necessary to have an accidental cancellation in the off-diagonal element of the Higgs mass matrix. 
Interestingly, the parameter space for the bino-like neutralino will soon be tested by direct-detection experiments, while the singlino-like neutralinos might escape direct detection even at the next-generation experiments.

We have also performed a parameter scan for the $b \bar b$-channel, focusing on bino-dominated neutralinos. 
Due to direct-detection constraints on the coupling strength of the neutralino LSP to the Higgs sector,  the annihilation cross section is typically not large enough to generate a sufficient $\gamma$-ray flux for the excess. We have therefore relied on a light pseudoscalar in the $s$-channel with mass chosen such that the cross section is resonantly enhanced ($\smash{m_a \simeq 2 m_{\tilde{\chi}}}$; this mass relation may require a mild tuning). This results in a large increase of the cross section during freeze-out if $2m_{\tilde\chi}< m_a$. In this regime, all our points that could otherwise explain the excess therefore do not have a sufficient abundance to account for the observed amount of DM. 
In the opposite regime $2m_{\tilde\chi}> m_a$, we do find points which can explain both the excess and have a sufficient DM abundance. 
The vast majority of these points will be tested soon by the currently running direct-detection experiments and by improvements in constraining the invisible decay width of the Higgs.

We have also studied an extension of the NMSSM with right-handed neutrinos that can explain the neutrino masses via the seesaw mechanism. 
The lightest right-handed sneutrino is then an interesting alternative DM candidate. It is straightforward to arrange for the sneutrino to annihilate dominantly into pseudoscalar-pairs. We have performed a fit for this channel and again found it to be just as good or even better than the $b \bar b$- and Higgs-channel for a relatively light pseudoscalar, with mass less than about $120\,\GeV$. The best-fit cross-section is $0.3 - 30\times 10^{-26}\,\text{cm}^3/\text{s}$, where the large range is again dominated by uncertainties in the DM halo profile. Again it is preferred that the annihilation happens close to the threshold, typically within $60\,\GeV$. We have performed a parameter scan of this extended NMSSM and found regions of parameter space where this new channel can be realised. It provides an interesting explanation of the excess as it can be easily obtained 
without relying on resonant enhancement of the annihilation cross-section. Finally, we have also considered a model where a right-handed sneutrino NLSP is metastable and decays after the freeze-out of the neutralino LSP. The relic abundance of the neutralino LSP can then be regenerated through these decays. The points for the $b \bar b$-channel with an underabundant neutralino LSP can thereby be resurrected.

It is exciting that the $\gamma$-ray excess from the galactic centre can be explained by DM annihilating into quark or scalar pairs and that this is straightforward to implement in the NMSSM with improved electroweak naturalness. Parts of the viable parameter space that we have found will soon be explored in direct-detection experiments. In addition, the $\gamma$-ray line from scalar or pseudoscalar decays into photons may possibly be detected in the future. This would be an astonishing signal for supersymmetry from the sky.

\section*{Acknowledgements}
We thank Iason Baldes, Brian Batell, Nick Rodd, and Alfredo Urbano for useful discussions. This work was supported in part by the Australian Research Council. The work of TG was supported by the Department of Energy grant DE-SC0011842 at the University of Minnesota. We acknowledge the use of \texttt{matplotlib}~\cite{Hunter:2007}, \texttt{ipython}~\cite{PER-GRA:2007} and \texttt{Julia}~\cite{2012arXiv1209.5145B,2014arXiv1411.1607B}.

\appendix

\section{(S)Neutrino couplings \label{app:couplings}}
The relevant couplings in the right-handed neutrino sector are given by
\bea
c_{ \tilde{N}_R\tilde{\chi}_k N}  &=&  -\sqrt{2} \, \lambda_N \, \mathcal{N}_{k5}\\
c_{\tilde{N}_I \tilde{\chi}_k N} &=&  -i\sqrt{2} \, \lambda_N \, \mathcal{N}_{k5}\\
c_{aNN}  &=&  -i\sqrt{2} \, \lambda_N \, \mathcal{P}_{13}\\
c_{hNN}  &=&  -\sqrt{2} \, \lambda_N \, \mathcal{S}_{13}\\
c_{h\tilde{N}_R\tilde{N}_R}  &=&  -\sqrt{2} \, \lambda_N \, (\mathcal{S}_{13}(A_{\lambda_N}+4v_s\lambda_N+2\kappa v_s)-\lambda \, v \, (\sin\beta \, \mathcal{S}_{11}+\cos\beta \, \mathcal{S}_{12}))\\
c_{h\tilde{N}_I\tilde{N}_I}  &=&  \sqrt{2} \, \lambda_N \, (\mathcal{S}_{13}(A_{\lambda_N}-4v_s\lambda_N+2\kappa v_s)-\lambda \, v \, (\sin\beta \, \mathcal{S}_{11}+\cos\beta \, \mathcal{S}_{12}))\label{cNNh}\\
c_{hh\tilde{N}_R\tilde{N}_R}   &=&  -2 \, \lambda_N\, ((2\lambda_N+\kappa)\mathcal{S}_{13}^2-\lambda \, \mathcal{S}_{12}\mathcal{S}_{11})\\
c_{hh\tilde{N}_I\tilde{N}_I}  &=&  -2\, \lambda_N \, ((2\lambda_N-\kappa)\mathcal{S}_{13}^2+\lambda \, \mathcal{S}_{12}\mathcal{S}_{11})\label{chhNN}\\
c_{a a\tilde{N}_R\tilde{N}_R}  &=&  -2\, \lambda_N\, ((2\lambda_N-\kappa)\mathcal{P}_{13}^2+\lambda \, \mathcal{P}_{12}\mathcal{P}_{11})\\
c_{a a\tilde{N}_I\tilde{N}_I}  &=&  -2\, \lambda_N \,((2\lambda_N+\kappa)\mathcal{P}_{13}^2-\lambda \, \mathcal{P}_{12}\mathcal{P}_{11})\label{cNNaa}\\
c_{a\tilde{N}_I\tilde{N}_R}  &=&  -\sqrt{2}\, \lambda_N\, ((A_{\lambda_N}-2\kappa v_s)\mathcal{P}_{13}+\lambda\,  v\, (\mathcal{P}_{12}\cos\beta +\mathcal{P}_{11}\sin\beta)) \, .\label{cNNa}
\eea
Our notation follows \cite{Ellwanger:2009dp} and \cite{Cerdeno:2009dv}. For NMSSM-specific couplings, see \cite{Ellwanger:2009dp}.

\inputencoding{latin1}

\newcommand{\eprint}[1]{arXiv: \href{http://arxiv.org/abs/#1}{\texttt{#1}}}

\bibliography{Hooperon_FixedUmlaut}

\end{document}